\documentclass[12pt]{article}
\usepackage{times}
\usepackage{natbib}
\bibpunct{(}{)}{;}{a}{,}{,}
\usepackage{graphicx}
\usepackage{amsmath,amsfonts,amssymb,mathrsfs,amsthm,array}
\usepackage{alltt,color,url}
\usepackage[ruled,vlined]{algorithm2e}
\usepackage{xy}
\usepackage{epsfig, rotating}
\usepackage{bm}
\usepackage{rotating}
\usepackage{multicol}
\usepackage{titlesec}
\usepackage{latexsym}
\usepackage[pdftex,colorlinks=true,linkcolor=blue,citecolor=blue,urlcolor=blue,bookmarks=false,pdfpagemode=None]{hyperref}
\usepackage{url}
\usepackage[vlined]{algorithm2e}
\usepackage{rotating}
\usepackage{verbatim}
\usepackage{fancyhdr}
\usepackage{setspace}
\usepackage{paralist}
\usepackage{boxedminipage}
\usepackage{color}
\definecolor{lightgrey}{rgb}{0.85,0.85,0.85}
\usepackage{lineno}
\usepackage[top=1in, bottom=1in, left=1in, right=1in]{geometry}

\titlespacing{\section}{0pt}{*0.9}{*0.9}
\titlespacing{\subsection}{0pt}{*0.8}{*0.8}
\titlespacing{\subsubsection}{0pt}{*0.8}{*0.8}

\newtheorem{prop}{Proposition}[section]
\newtheorem{definition}{Definition}[section]

\begin{document}

\title{Modeling Network Populations via Graph Distances
\protect\thanks{ Sim\'on Lunag\'omez is a lecturer at the Department of Mathematics and Statistics, Lancaster University, Sofia C. Olhede is a Professor at the Institute of Mathematics, \'{E}cole polytechnique f\'{e}d\'{e}rale de Lausanne, and Patrick J. Wolfe is a Professor at the Departments of Statistics, Computer Science, and Electrical \& Computer Engineering, Purdue University.  This work was supported  European Research Council Fellowship via Grant CoG 2015-682172NETS (Olhede). We are grateful to Peter M\"{u}ller, Edoardo M. Airoldi, Christopher Nemeth and Anastasia Mantziou for helpful conversations.}}
\author{ Sim\'on Lunag\'omez, Sofia C. Olhede, Patrick J. Wolfe}
\maketitle

\begin{abstract}
This article introduces a new class of models for multiple networks. The core idea is to parametrize a distribution on labelled graphs in terms of a Fr\'{e}chet mean graph (which depends on a user-specified choice of metric or graph distance) and a parameter that controls the concentration of this distribution about its mean. Entropy is the natural parameter for such control, varying from a point mass concentrated on the Fr\'{e}chet mean itself to a uniform distribution over all graphs on a given vertex set. We provide a hierarchical Bayesian approach for exploiting this construction, along with straightforward strategies for sampling from the resultant posterior distribution. We conclude by demonstrating the efficacy of our approach via simulation studies and two multiple-network data analysis examples: one drawn from systems biology and the other from neuroscience.

\bigskip

\end{abstract}

\onehalfspacing


\begin{figure}[!t]
\begin{center}
  \includegraphics[height=40mm]{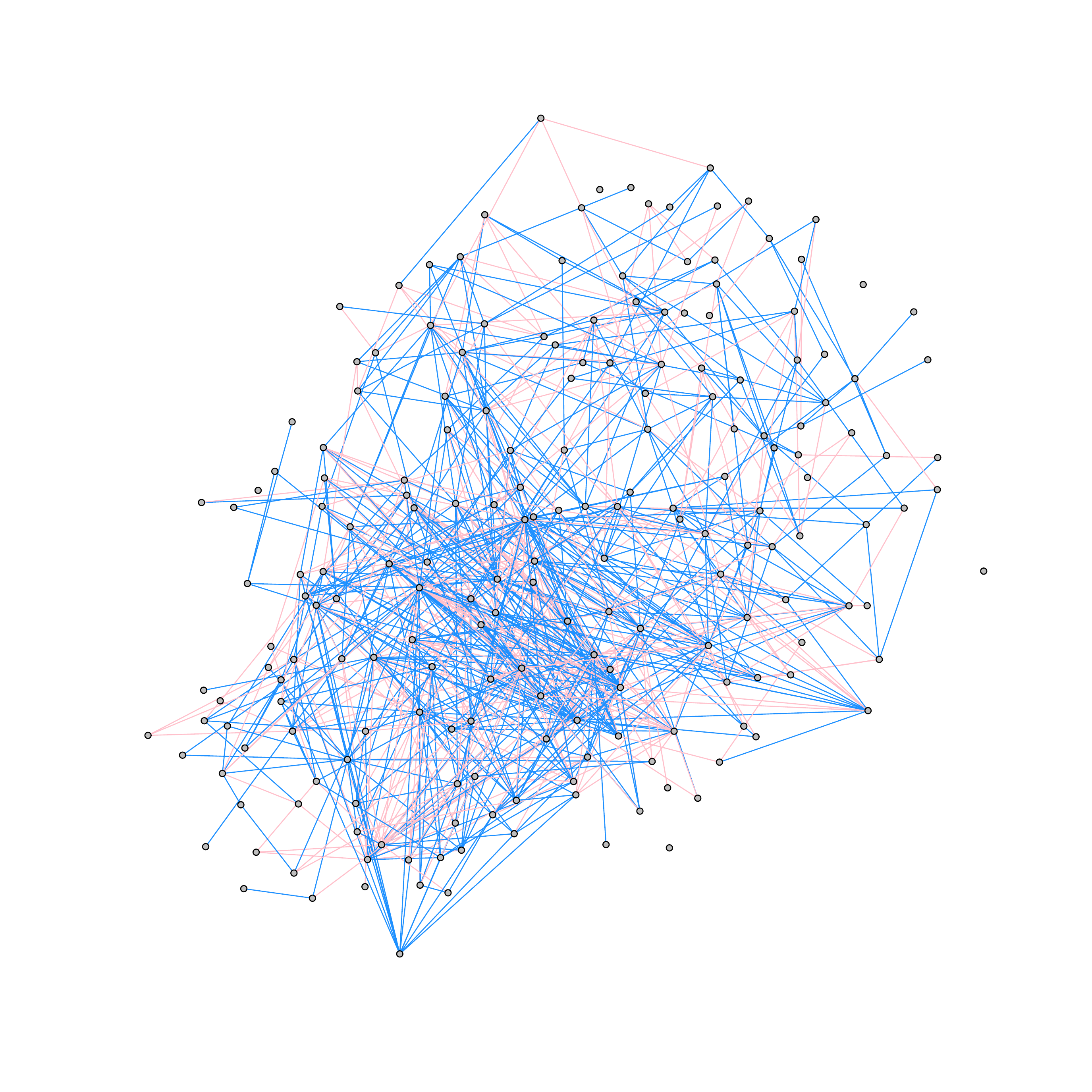}
  \includegraphics[height=40mm]{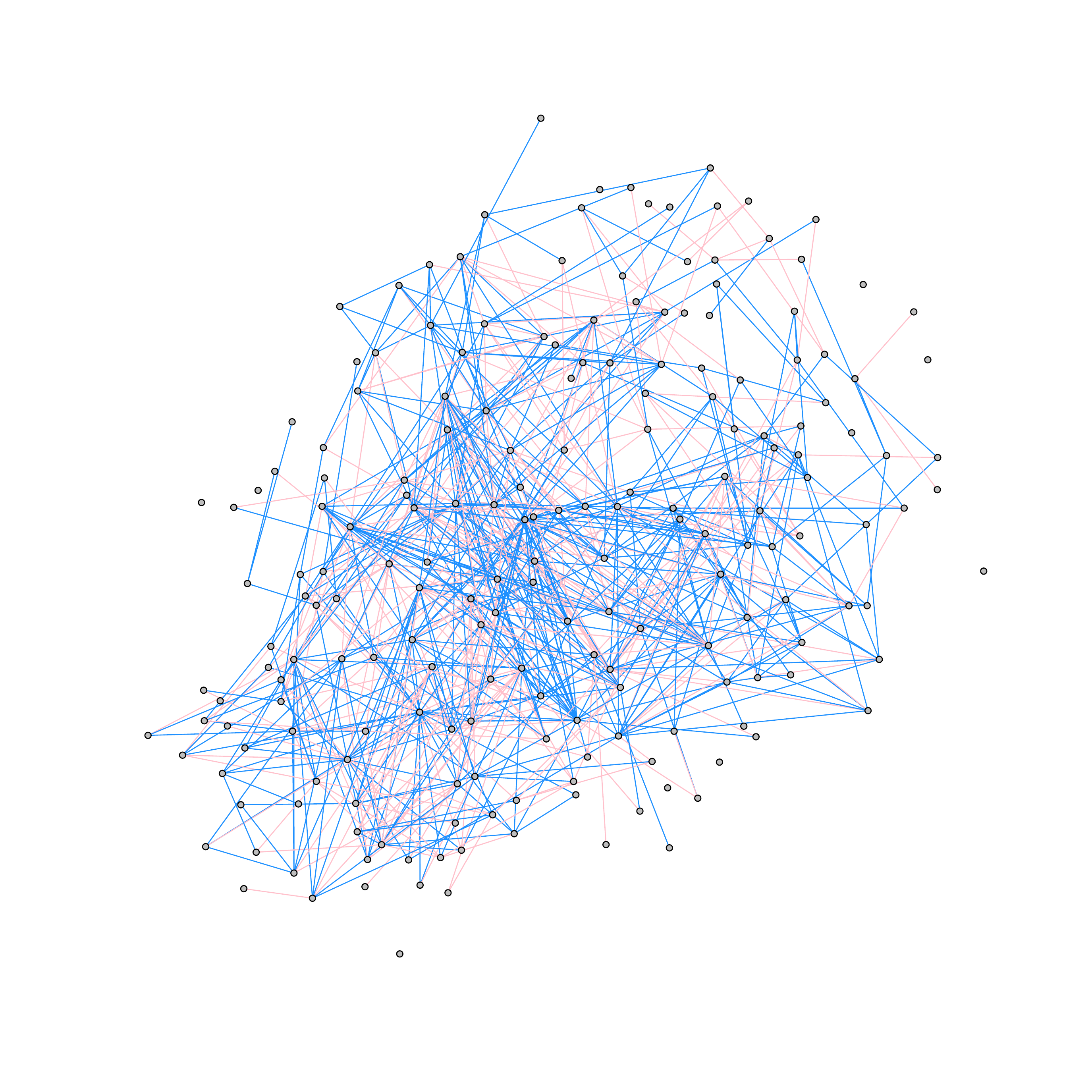}
  \includegraphics[height=40mm]{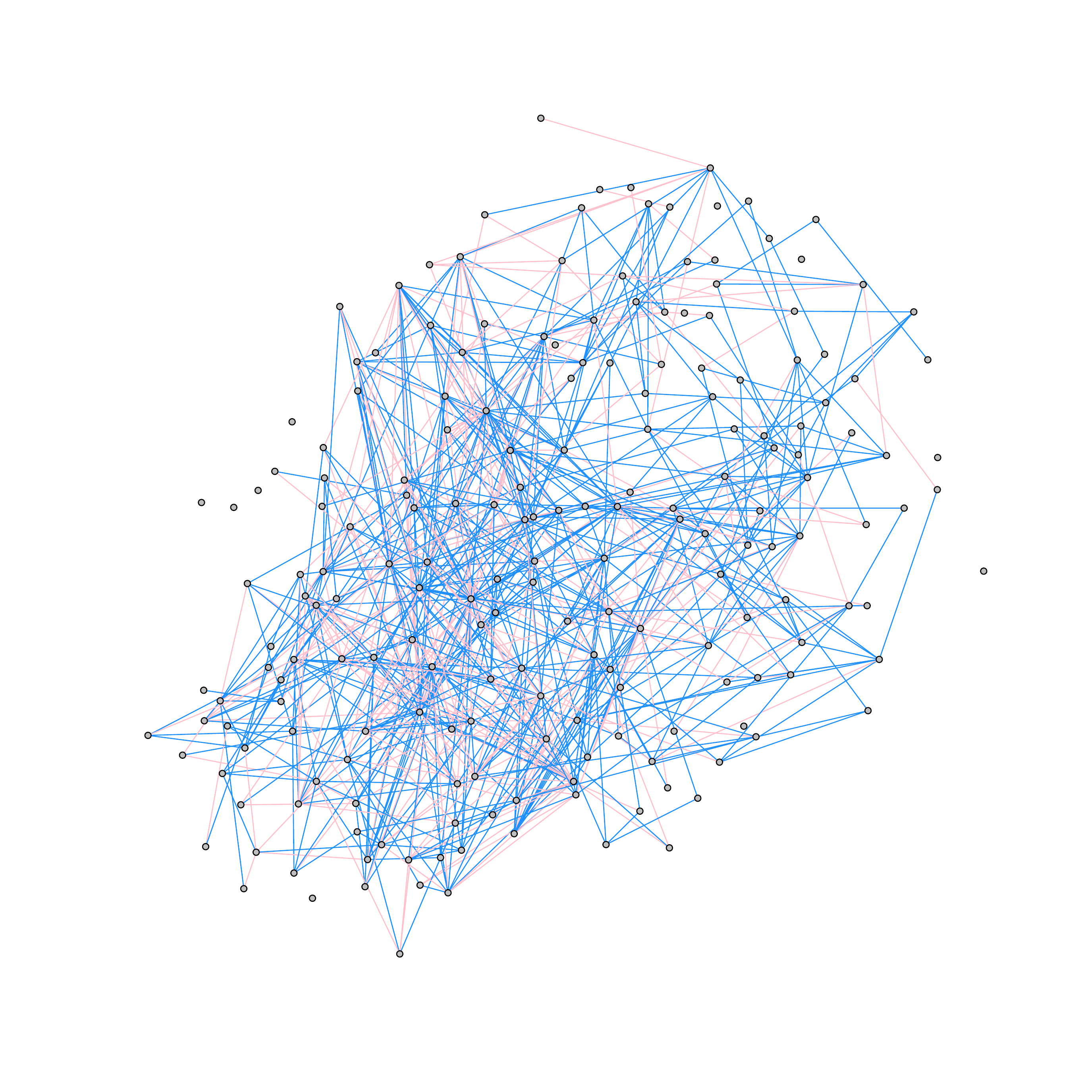}
  \includegraphics[height=40mm]{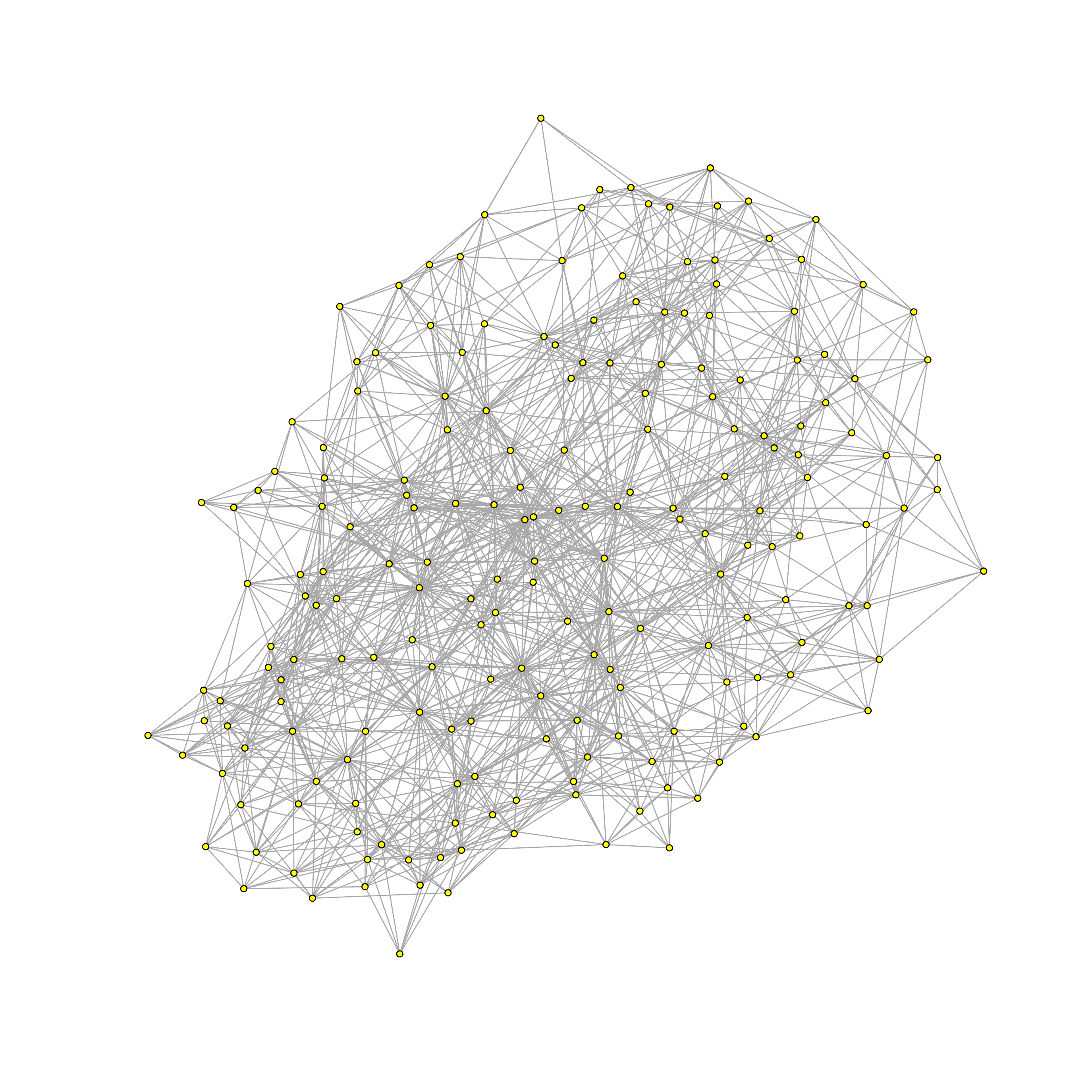}
\caption{Example of multiple network data in the context of neuroscience, with each node representing a region of the brain (see \cite{ZuoZhou} and \cite{ArroAthr}). The networks are defined over the same set of $200$ nodes. First three figures (from Left to Right): Discrepancies of three observed brain networks with respect to the point estimate of the Frech\'{e}t mean. Edges only present in the observed network are colored in blue, while edges present in the point estimate but not in the data point are colored in pink. Right: Posterior mode estimate $\mathcal{G}^{FM}$ of the Frech\'{e}t mean of $300$ brain networks using a metric on graph space based on diffusion.}
  \label{fig:MotivExample}
\end{center}
\end{figure}

\section{Introduction}\label{s:intro}

This article introduces a new class of models for data consisting of observations of multiple networks. With advances in measurement technology, these types of data are rapidly becoming prominent in fields such as systems biology and neuroscience, among others. In systems biology, inferences must often be combined on the same gene interaction network, where different inferences correspond to different data sets or to different analysis procedures applied to the same data \citep{BartOlhe}. In neuroscience, a population of networks encodes the way different regions of the brain interact when individuals perform a given task \citep{BiswMenn}, or characterises a population of individuals suffering from a neurological or psychiatric disorder \citep{LynaBass,NelsBass}. 

The developments proposed herein are therefore motivated by the problem of modelling populations of networks. The class of models we propose is based on the idea that distributions on graph space are naturally parameterised in terms of a mean---the Fr\'{e}chet mean, which is itself a network---and a measure of how concentrated the distribution is about this mean. A benefit of our approach is that the Fr\'{e}chet mean itself can be interpreted as the representative of a population of networks, relative to a user-specified choice of metric or graph distance. To specify concentration around the Fr\'{e}chet mean, we use entropy as described below. We then provide general strategies for performing Bayesian inference for these new models, allowing for the modeller to decide which metric is most suitable for the given application at hand. 


By multiple networks we mean two or more networks comprising a set of independent observations, which we assume here are defined over the same vertex set. Generalizing this problem to map networks of arbitrary different sizes  to a common reference or with scrambled order of nodes is a non-trivial extension, solving a potentially computationally intractable problem (we elaborate more on this point in the discussion). In for example medical imaging and bioinformatics this assumption is not unreasonable, if admittedly restrictive. A brain connectome example (which we study later in Section \ref{Sec:Data}) drawn from neuroscience is displayed in Fig.~\ref{fig:MotivExample}, with regions of the brain assigned to nodes according to the CC200 atlas, which was proposed by \cite{CradJame}. Note from Fig.~\ref{fig:MotivExample} that if we consider each possible pair of observations (first three figures, from left to right), any member of such pair can be seen as a modification of the other member or as a modification from a representative of the population (Fig.~\ref{fig:MotivExample}, on the right). Thus, although one modelling approach would be to treat such networks as realisations from a single random graph model based on global features, such as a stochastic block model, this limits the inferential insights that can be gained from multiple networks as opposed to a single one.

Indeed, the questions arising from multiple network data demand a different perspective:
 
\begin{enumerate}
 
 \item How does one find a summary or representative (at the population level) for multiple observed networks? In other words, what type of structure must the modeller impose on the space of labelled graphs to define a suitable estimand? Without such an estimand (\emph{e.g.}, in the case of block modelling or link prediction) we risk our inference yielding a summary of the population that does not look like any of its elements, and cannot be used in place of them.
 
 \item In the Bayesian setting, if we have multiple networks (such as those in Fig.~\ref{fig:MotivExample}) as historical data, how do we perform prior elicitation without resorting to global assumptions on the network structure? For example, in systems biology it is typical that past inferences regarding a given gene interaction network may provide a very accurate idea about what a newly inferred network might be expected to look like, when obtained using a new measurement technology. This is illustrated in Section \ref{Sec:SysBio}.
 
 \end{enumerate} 

We show here that both questions can be answered by first assuming that the observed networks are perturbations of a ``typical'' network, and then characterising the variability of the data in those terms. Specifically, the Fr\'{e}chet mean implied by a given metric will parametrize a generative model, under the assumption that the probability of generating a specific network is given by a strictly decreasing function only of its distance from this Fr\'{e}chet mean.
  
To construct our models, we borrow ideas from the graphical models and shape theory literatures, where authors have considered the notion of a ``typical'' non-Euclidean observation, and random perturbations from that observation. Previous work on multiple networks in the statistics literature includes the following: The approaches proposed by \cite{BalaKolaVile} and \cite{ChanKola} for estimating features (subgraph counts and density, respectively) from network data; the model proposed by \cite{GollMurp} (which is an extension of the latent space model proposed by \cite{HoffRaftHand}) for describing the variability of a homogeneous population of networks; the Bayesian nonparametric approach proposed by \cite{DurDunVog} for modelling heterogeneous populations of networks; and the approach for comparing populations of networks via testing by \cite{GineBalaKola2017}. The methodology of the last paper is based on the asymptotic theory for the space of unlabelled networks developed by \cite{KolaLinRoseWalt2019}, which serves to quantify how concentrated the distribution is around a mean network when formulated in terms of a very specific metric. \cite{KolaLinRoseWalt2019} and earlier \cite{FeraH} discuss the problem of estimating a mean and the geometry associated to the space of possible values for that estimand, the former for the space of graphs while the latter for the space of trees. 

Recently, \cite{NielWitt} proposed a multiple network model based on the random dot product graph model; their approach builds on work by \cite{WangVogePrieb}, who proposed a gradient-descent method to compute the simultaneous embedding of a set of graphs. In terms of inference, \cite{NielWitt} focus on the problem of comparing populations of networks. \cite{TangAthrSuss} focus on the problem of testing for the difference of two populations of networks; the authors assume a random dot product graph model, as in \cite{WangVogePrieb}, but computation is done using the bootstrap. We also note the model-based approach for estimating the generating mechanism of multiple networks given by \cite{bhattacharyya2018spectral}. Finally, some of the ideas developed in this paper have parallels in the literature for modelling measurement error for networks, including recent work by \cite{Newm}, \cite{Peix2018} and \cite{LeLevLevi2018}. 

In a different direction, similarity measures on the local structure of a network have been used to perform prior elicitation on graph space, particularly in the graphical models literature; this idea has been discussed by  \cite{Mukh:Spee:2008} as well as \cite{MitrMull}. Our approach can also be related to work by \cite{TanJas} and the work by \cite{NuMuZhuJi} in the graphical models literature, who propose  hierarchical models on graph space. From the shape theory literature, we borrow insight from the work of \cite{MardDryd}, which uses the idea of modelling a set of non-Euclidean objects (shapes) in terms of a centroid and parameters that control how concentrated the distribution will be around that centroid. 

From a Bayesian point of view, computing a Fr\'{e}chet mean at the population level is analogous to minimising the posterior expected loss, and becomes the same problem when the loss is a metric. \cite{WadeGhah} exploit this idea, in the context of cluster analysis. In our work, we use entropy in conjunction with the Fr\'{e}chet mean to define a distribution for non-Euclidean data, and in that sense, our work relates to the methodology developed by \cite{Penn}.
 
Distinct from the literature discussed above, the methodology we propose here achieves different goals: (1) It enables the modeller to characterise the variability of a set of observed networks in terms of a Fr\'{e}chet mean and a measure of how concentrated the distribution is around this mean, and to perform Bayesian inference, without resorting to asymptotics; (2) It enables the practitioner interested in network data to perform prior elicitation on graph space by using an observed network as starting point; and (3) It provides tools for incorporating different metrics on graph space into the modelling procedure, enabling the encoding of different assumptions the practitioner may have regarding similarity among graphs.
We also are able to discuss a looser notion of a location-scale family for random graph models using this set of technology, taking inspiration from~\citep{FangKotzNg}, we use the functional form of a symmetric multivariate distribution whose both location and scale need to be relaxed to apply. Concrete examples of this notion are provided, along with theoretical results that show that these examples are legitimate. We show how these examples relate to one another.

The remainder of this article is organised as follows. Section \ref{Sec:pre} first introduces the necessary preliminaries, including metrics on graph space and the Fr\'{e}chet mean. Section \ref{Sec:TheModel} then details the general concepts on which the generative models proposed in this paper are be based on, along some examples. The corresponding strategies for Bayesian modelling and computation are presented in Section \ref{Sec:Inference}. Section \ref{sec:Simul} documents the behaviour of our models via simulation studies, and Section \ref{Sec:Data} describes fully the fitting of our models to the multiple-network data introduced in Fig.~\ref{fig:MotivExample} above. Finally, Section \ref{Sec:Discussion} discusses briefly the contributions of our approach, placing it in context and outlining limitations as well as future possibilities.

\section{Preliminaries}\label{Sec:pre}

A simple labelled graph $\mathcal{G}=(\mathcal{V},\mathcal{E})$ comprises a set of vertices $\mathcal{V}$ and a set of edges $\mathcal{E}\subset \left\{ E \subset \mathcal{V}: |E|=2 \right\} $. Letting $N=|\mathcal{V}|$, we may represent $\mathcal{G}$ by an $N\times N$ adjacency matrix $A_{\mathcal{G}}$ such that
\begin{displaymath}
A_{\mathcal{G}}(i,j)= 
\begin{cases} 
1  & \text{if there is an edge between nodes } i \text{ and }j,\\
 0 & \text{otherwise}.
 \end{cases}
\end{displaymath}
The models and methods we propose can all be applied equally to directed graphs (with $A_{\mathcal{G}}(i,j)$ distinct from $A_{\mathcal{G}}(j,i)$ for $i<j$) and those having self-loops ($A_{\mathcal{G}}(i,i) = 1$), as well as more generally any weighted graph such that each $A_{\mathcal{G}}(i,j)$ takes values in some finite, discrete set. We write that a graph $\mathcal{G}_1=(\mathcal{V}_1,\mathcal{E}_1)$ is a subgraph of $\mathcal{G}_2=(\mathcal{V}_2,\mathcal{E}_2)$ if $\mathcal{V}_1 \subseteq \mathcal{V}_2$ and $\mathcal{E}_1 \subseteq \mathcal{E}_2$. For a set $\left\{ \mathcal{G}_s \right\}_{s\in S}$, we denote by $A_{\mathcal{G}_k}(i,j)$ the $(i,j)$th entry of the adjacency matrix of $\mathcal{G}_k$, $k\in S$.

For $N\in \mathbb{N}$ and $\mathcal{V} := \{1,2,\ldots,N\}$, define
\begin{displaymath}
{\mathscr{G}}:=\left\{\mathcal{G}_{[N]} \right\}:=\left\{ \mathcal{G}=(\mathcal{V},\mathcal{E}) :  | \mathcal{V} | =N   \right\},
\end{displaymath}
so that $\left\{\mathcal{G}_{[N]} \right\}$ represents the set of all $N$-node labelled networks of a given type (simple, directed, etc.). If we consider simple graphs, for example, then $\left| \left\{\mathcal{G}_{[N]} \right\} \right| = 2^{\binom{N}{2}}$. We refer to  $\left\{\mathcal{G}_{[N]} \right\}$ for the simple directed case as a \emph{graph space}. This term tends to be used in this way (rather informally) in the graphical models literature. 

Metrics on graph spaces in turn allow for an appropriate definition of network structural similarity \citep{DonnHolm}. We are interested in developing probability models on $\left\{\mathcal{G}_{[N]} \right\}$ given the choice of a metric $d_G(\cdot,\cdot)$ on $\left\{\mathcal{G}_{[N]} \right\}$. Two examples of metrics which can be used to formulate the models introduced in Section \ref{Sec:TheModel} below are as follows:

\begin{enumerate}

\item The \emph{Hamming distance} between two graphs when their adjacency matrices are treated as strings, which is given by the number of entries that disagree. We will use the notation
\begin{displaymath}
d_H (\mathcal{G}_1 , \mathcal{G}_2 ) = |A_{\mathcal{G}_1} - A_{\mathcal{G}_2} |_H,
\end{displaymath}
to denote this distance independently of the type of network under consideration by the modeller (\emph{e.g.}, simple, directed).

\item A \emph{Diffusion distance} based on the graph Laplacian, for example the choice made by \cite{Hamm}:
\begin{displaymath}
d_L(\mathcal{G}_1,\mathcal{G}_2; t)=\| \exp(-tL_{\mathcal{G}_1}) -\exp(-tL_{\mathcal{G}_2})\|^2_{F}, \quad t > 0;
\end{displaymath}
where $\| \cdot \|_{F}$ is the Frobenius norm and $L_{\mathcal{G}}$ is the combinatorial Laplacian matrix associated to an undirected graph $\mathcal{G}$:
\begin{displaymath}
L_{\mathcal{G}}(i,j)= 
\begin{cases} 
\sum_{k=1}^N A_{\mathcal{G}}(i,k) & \text{if } i = j,\\
 - A_{\mathcal{G}}(i,j) & \text{otherwise}.
 \end{cases}
\end{displaymath}
Note that this is referred to by the letter $L$ in \citep{Chung}, whilst Chung (unlike Hammond et al.) defines the Laplacian to be 
normalized. The normalized version of the matrix is an operator related to the Laplace-Beltrami operator for objects different than networks via the discretization of a derivative.

Hammond and coauthors \citep{Hamm} argued that the diffusion distance is natural as two graphs are similar if they transmit information in the same way. Generic transmission is by them modelled using heat diffusion on the network.
The distance therefore arises as $\exp(-tL_{\mathcal{G}})$ is the kernel associated with (\emph{e.g.}, classical heat) diffusion on a graph $\mathcal{G}$ via the discrete Laplace operator $L_{\mathcal{G}}$. The value of $t$ is here the time of diffusion. As $t\rightarrow 0$ we should return to whatever initial conditions were specified, and at $t\rightarrow \infty$ 
equal proportions of diffused ``stuff'' should be at each node.

The value of $d_L(\mathcal{G}_1,\mathcal{G}_2)$ measures the discrepancy after $t$ units of time between the diffusion on $\mathcal{G}_1$ versus that on $\mathcal{G}_2$. For our purposes $t$ may be regarded as a parameter whose value can be elicited \emph{a priori} using information from the application domain under consideration. As $t$ decreases, it becomes harder to distinguish between diffusion patterns (no diffusion has happened yet) and therefore to distinguish between different elements of $\left\{\mathcal{G}_{[N]} \right\}$.
Finally the graph Laplacian is discussed in detail in \citep{Chung}, and we use the unnormalized version. \cite{Hamm} argues that this captures the temporal evolution of the vector representing the diffusion. Thus this metric captured how differently things have flowed up to time $t$.
\end{enumerate}
While the Hamming distance focuses on simple flips of edges into non-edges (changes in very local structure), the diffusion distance is treating the objects functionally (\emph{i.e.} it focuses on changes that may impact the global structure).

One might ask what choice of metric should be made? \cite{DonnHolm} provide some guidelines in this choice of metric. The Hamming distance can be interpreted as simple flips of edges (a local modification of the network). The diffusion distance allow information to diffuse on the network, and then lets us compare that diffusion.  As~\cite{DonnHolm} discuss, the Hamming distance assume  deletions and additions carry the same weight, even if their structural impact may not be equivalent. The Hamming distance is strongly affected by the sparsity of the graph, as already pointed out by the aforementioned authors. To take into account the sparsity into the metric, the Jaccard distance is used~\citep{DonnHolm}.
The authors additionally discuss global metrics based on spectral distances. The diffusion distance balances information differently taking the diffusion of information on the network into account. Other balances between global and local can be made~\citep{DonnHolm}. 
Computationcal cost factor into our usage of these two metrics. Later on (see Section \ref{Sec:EGM}) we will illustrate how a model based on a simple metric can aid the computation of the posterior for a model based on a more sophisticated metric.

We conclude this section by introducing the  \citet{Frech} mean for use in the context of the metric spaces $(\left\{\mathcal{G}_{[N]}\right\},d_G)$ and associated probability models that we will consider below. Given an arbitrary metric space $(\mathcal{Y},d)$ and a probability measure on $\mathcal{Y}$, the Fr\'{e}chet mean provides the notion of an average or measure of central tendency with respect to $d$. It generalizes the first moment to non-Euclidean settings and has seen wide use in areas such as shape theory.

\begin{definition} [Fr\'{e}chet mean]\label{def:Frechet}  Let $Y$ be a random element defined on sample space $\mathcal{Y}$ and let $d(\cdot,\cdot)$ be a metric on $\mathcal{Y}$. The set
\begin{equation}\label{Eq:FrechMean}
\psi^m=\arg\inf_{\psi \in \mathcal{Y}} \mathbb{E}_{Y}[d^2(Y,\psi)]
\end{equation} 
is called the Fr\'{e}chet mean set of $Y$. 
\end{definition}

We will use the Fr\'{e}chet mean in conjunction with unimodality to formulate natural and intuitive models on the space $\left\{\mathcal{G}_{[N]}\right\}$ of labelled $N$-node networks.

\section{Modelling Approach}\label{Sec:TheModel}

In this section, we propose a generative modelling approach for data sets consisting of multiple networks. Our models are parametrized in terms of a unique mode and a univariate measure of dispersion around that mode. The mode in the space of labelled $N$-node networks $\left\{\mathcal{G}_{[N]}\right\}$ is itself a network defined on the same vertex set as each individual observation, allowing us to define a suitable estimand to obtain directly a population-level summary of multiple networks.

In analogy to a location--scale family, we provide concepts that enable us to propose probability models on $\left\{\mathcal{G}_{[N]}\right\}$ in terms of a central graph (location) and concentration around that central graph (scale of variation).  We use the terms loosely given that we are working in a non-Euclidean setting. In contrast with the location--scale family, which takes the vector space structure for granted, we are constrained by the structure entailed by a metric in $\left\{\mathcal{G}_{[N]}\right\}$ and the fact that the space is finite.


\begin{definition}[Unimodal network distribution based on location]\label{def:RGDis}
Fix a metric $d_G$ on $\left\{\mathcal{G}_{[N]} \right\}$ for $N\in \mathbb{N}$, and consider a family of probability mass functions $\left\{ p(\cdot\mid \mathcal{G}^{m}) \right\}_{\mathcal{G}^{m}\in \left\{\mathcal{G}_{[N]} \right\}}$ on $\left\{\mathcal{G}_{[N]} \right\}$ such that 
\begin{enumerate}
\item Each $p(\cdot\mid \mathcal{G}^{m})$ is unimodal with mode $\mathcal{G}^{m}\in \left\{\mathcal{G}_{[N]} \right\}$;
\item For $\mathcal{G}_1,\mathcal{G}_2 \in  \left\{\mathcal{G}_{[N]} \right\}$, we have that $d_G(\mathcal{G}_1,\mathcal{G}^{m})>d_G(\mathcal{G}_2,\mathcal{G}^{m})$ implies
$p(\mathcal{G}_2)>p(\mathcal{G}_1)$, while $d_G(\mathcal{G}_1,\mathcal{G}^{m})=d_G(\mathcal{G}_2,\mathcal{G}^{m})$ implies
$p(\mathcal{G}_2)=p(\mathcal{G}_1)$. 
\end{enumerate}
\end{definition}

The most straightforward example is as follows: the Centred Erd\"{o}s--R\'{e}nyi Model, which will be introduced later on this section. 
We shall now need another important concept from information theory, namely that of entropy, see~\citet{MezaMont}. This is used to measure the uncertainty of a random variable  and takes the form of
\begin{equation}
    H_{\cal G}=-\sum_{{\cal G}\in{\mathscr{G}}}
    p({\cal G}\mid \mathcal{G}^{m})\cdot \log\left(  p({\cal G}\mid \mathcal{G}^{m})\right).
\end{equation}
Sometimes $\log(\cdot)$ in the above expression is replaced by
$\log_2(\cdot)$. We set $0\cdot \log(0)$ to equate to zero, as usual. 

Building from unimodality we also need to introduce scale, which is our next step.
%
%
\begin{definition}[Unimodal network distribution with location \& scale]\label{def:RGDisEnt}
Fix a metric $d_G(\cdot,\cdot)$ on $\left\{\mathcal{G}_{[N]} \right\}$ for $N\in \mathbb{N},$ nonempty set \bigskip $\Gamma \subset \mathbb{R_+}$, and consider a family
$\left\{  p(\cdot\mid \mathcal{G}^{m},\gamma) \right\}_{\mathcal{G}^{m}\in \left\{\mathcal{G}_{[N]} \right\},\gamma\in\Gamma}$: 
\begin{enumerate}
\item For every fixed scale parameter  $\gamma^{*} \in \Gamma$, the family $\left\{ p(\cdot\mid \mathcal{G}^{m},\gamma^{*}) \right\}_{\mathcal{G}^{m}\in \left\{\mathcal{G}_{[N]} \right\}}$ satisfies Definition \ref{def:RGDis} with respect to the metric $d_G$.
\item For every fixed location parameter $\mathcal{G}^{*}\in \left\{\mathcal{G}_{[N]} \right\}$, the entropy associated to the family $\left\{  p(\cdot\mid \mathcal{G}^{*},\gamma)\right\}_{\gamma \in \Gamma}$ is a strictly monotone function of $\gamma \in \Gamma$.
\end{enumerate}
\end{definition}

For finite, discrete sets such as $\left\{\mathcal{G}_{[N]} \right\}$ and associated probability mass function $p(\cdot)$, entropy $-\mathbb{E} \left\{\log p(\cdot)\right\} $ provides a convenient characterization akin to variance, ranging from $0$ for a point mass to $\log(|\left\{\mathcal{G}_{[N]} \right\}|)$ for the uniform distribution on $\left\{\mathcal{G}_{[N]} \right\}$. Entropy can thus be used to parametrize a family of discrete distributions on $\left\{\mathcal{G}_{[N]} \right\}$ with the same unique mode, in an analogous way to how the scale parameter would parametrize a member of the location--scale family when the location parameter has been specified. The metric provides a ranking of the elements $\left\{\mathcal{G}_{[N]} \right\}$ given the mode, the entropy enables the statistician to control the decay of the values of the probability mass function given that ranking. To take the analogy with a Gaussian distribution $\gamma$ plays the role of $1/\sigma$ for the Gaussian, where $\sigma^2$ is the variance. Therefore we expect $\gamma\rightarrow 0$ to play the role of $\sigma\rightarrow \infty$, or the maximum entropy solution that should be the least peaked. In contrast, $\gamma\rightarrow \infty$, we expect to correspond to 
the minimum entropy solution, and be the most concentrated distribution. Therefore intuitively, we expect the entropy to decay in $\gamma$. This allows us to consider the analogy of ``peaked'' versus ``flat'' distributions where $\gamma$ controls the peak.

We now provide two examples for the random graph distribution based on distance and entropy. These examples will be discussed in detail in Sections \ref{Sec:SER} and \ref{Sec:ENF}, respectively.

We will now introduce a first example of a random graph distribution based on distance and entropy; we call it the Centred Erd\"{o}s--R\'{e}nyi Model. The intuition behind this model is that  noisy versions of the centroid (which is denoted by $\mathcal{G}^{m}$) are generated by flipping edges independently at random with probability $\alpha$. From a modelling perspective, it is sensible to penalize (or constrain) $\alpha$ so it takes values much smaller than the density of $\mathcal{G}^{m}$; there is little utility for a model where the trend is overwhelmed by noise.
\begin{definition}[Centred Erd\"{o}s--R\'{e}nyi Model]\label{def:SER} Given a graph  $\mathcal{G}^{m} \in \left\{\mathcal{G}_{[N]} \right\} $ and $1/2> \alpha>0$, consider a model $p(\cdot \mid \mathcal{G}^{m}, \alpha )$ on $ \left\{\mathcal{G}_{[N]} \right\} $ of the form :
\begin{equation}
    \Pr\left(A_{\mathcal{G}}(i,j)= A_{\mathcal{G}^{m}}(i,j)\right)=1-\alpha.
\end{equation}
We call this the Centred Erd\"{o}s--R\'{e}nyi Model (CER) with mode $\mathcal{G}^{m}$ and parameter $\alpha$.
\end{definition}
Note that $A_{\mathcal{G}}(i,j)$ generating mechanism can also be written as
\begin{displaymath}
A_{\mathcal{G}}(i,j)\mid A_{\mathcal{G}^{m}}(i,j),\alpha = |A_{\mathcal{G}^{m}}(i,j)-Z(i,j)|,
\end{displaymath}
where the $Z(i,j)$'s are $\mathrm{iid}$ $\mathrm{Ber}(\alpha)$ for $1\leq i<j \leq N$.\\

This way of describing $A_{\mathcal{G}}(i,j)$'s generating mechanism highlights that edges or flipped to non-edges, or non-edges to edges, with probability $\alpha$. This clarifies why we expect $\alpha\leq \frac{1}{2}$, as otherwise we are more likely to flip all edges and not be ``centered'' at $A_{\mathcal{G}^{m}}(i,j)$. This condition also will be required in the proofs of Proposition 3.1, which exactly establishes mode etc. Thus, for the Centred Erd\"{o}s--R\'{e}nyi to effectively serve as an error measurement model, the $\alpha$ parameter should be constrained to be be smaller than the edge density parameter of $\mathcal{G}^m$. The condition $1/2> \alpha>0$ furthermore ensures that the maximum likelihood estimator (of $\mathcal{G}^m$)  will be the graph that minimises the average number of mismatches with respect to the observed networks. For this model, we do not expect the observed graphs to be, on average, of different density than $\mathcal{G}^{m}$; this is because the error model affects edges and non-edges equally. Observe that if a parametric random graph model is further imposed upon $\mathcal{G}^{m}$ (\emph{e.g.}, Erd\"{o}s--R\'{e}nyi), then this model can be cast into the approach proposed by \cite{Newm}.

\begin{prop}\label{Prop:CER}
We let $d_G(\cdot,\cdot)$ denote the Hamming distance on  $\left\{\mathcal{G}_{[N]} \right\}$.
If two graphs ${\mathcal{G}}_1$ and  ${\mathcal{G}}_2$ are generated from 
the Centred Erd\"{o}s--R\'{e}nyi model with centroid $\mathcal{G}^m \in \left\{\mathcal{G}_{[N]} \right\} $
and $0<\alpha\leq 1/2$ then we have that $d_G(\mathcal{G}_1,\mathcal{G}^{m})>d_G(\mathcal{G}_2,\mathcal{G}^{m})$ implies
$p(\mathcal{G}_2)>p(\mathcal{G}_1)$, while $d_G(\mathcal{G}_1,\mathcal{G}^{m})=d_G(\mathcal{G}_2,\mathcal{G}^{m})$ implies
$p(\mathcal{G}_2)=p(\mathcal{G}_1)$. 
We deduce that $p(\mathcal{G})$ is unimodal, and that the Centred Erd\"{o}s--R\'{e}nyi model is a unimodal network distribution based on location and scale.
\end{prop}

As a second example of a unimodal network distribution based on location and scale, we introduce a model motivated by the notion that the similarity with respect to the centroid is made concrete by the choice of $d_G(\cdot,\cdot)$ (\emph{e.g.} the metrics proposed by \cite{Zelin}, \cite{Hamm}, or the ones discussed in \cite{DonnHolm}), and covered by our discussion in Section 2 earlier in the paper. 
\begin{definition}[Spherical Network Family]\label{def:ENF} 
Given a graph $\mathcal{G}^{m}\in \left\{\mathcal{G}_{[N]} \right\} $, a metric $d_G(\cdot,\cdot)$ on $ \left\{\mathcal{G}_{[N]} \right\} $, and $ \gamma>0$, we propose:
\begin{equation}
\label{PG}
p(\mathcal{G}\mid \mathcal{G}^{m},\gamma)\propto \exp\left\{ -\gamma \phi(d_G (\mathcal{G},\mathcal{G}^{m}))   \right\},
\end{equation}
where $\phi(\cdot)$ is a non-negative strictly increasing function such that $\phi(0)=0$. This is the Spherical Network Family with parameters $\mathcal{G}^{m}$ and $\gamma$.
\end{definition}
This model is related to the prior introduced by \cite{MitrMull}, which was introduced in the context of graphical modelling. A main difference with respect to their approach is that the Spherical Network Family is aimed to serve as the functional form for both the likelihood and the prior. This model also relates to the similarity measure proposed by \cite{DahlDayTsai} for random partitions. The normalizing constant for this model is the reciprocal of:
\begin{equation}\label{Eq:Partit}
Z(\mathcal{G}^m,\gamma)=\sum_{\mathcal{G}\in \left\{ \mathcal{G}_{[N]} \right\} }\exp\left\{ -\gamma \phi(d_G (\mathcal{G},\mathcal{G}^{m}))\right\},
\end{equation}
here $Z(\mathcal{G}^m,\gamma)$ is known as the \emph{partition function} of $p(\mathcal{G}\mid \mathcal{G}^m,\gamma)$.
We observe directly that $Z(\gamma)>0$ as it is a sum of positive terms. Just like the normalizing constant of any probability mass function,  as
~\eqref{Eq:Partit} aggregates over $\mathcal{G}\in \left\{ \mathcal{G}_{[N]} \right\}$, the sum will not be a function directly of $\phi(d_G (\mathcal{G},\mathcal{G}^{m}))$, only implicitly as the sum will vary depending on the functional form. Therefore  $Z(\mathcal{G}^m,\gamma)$ is a positive constant that does not depend on $d_G(\mathcal{G},\mathcal{G}^m)$.\\

The functional form proposed for the Spherical Network Family (SNF) is inspired by the notion of symmetry of the density discussed in \cite{FangKotzNg}. A random variable $X$ on $\mathcal{X}$ has the symmetry of the density property if its density $p(\cdot\mid \mu,\gamma)$ is of the form
\begin{displaymath}
p(X\mid \mu,\gamma)=Z^{-1}(\gamma) \cdot \exp \left[ - \gamma \phi(d(X,\mu)) \right],
\end{displaymath}
where $\mu\in \mathcal{X}$, $\gamma>0$, $\phi(\cdot)\geq 0$ is a non-decreasing function, $d(\cdot,\cdot)$ is a metric on $\mathcal{X}$.

\begin{prop}\label{CERSNF}
The CER is a member of the SNF.
\end{prop}

The above proposition demonstrates that the SNF is not empty. There are some other properties we would like to see, and to be clear on what properties that we desire, let us show that they hold in the following proposition.  
\begin{prop}\label{Prop:Spher}
We let $d_G(\cdot,\cdot)$ denote a graph metric on  $\left\{\mathcal{G}_{[N]} \right\} $.
If two graphs ${\mathcal{G}}_1$ and  ${\mathcal{G}}_2$ are generated from 
the Spherical Network Family with centroid $\mathcal{G}^m$
and $\gamma \in \mathbb{R}^+$ then we have that $d_G(\mathcal{G}_1,\mathcal{G}^{m})>d_G(\mathcal{G}_2,\mathcal{G}^{m})$ implies
$p(\mathcal{G}_2)>p(\mathcal{G}_1)$, while $d_G(\mathcal{G}_1,\mathcal{G}^{m})=d_G(\mathcal{G}_2,\mathcal{G}^{m})$ implies
$p(\mathcal{G}_2)=p(\mathcal{G}_1)$. 
As a consequence $p(\mathcal{G})$ is unimodal, and the Spherical Network Family is a unimodal network distribution based on location. In addition, the Spherical Network Family is unimodal network distribution based on location and scale if $\mathbb{V}\text{ar}\left\{\phi\left[d(\mathcal{G},\mathcal{G}^m)\right]\right\} >0$. 
\end{prop}

The next step consists in verifying if the examples we have presented fulfill the condition stated in Proposition \ref{Prop:Spher}.

\begin{prop}\label{Prop:CheckPropp}
The CER  and SNF equipped with the diffusion distance fulfill the condition $\mathbb{V}\text{ar}\left\{\phi\left[d(\mathcal{G},\mathcal{G}^m)\right]\right\} >0$ when $\phi(\cdot)$ is the identity function.
\end{prop}





The following property of the sample Fr\'{e}chet mean will provide insight regarding the behavior of the MLE for both models defined above and supports our intuition that the posterior mode will tend to the true value of the Fr\'{e}chet mean as the sample size increases.
\begin{prop}\label{Prop:Frechet}
The sample Fr\'{e}chet mean in  $\left\{\mathcal{G}_{[N]} \right\}$  converges to the true Fr\'{e}chet mean when the later exists and is unique, for $N\in \mathbb{N}$.
\end{prop}
Definition \ref{def:RGDis} is expressed in terms of the mode of the distribution. The following result indicates how the Fr\'{e}chet mean and mode relate for the Centred Erd\"{o}s--R\'{e}nyi Model:
\begin{prop}\label{Prop:SER}
The mode and Fr\'{e}chet mean coincide for the Centred Erd\"{o}s--R\'{e}nyi Model defined on $\left\{\mathcal{G}_{[N]} \right\}$, $N\in \mathbb{N}$. 
\end{prop}
For the Spherical Network Family, the Fr\'{e}chet mean maximises the kernel of the Boltzmann distribution in Eqn~\eqref{PG}. This is a direct consequence of Definitions \ref{def:Frechet} and \ref{def:ENF}.

Now, we have enough elements for presenting our model for multivariate network data. Let $N\in \mathbb{N}$ and $d_G$ a metric on $\left\{\mathcal{G}_{[N]} \right\}$. To describe the variability of a set observations $\left\{  \mathcal{G}_1,\mathcal{G}_2,\dots,\mathcal{G}_n  \right\}$ in $\left\{\mathcal{G}_{[N]} \right\}^n$, we propose a model of the form:
\begin{equation}\label{Eq:BayesModel}
p(\mathcal{G}_1,\mathcal{G}_2,\dots,\mathcal{G}_n\mid \mathcal{G}^{m},\gamma)=p(\mathcal{G}^{m}\mid \mathcal{G}^{0},\gamma^{0})
p(\gamma)
\prod_{i=1}^n p(\mathcal{G}_i\mid\mathcal{G}^{m},\gamma),
\end{equation}
where $p(\cdot \mid\mathcal{G}^{m},\gamma)$ is the likelihood, which is given by a unimodal network distribution based on location and scale; $p(\cdot \mid \mathcal{G}^{0},\gamma^{0})$ is the prior on the mode of the distribution, such prior is also given by a distribution with the same functional form as the likelihood; finally, $p(\gamma)$ is the prior on the entropy of the distribution. One implication of choosing this parametrization is that the inference will be in terms of the population centroid, which is a network by itself. This enables the statistician to perform an operation equivalent to smoothing in graph space.

We propose this model with the aim to represent the variability of a set of observations $\left\{  \mathcal{G}_1,\mathcal{G}_2,\right.$ $\left. \dots,\mathcal{G}_n  \right\}$ in $\left\{\mathcal{G}_{[N]} \right\}^n$ such that, for every pair $\left\{\mathcal{G}_i,\mathcal{G}_j \right\}$ with $1\leq i<j\leq n $, $\mathcal{G}_i$ is a small perturbation of $\mathcal{G}_j$ according to $d_G$. The main assumptions encoded by the model presented in Equation \ref{Eq:BayesModel} are:
\begin{enumerate}
\item The distribution of the observations is assumed to be unimodal \emph{a priori};
\item The variability of the observations is characterised in terms of the dispersion around the mode. Such dispersion is defined in terms of $d_G$ a metric on $\left\{\mathcal{G}_{[N]} \right\}$;
\item The prior distribution for the mode is assumed to have the same functional form as the likelihood. This implies that it will be unimodal; its mode will be denoted by $\mathcal{G}^{0}$. We will not assume any structure on $\mathcal{G}^{0}$, unless we state otherwise.
\end{enumerate}
The first condition is set to guarantee identifiability of the model. The second condition enables the statistician to use the notion of similarity between networks, which can be subject to elicitation, to define variability in the space of graphs, which is, in contrast, very challenging to elicit. The third condition has parallel versions in the functional data analysis literature: we assume a parametric model for the error, with very simple structure, while allowing the trend to be as complex as it needs to be. An alternative approach would be to assume a trend with more defined structure and allow for a richer error structure. We elaborate more on this point in the discussion.

\section{Bayesian Modelling and Computation}\label{Sec:Inference}
In this section, we introduce Bayesian hierarchical models based on the distributions presented in Section \ref{Sec:TheModel}. For these models, we assume the same functional form for the sampling distribution and for the prior on the Fr\'{e}chet mean. We also discuss strategies for sampling from the posterior, with emphasis on the case when the normalising constant depends on the  Fr\'{e}chet mean.

\subsection{Bayesian Inference for the Centred Erd\"{o}s--R\'{e}nyi Model}\label{Sec:SER}
We now discuss a model of the form given in \eqref{Eq:BayesModel} that is inspired by the Centred Erd\"{o}s--R\'{e}nyi Model (CER). The intuition behind this model is the following: given a set of observed networks $\left\{  \mathcal{G}_1,\mathcal{G}_2,\dots,\mathcal{G}_n  \right\}$ in $\left\{\mathcal{G}_{[N]} \right\}^n$, their variability can be characterised in terms of the network $\mathcal{G}^{m}$ that serves as the mode of the distribution and the dispersion around that network. The network $\mathcal{G}^{m}$ can also be interpreted as the Fr\'{e}chet mean of $\left\{\mathcal{G}_{[N]} \right\}^n$ implied by the metric and the probability model.

Within this context, the contribution to the likelihood by each observation $\mathcal{G}_i$ is therefore given by:
\begin{equation}\label{Eq:logObSER}
p(\mathcal{G}_i\mid\mathcal{G}^m,\alpha) = \alpha^{d_H(\mathcal{G}_i,\mathcal{G}^m)} (1-\alpha)^{\frac{(N-1)N}{2}-d_H(\mathcal{G}_i,\mathcal{G}^m)},
\end{equation}
where $d_H(\cdot,\cdot)$ is the Hamming norm for matrices. Expressions \ref{Eq:BayesModel} and \ref{Eq:logObSER} provide the elements we need to propose the following Bayesian model:

\begin{definition}[CER\slash CER Model]\label{def:SERSER} Let $N$ and $n$ be elements of $\mathbb{N}$, and take $0<\alpha_0<1/2$. The CER\slash CER Model is a multivariate network model on $\left\{\mathcal{G}_{[N]} \right\}^n$ of the form
\begin{eqnarray}\label{Eq:SERSER}
p(\mathcal{G}_1,\mathcal{G}_2,\dots,\mathcal{G}_n\mid \mathcal{G}^{m},\alpha) & = &
\alpha_{0}^{d_H(\mathcal{G}^m,\mathcal{G}_0)} (1-\alpha_0)^{\frac{(N-1)N}{2}-d_H(\mathcal{G}^m,\mathcal{G}_0)} p(\alpha) \nonumber\\
 &  & \times
\prod_{i=1}^n \alpha^{d_H(\mathcal{G}_i,\mathcal{G}^m)} (1-\alpha)^{\frac{(N-1)N}{2}-d_H(\mathcal{G}_i,\mathcal{G}^m)} ,
\end{eqnarray}
where, the prior $p(\cdot)$ for $\alpha$ is a scaled Beta on $(0,\frac{1}{2})$. Here, $\mathcal{G}_{0} \in \left\{\mathcal{G}_{[N]} \right\}$ and $\alpha_{0}\in (0,1)$ are the hyperparameters of the model.
\end{definition}
We make no assumptions regarding $N$ and $n$. Expression \eqref{Eq:SERSER} is a consequence of the independence of the error, which should be noted. We assume a Beta distribution for $\alpha$ is reasonable, since it can be specified in such a way that it is unimodal and favours values close to zero.

Equation \ref{Eq:logObSER} proves helpful for understanding the properties of an Erd\"{o}s--R\'{e}nyi random graph as an measurement error model. This implies the following properties for the CER\slash CER Model:
\begin{enumerate}
\item The log-likelihood can be computed using $O(N^2n)$ operations; this should be kept in  mind when performing Bayesian computations, such as MCMC.
\item For $\alpha$ specified, the MLE is the graph $\widehat{\mathcal{G}^{m}}$ that minimises the average number of mismatches with respect to the observed networks. 
\end{enumerate}

The prior for $\mathcal{G}^{m}$ has $\mathcal{G}_{0}$ as its mode and its entropy is determined by the Hamming norm and $\alpha_0$. For the CER\slash CER model, the normalizing constant does not depend on either $\mathcal{G}^{m}$ or $\alpha$, therefore, samples of the posterior for $(\mathcal{G}^{m},\alpha)$ can be obtained via a Metropolis\slash Hastings algorithm with a mixture of kernels. To update $A_{\mathcal{G}^{m}}$, the adjacency matrix associated to $\mathcal{G}^{m}$, we use the following proposals:
\begin{enumerate}
 \item Each $A_{\mathcal{G}^{m}}(i,j)$ changes its value independently to $1-A_{\mathcal{G}^{m}}(i,j)$ with probability $0<\tau<1$, or stays fixed with probability $1-\tau$.
 \item Each $A_{\mathcal{G}^{m}}(i,j)$ is sampled independently from a $\mathrm{Ber}\left\{\frac{1}{n}\sum_{k=1}^{n}A_{\mathcal{G}_k}(i,j)\right\}$.
\end{enumerate} 
To update $\alpha$ we use a mixture of random walks that reflect at $0$ and $0.5$. For each of these random walks (indexed by $k$), the proposed value $\alpha^{*}$ for $\alpha^{(i+1)}$ is given by:
\begin{enumerate}
\item $y=\alpha^{(i)}+\zeta^{(i+1)}$, with $\zeta \sim \mathrm{Unif}(-\upsilon_k,\upsilon_k)$, if $0<y<0.5$;
\item $-y$, if $y<0$;
\item $1-y$, if $y>0.5$.
\end{enumerate}
The mixture is over $\left\{ \upsilon_1,\upsilon_2,\cdots,\upsilon_K\right\}  $.

\subsection{Bayesian Inference for Models in the Spherical Network Family}\label{Sec:ENF}
\label{Sec:EGM}

The Spherical Network Family was defined following the intuition that the likelihood should decrease as a function of the distance $d_G(\cdot,\cdot)$ with respect to a graph $\mathcal{G}^{m}$ that serves as the Fr\'{e}chet mean. When proposing the functional form, we adopted concepts from the Rotationally Symmetric Family, proposed by \cite{MardDryd}. In contrast to the CER/CER model discussed in Section \ref{Sec:SER}, more structure is left unspecified and the model presented in this section allows us to specify $d_G(\cdot,\cdot)$. To perform Bayesian inference for $(\mathcal{G}^{m},\gamma)$ as described in Definition \ref{def:ENF} , we propose to use a hierarchical model, following the form proposed in Equation \ref{Eq:BayesModel}:

\begin{definition}[SN\slash SN Model]\label{def:ENEN} Let $N$ and $n$ be elements of $\mathbb{N}$ and $d_G(\cdot,\cdot)$ a metric on $\left\{\mathcal{G}_{[N]} \right\}$. The SN\slash SN Model is a multivariate network model on $\left\{\mathcal{G}_{[N]} \right\}^n$ of the form
\begin{eqnarray}\label{Eq:SNSN}
p(\mathcal{G}_1,\mathcal{G}_2,\dots,\mathcal{G}_n\mid \mathcal{G}^{m},\gamma) &\propto & \exp\left\{ -\gamma_0 \phi(d_G (\mathcal{G}^{m},\mathcal{G}_0))   \right\}
p(\gamma)\nonumber \\ 
&  &\times \exp\left\{ -\gamma\sum_{i=1}^{n} \phi(d_G(\mathcal{G}_i,\mathcal{G}^{m}))\right\},
\end{eqnarray}
where, $p(\cdot)$ is the prior on $\gamma$, which has support on $\mathbb{R}^{+}$. Here, $\mathcal{G}_{0} \in \left\{\mathcal{G}_{[N]} \right\}$ and $\gamma_0 \in \mathbb{R}^{+}$ are the hyperparameters of the model.
\end{definition}

Some features of this model are:
\begin{enumerate}
\item The model allows for different specifications of the metric $d_G(\cdot,\cdot)$, which can be chosen with flexibility, for concreteness, \emph{e.g.} distance based on the graph Laplacian, or a metric based on subgraph counts.
\item It is straightforward to set up a Metropolis\slash Hastings algorithm to sample from the prior. The Metropolis ratio for updating $\mathcal{G}^{(\cdot)}$ is of the form:
\begin{equation}\label{Eq:SamplePriorSN}
H^{(t,t+1)}=\frac{\exp\left\{ -\gamma^{0}  \phi(d_G(\mathcal{G}^{(t+1)},\mathcal{G}^0)) \right\} }{\exp\left\{ -\gamma^{0} \phi(d_G(\mathcal{G}^{(t)},\mathcal{G}^0))  \right\} } \times \frac{q(\mathcal{G}^{(t)}\mid \mathcal{G}^{(t+1)})}{q(\mathcal{G}^{(t+1)}\mid \mathcal{G}^{(t)})},
\end{equation}
where $q$ is the proposal distribution; here, we are conditioning on the value of $\gamma^{0}$. 
\item The argument $\hat{\mathcal{G}}^{m}$ that maximises the log of the function: 
\begin{equation}\label{Eq:logLik}
\log \left( \prod_{i=1}^{n}Z(\mathcal{G}^{m},\gamma)\times p(\mathcal{G}_i\mid \mathcal{G}^{m},\gamma) \right)=-\gamma\sum_{i=1}^{n} \phi(d_G(\mathcal{G}_i,\mathcal{G}^{m})),
\end{equation}
where $\gamma$ is specified, coincides with the Fr\'echet mean of the observed networks when $\phi(x)=x^2$. This follows from applying the definition of a centroid directly.
\end{enumerate}

From a computational perspective, the fact that the normalizing constant for the observations (\emph{i.e.}, the reciprocal of $Z(\cdot)$ in Equation \ref{Eq:Partit}) depends on $\mathcal{G}^{m}$ implies that the Metropolis\slash Hastings algorithm cannot be implemented directly for sampling from the posterior of $(\mathcal{G}^{m},\gamma)$. For $\mathcal{G}^{m}$ unspecified, this model falls into the double-intractable constant distributions. Fortunately, sampling from the posterior for the SN\slash SN model falls into the setup discussed by \cite{Moller}. Therefore, the techniques proposed by  \cite{Moller}   and \cite{AndrieuRob} can be implemented to sample from the posterior. 

The MCMC scheme proposed by \cite{Moller} is based on the idea of simulating auxiliary variables $\mathcal{G}_{\ast,i}$, which are defined on the same sample space as the data $\mathcal{G}_i$, $1\leq i \leq n$. These variables are sampled so the factors $Z(\mathcal{G}^m,\gamma)^{-n}$ cancel from the Metropolis ratio. We now introduce some additional notation:
\begin{displaymath}
\vec{\mathcal{G}}=\left\{  \mathcal{G}_{1},\dots,\mathcal{G}_{n} \right\}\quad \text{ and } \quad \vec{\mathcal{G}_{\ast}}=\left\{  \mathcal{G}_{\ast,1},\dots,\mathcal{G}_{\ast, n} \right\}.
\end{displaymath}
When applied to the SN\slash SN Model, the Metropolis ratio for the scheme proposed by \cite{Moller} takes the form:
\begin{eqnarray}
H_{(\mathcal{G}^{m,(t+1)},\gamma^{(t+1)}\mid \mathcal{G}^{m,(t)},\gamma^{(t)})} & = & \frac{f(\vec{\mathcal{G}_{\ast}}^{(t+1)}\mid  \mathcal{G}^{m, (t+1)} ,\tilde{\alpha}) } {f(\vec{\mathcal{G}_{\ast}}^{(t)}\mid  \mathcal{G}^{m, (t)} ,\tilde{\alpha} ) }
       \times   \frac{p(\mathcal{G}^{m,(t+1)}\mid \mathcal{G}_0,\gamma_0)} { p(\mathcal{G}^{m,(t)}\mid \mathcal{G}_0,\gamma_0)}
       \times   \frac{p(\vec{\mathcal{G}}\mid \mathcal{G}^{m,(t+1)},\gamma^{(t+1)} )}{p(\vec{\mathcal{G}}\mid \mathcal{G}^{m,(t)},\gamma^{(t)} )}\nonumber \\
  &  &\times \frac{ p(\vec{\mathcal{G}_{\ast}}^{(t)}\mid \mathcal{G}^{m,(t)},\gamma^{(t)} )}{p(\vec{\mathcal{G}_{\ast}}^{(t+1)}\mid \mathcal{G}^{m,(t+1)},\gamma^{(t+1)} ) }
      \times \frac{q(\mathcal{G}^{m,(t)},\gamma^{(t)}\mid \mathcal{G}^{m,(t+1)},\gamma^{(t+1)})}
            {q(\mathcal{G}^{m,(t+1)},\gamma^{(t+1)} \mid \mathcal{G}^{m,(t)},\gamma^{(t)} )} ,             
\end{eqnarray}
where the terms of the form:
\begin{enumerate}
\item $p(\vec{\mathcal{G}}\mid \mathcal{G}^{m,(\cdot)},\gamma^{(\cdot)} )$ correspond to the product of kernel of the Boltzmann distribution evaluated at the data $\vec{\mathcal{G}}$, \emph{i.e.},
\begin{displaymath}
   p(\vec{\mathcal{G}}\mid \mathcal{G}^{m,(\cdot)},\gamma^{(\cdot)} )=\exp\left\{  -\gamma^{(\cdot)}\sum_{i=1}^n \phi\left[ d(\mathcal{G}_i,\mathcal{G}^{m,(\cdot)}) \right]  \right\}.
\end{displaymath}
 The notation $(\mathcal{G}^{m,(\cdot)},\gamma^{(\cdot)})$ means that are specified by the state of the chain.
\item $p(\vec{\mathcal{G}_{\ast}}^{(\cdot)}\mid \mathcal{G}^{m,(\cdot)},\gamma^{(\cdot)} )$ correspond to the kernel of the Boltzmann distribution evaluated at the auxiliary variables $\vec{\mathcal{G}_{\ast}}$. These auxiliary variables are obtained via a Metropolis-Hastings scheme. This scheme is the same as the one used for sampling from the prior (Equation \ref{Eq:SamplePriorSN}). 
\item $p(\mathcal{G}^{m,(\cdot)},\gamma^{(\cdot)}\mid \mathcal{G}_0,\gamma_0)$ correspond to the prior for $(\mathcal{G}^{m},\gamma)$ evaluated at the state of the chain.
\item $q(\mathcal{G}^{m,(\cdot)},\gamma^{(\cdot)} \mid \mathcal{G}^{m,(\cdot)},\gamma^{(\cdot)} )$ correspond to the proposal distribution for $(\mathcal{G}^{m},\gamma)$. To update $A_{\mathcal{G}^{m}}$, we use the same hybrid kernel as the one described in Section \ref{Sec:SER}. To update $\gamma$, the parameter that controls the entropy of the distribution, we use a hybrid kernel formed by a collection of random walks that reflect at 0. 
\item $f(\vec{\mathcal{G}_{\ast}}^{(\cdot)}\mid  \mathcal{G}^{m, (\cdot)} ,\tilde{\alpha})$ correspond to the conditional density of the auxiliary variables. We adopted the probability mass function of the Centred Erd\"{o}s--R\'{e}nyi Model as the conditional density for the auxiliary variables, which are denoted by $(\mathcal{G}^{1}_{\ast},\dots,\mathcal{G}^{n}_{\ast})$, \emph{i.e.}, 
\begin{displaymath}
f(\mathcal{G}^{1}_{\ast},\dots,\mathcal{G}^{n}_{\ast} \mid \mathcal{G}^{m} ,\tilde{\alpha})= \tilde{\alpha}^{ \sum_{i=1}^{n} d_H(\mathcal{G}_i,\mathcal{G}^m)} (1-\tilde{\alpha})^{\frac{(N-1)N}{2}-\sum_{i=1}^{n} d_H(\mathcal{G}_i,\mathcal{G}^m)},
\end{displaymath}
as in Section 2 of \cite{Moller}. Here, $\tilde{\alpha}$ is the posterior mean of the dispersion parameter of a CER\slash CER model, which can be estimated as described in Section \ref{Sec:SER}. This is the strategy suggested in Equation 7 of \cite{Moller}. 
\end{enumerate}

Some of these terms involve tuning parameters; for instance, $q(\mathcal{G}^{m,(\cdot)},\gamma^{(\cdot)} \mid \mathcal{G}^{m,(\cdot)},\gamma^{(\cdot)} )$ requires us to define a mixture of random walks (indexed by $k$), to propose a value $\gamma^{*}$ for $\gamma^{(i+1)}$. One way to define such random walks is given by:
\begin{enumerate}
\item $y=\gamma^{(i)}+\zeta^{(i+1)}$, with $\zeta \sim \mathrm{Unif}(-\upsilon_k,\upsilon_k)$, if $0<y$;
\item $-y$, if $y<0$;
\end{enumerate}
The mixture is over $\left\{ \upsilon_1,\upsilon_2,\cdots,\upsilon_K\right\}  $. Here $\upsilon$ is a tuning parameter. Specifying a prior for $\gamma$ also presents a challenge, since its behaviour will depend drastically on the metric.


\begin{figure}[!t]
\begin{center}
  \includegraphics[height=65mm]{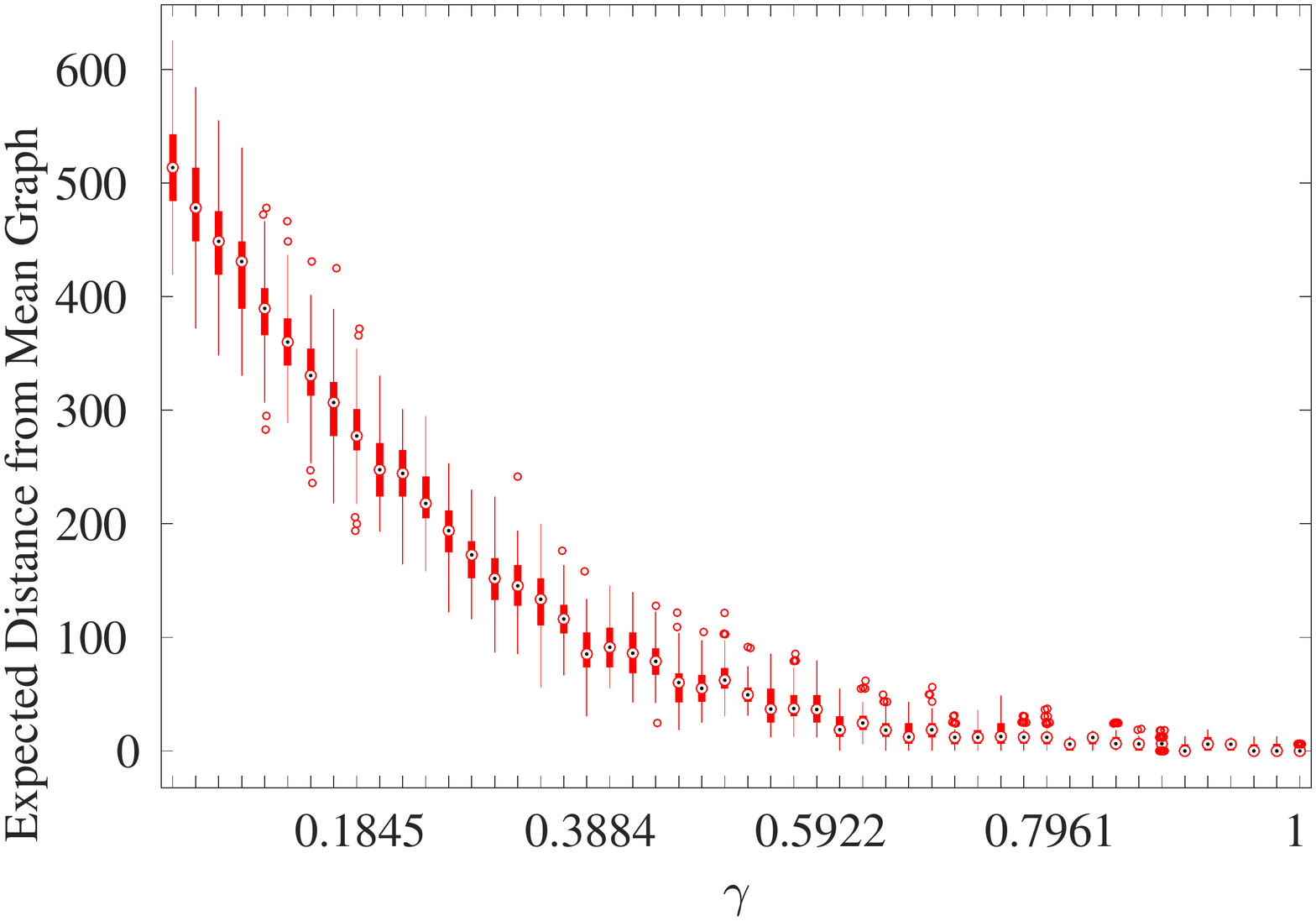}
  \caption{Distribution of the distance (summarised as a boxplot) to $\mathcal{G}^m$ for the SN model as a function of $\gamma$. Here $N=19$, $t=1$ and $\mathcal{G}^m$ was specified as the network displayed in Figure \ref{fig:GData1}. The figure shows that, for $\gamma>1$ increments on that parameter do not have a perceptible effect on the distribution. This plot serves to inform the scales that are relevant for defining random walk proposals for $\gamma$. }
  \label{fig:MetricsAndGamma}
\end{center}
\end{figure}

In Figure \ref{fig:MetricsAndGamma} we display the results of a simulation aimed to show the relationship between $\gamma$ and  $\mathbb{E}\left\{ \phi\left[d(\mathcal{G},\mathcal{G}^m)\right]\right\}$ for the SN model. These figures serve multiple purposes: (i) to provide information regarding which scales are reasonable for $\upsilon$, since they provide intuition of how a local change in $\gamma$ would impact a value that is easier to interpret; (ii) to inform where in $\mathbb{R}^{+}$ the practitioner should allocate most of the mass of the prior for $\gamma$; (iii) to make informed decisions of how to specify $\gamma^0$; (iv) to determine if it is more sensible to keep the prior support for $\gamma$ as $\mathbb{R}^{+}$, or to constrain it to an interval $(0,\kappa)$. The same applies to the random walks to update $\gamma$.

\section{Simulation Studies}\label{sec:Simul} 
In this section, we explore the behaviour of the CER\slash CER model and the SN\slash SN model via simulation studies. We consider that it should be of interest to practitioners to know: i) How precise the inferences become as a function of the number of networks analysed (we will refer to this number as the sample size); ii) To what extent samples from the posterior predictive resemble the data used to obtain the posterior; iii) How sensitive are the inferences with respect to model misspecification. With this in mind, we designed the simulation studies to investigate how the posterior concentrates around the true Fr\'{e}chet mean as a function of sample size, how regions of high mass of the predictive resemble a neighborhood of the data and robustness.

\subsection{Concentration of the Posterior as a Function of Sample Size}\label{ss:samp}
\begin{figure}[!b]
\begin{center}
  \includegraphics[height=75mm]{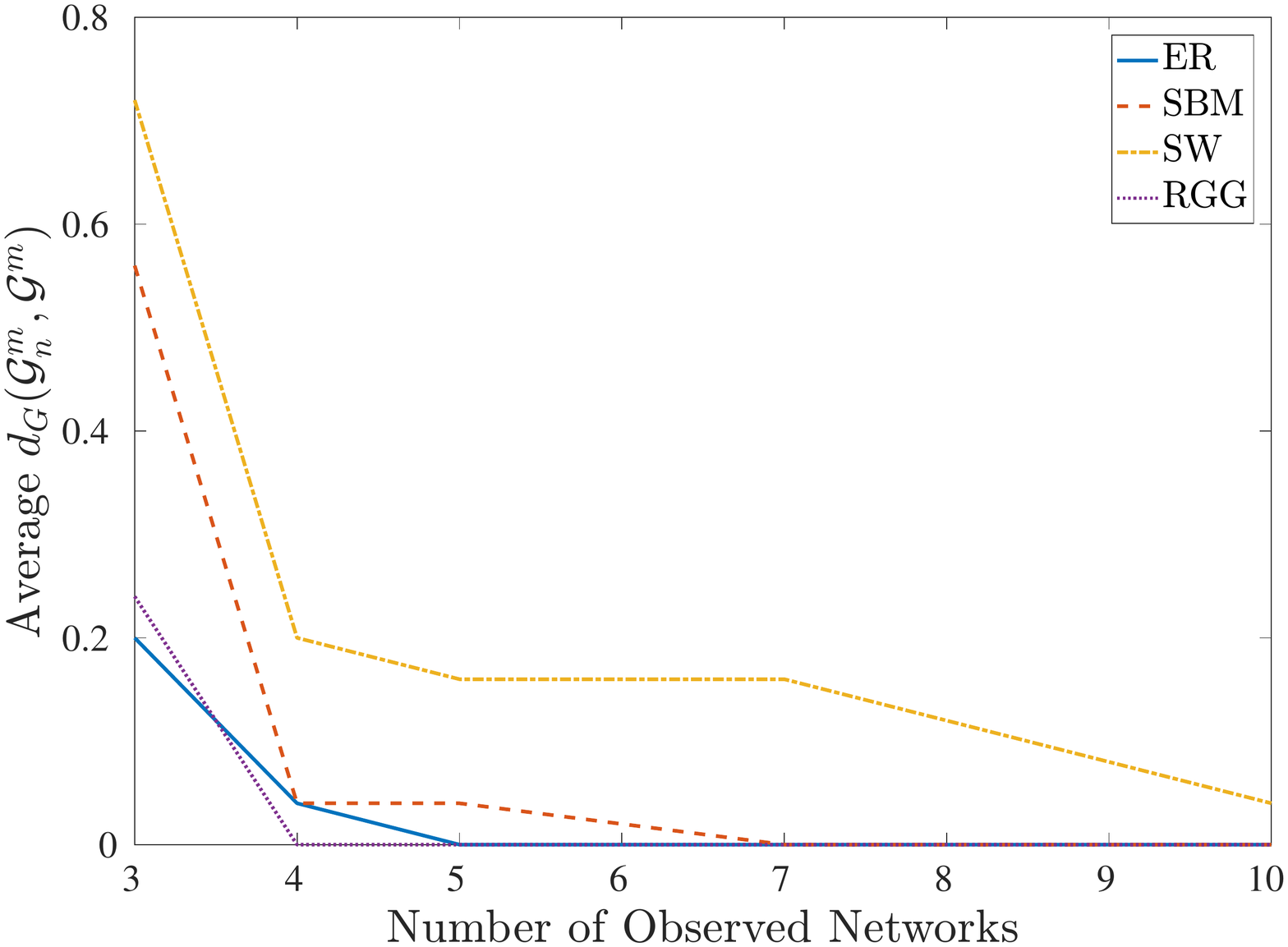}
  \caption{Average distance of posterior mode to $\mathcal{G}^{m}$ as a function of sample size.}
  \label{fig:Concentrate_SER}
\end{center}
\end{figure}

In this section, we propose simulation experiments to obtain better understanding of how the posterior for $\mathcal{G}^{m}$ concentrates around its true value as a function of sample size. Ideally, we would like to investigate if the limit
\begin{equation}\label{Eq:consistent}
\Pr \left\{ d_G(\mathcal{G},\mathcal{G}^{m} ) >\epsilon \mid  \mathcal{G}_1,\mathcal{G}_2,\dots,\mathcal{G}_n \right\} \to 0,
\end{equation}
holds almost surely as $n \to \infty$, given $\epsilon>0$, as $N$ is assumed fixed. This is equivalent to asking about the concentration of the posterior, as explained in Section 13.4.1  of \cite{GhosVaar}. The intuition behind Equation \ref{Eq:consistent} is that, as the sample size $n$ increases, the probability mass of the posterior tends to concentrate on a neighborhood of the true value of the parameter. This statement should be valid for every size of the neighbourhood $\epsilon>0$. In Equation \ref{Eq:consistent}, $\mathcal{G}^{m}$ is the true value of the mode and $\mathcal{G}$ is a sampled value from the posterior distribution implied by $\left\{ \mathcal{G}_1,\mathcal{G}_2,\dots,\mathcal{G}_n \right\}$. Equation \ref{Eq:consistent} provides the principle behind the following simulation experiments:
\begin{enumerate}
\item Explore how the distance between the point estimate  $\widehat{\mathcal{G}}^m$ given by the posterior mode and $\mathcal{G}^m$ behaves as a function of sample size.
\item Investigate how the probability 
\begin{equation}
\Pr \left\{ d_G(\mathcal{G},\mathcal{G}^{m} ) >\epsilon \mid  \mathcal{G}_1,\mathcal{G}_2,\dots,\mathcal{G}_n \right\}<\delta,
\end{equation}
behaves as a function of $n\in \mathbb{N}^{+}$, here $\epsilon>0$, $\delta>0$ are in turn fixed.
\end{enumerate}
The first simulation provides insight about the speed of convergence of a point estimate (see Fig.~\ref{fig:Concentrate_SER}), while the second simulation investigates how the posterior mass becomes contained in a neighborhood of size $\epsilon$ of $\mathcal{G}^{m}$ as the sample size increases
(see Table~\ref{tab:Concentrate}). Here, the size of the neighborhood is controlled by $\epsilon$, and $\delta$ serves as a threshold for the amount of posterior mass to be allowed outside the neighborhood. 

The simulation regimes are given by:
\begin{enumerate}
\item The type of hierarchical model under study (CER\slash CER, SN\slash SN);
\item The structure imposed on $\mathcal{G}^{m}$, the centroid of the distribution. These were generated from the Erd\"{o}s--R\'{e}nyi model (ER), the Stochastic Block model (SBM), the Small World model (SW),  or as a Random Geometric Graph (RGG). The specification of the parameters for these models is displayed in Table \ref{tab:SimRegimes}.
\end{enumerate}

\begin{table}[h]
\begin{center}
\begin{tabular}{|c| l|}
\hline
\text{Random Graph Model} & \text{Specification} \\
\hline
  \text{ER} & \text{Probability of inclusion was set to }$0.1$. \\
  \text{RGG} & \text{This is a proximity graph defined on the unit square. }\\
             &\text{The radius of the ball was set to} $r=0.175$.\\
  \text{SBM} & \text{We set the number of blocks } $K=3$ \text{, with all membership}\\
             &\text{probabilities equal to } $0.333$ \text{the inclusion probabilities}\\
             &\text{were set as} $0.16$ \text{and} $0.075$ \text{for diagonal and}\\
             &\text{non-diagonal blocks, respectively}.  \\
 \text{SW}  & \text{We set the degree of the lattice to }$2$\text{ and the} \\
            &\text{probability of re-wiring to } $0.2$.\\
\hline
\end{tabular}
\caption{Random graph models and the corresponding parameter specification used to define the simulation regimes. The parameters were chose so realisations would have approximately the same density across different models. }\label{tab:SimRegimes}
\end{center}
\end{table}

For both the CER\slash CER model and the SN\slash SN model, we used $250$ samples after a burn-in of $100,000$, and a lag of $50$. The size of the networks we considered was $N=50$. The value for $\gamma_0$ was specified as $0.01$ (for the CER\slash CER model, we set $\alpha_0=0.01$). A different value of $\mathcal{G}_0$ was obtained for each $\mathcal{G}^m$; it was sampled from $p(\cdot \mid \mathcal{G}^m,\gamma_0)$. This way, we were able to make $\mathcal{G}^m$ exhibit the different types of structure we needed while keeping it as a perturbation of $\mathcal{G}_0$, with concentration given by $\gamma_0$ ($\alpha_0$).

Results from the first and second simulation for the CER\slash CER model are summarised in Fig.~\ref{fig:Concentrate_SER} and Table~\ref{tab:Concentrate}, respectively. These results suggest that, the more homogeneous the adjacency matrix is in terms of inclusion probabilities, the faster the posterior concentrates around the true value. (The Small World and the Stochastic Block models take longer to converge then the Erd\"{o}s--R\'{e}nyi and the Random Geometric Graph do.)  The Small World model turned out to be especially challenging for our approach, as a consequence of the choice for $\mathcal{G}^0$, which favors a lattice structure. Since the Small World graph is obtained from a re-wiring on a lattice, it takes a larger sample size to disambiguate between the outcomes of the re-wiring process and the perturbation induced by the SN Model.

\begin{table}[h]
\begin{center}
\begin{tabular}{|l |l | c c c|}
\hline
n & $\text{Generative Model for }\mathcal{G}^m $& $\epsilon=1$ & $\epsilon=2 $&$ \epsilon=3 $ \\
\hline
  &                                      &             & $\delta=0.05$ &            \\   
\hline
3 & \text{ RGG }      &       0.92      &       1      &     1        \\
5 & \text{ RGG }      &        1      &       1      &     1        \\
3 & \text{ ER }      &        0.66      &       0.97      &     1        \\
5 & \text{ ER }      &       0.93     &       1      &     1        \\
7 & \text{ ER }      &       1      &       1      &     1        \\
3 & \text{ SBM }      &       0.63      &       0.87      &     0.96        \\
5 & \text{ SBM }      &        0.83     &       0.98      &     1        \\
7 & \text{ SBM }      &        0.91     &       1      &     1        \\
10 & \text{ SBM }      &        1      &       1      &     1        \\
3 & \text{ SW }      &        0.43      &       0.61      &     0.73        \\
5 & \text{ SW }      &       0.62     &       0.77     &     0.89        \\
7 & \text{ SW }      &        0.74     &       0.86      &     0.98        \\
10 & \text{ SW }      &       0.81     &       0.96      &     1        \\
\hline
\end{tabular}
\caption{Proportion of replications where $1-\delta$ of the posterior mass for $\mathcal{G}^m$ is within a ball of radius $\epsilon$ of the true value. We used 100 replications. }\label{tab:Concentrate}
\end{center}
\end{table}

We compared the performance of our method to the point estimate $\widehat{\mathcal{G}^m}$ we would obtain by computing the majority vote of the data $\left\{ \mathcal{G}_1,  \mathcal{G}_2,\dots,  \mathcal{G}_n \right\}$. We used the posterior mode implied by the CER\slash CER model to illustrate our method. Results are summarised in Table \ref{tab:ConcentrateCompare}.

\begin{table}[h]
\begin{center}
\begin{tabular}{|l |l | c c c| c c c|}
\hline
n & $\text{Generative Model for }\mathcal{G}^m $& $\epsilon=1$ & $\epsilon=2 $&$ \epsilon=3 $ & $\epsilon=1$ & $\epsilon=2 $&$ \epsilon=3 $ \\
\hline
  &                                      &             & $\text{Majority Vote}$ &   &             & $\text{CER}$ &     \\   
\hline
3 & \text{ RGG }      & 0.91   &   1    &     1  & 0.94   &   0.99    &     1          \\
5 & \text{ RGG }      & 1   &   1    &     1  & 1   &   1    &     1    \\
3 & \text{ ER }       & 0.99   &   1    &     1  & 0.96   &   0.99    &     1     \\
5 & \text{ ER }       & 1   &   1    &     1  &  1  &   1    &     1    \\
3 & \text{ SBM }      & 0.95   &   0.99    &  1  &  0.95   &   0.99    &     1     \\
5 & \text{ SBM }      & 1   &   1    &     1  & 1   &   1    &     1     \\
3 & \text{ SW }       & 0.97   &   1    &    1  & 0.91   &   0.99    &     1   \\
5 & \text{ SW }       & 1   &   1    &     1  & 0.99   &   1    &     1   \\
\hline
\end{tabular}
\caption{Proportion of replications where $1-\delta$ of the posterior mass for $\mathcal{G}^m$ is within a ball of radius $\epsilon$ of the true value. We used 100 replications. }\label{tab:ConcentrateCompare}
\end{center}
\end{table}

\subsection{Network Prediction}\label{sss:priorSamp}
In this section, we investigate the behaviour of our methodology in terms of prediction. We do this according to the following intuition: Given a sample $\left\{  \mathcal{G}_1,\mathcal{G}_2, \dots ,\mathcal{G}_n \right\}$, the posterior predictive distribution should satisfy the criterion that regions with highest posterior density tend to be contained in an open covering of the original sample. To protect ourselves against artifacts due to overfitting, we let the sample used to compute the posterior predictive be distinct from the sample used to compute the open covering. Both from conceptual and computational perspectives, the use of an open covering is valid in this context, since we are working on a metric space of graphs.

We now use this intuition to propose a simulation study. We first generate a sample 
\begin{displaymath}
\left\{  \mathcal{G}_1,\mathcal{G}_2, \dots ,\mathcal{G}_n,\mathcal{G}_{n+1},\dots,\mathcal{G}_{n_t} \right\}
\end{displaymath}
for $(\mathcal{G}^m, \gamma)$ specified, and then we partition this sample into a training set $\left\{ \mathcal{G}_i \right\}_{i \leq n}$ and a test set $\left\{ \mathcal{G}_i \right\}_{n< i \leq n_t}$. Here, $n_t-n$ may be considered a tuning parameter for the simulation, specified by the statistician. Here, the training set will be used to obtain the posterior predictive distribution, while the test set will be used to compute the envelope. In the context of the models we have presented, the assumptions regarding similarity are encoded by the metric $d_G(\cdot,\cdot)$. To make these notions precise, we introduce the tuning parameters $\delta \in (0,1)$ and $\psi_{\delta}>0$. Here, $\psi_{\delta}$ is the infimum of $\left\{ \psi :\psi >0  \right\}$ for which
\begin{equation}\label{Eq:Covering}
\Pr\left\{  \mathcal{G} \in \bigcup_{k=n+1}^{n_t} \mathbb{B}(\mathcal{G}_k; \psi)  \mid \mathcal{G}_1,\mathcal{G}_2,\dots,\mathcal{G}_n \right\}\geq 1-\delta,
\end{equation}
holds. In Equation \ref{Eq:Covering}, $\mathbb{B}(\mathcal{G}_k; \psi)$ denotes the ball with centre $\mathcal{G}_k$ and radius $\psi$ corresponding to $d_G(\cdot,\cdot)$, and $\mathcal{G}$ is a sampled value from the predictive distribution implied by the model and $\left\{ \mathcal{G}_1,\mathcal{G}_2,\dots,\mathcal{G}_n \right\}$. The larger $\psi_{\delta}$ is, the less concentrated the posterior predictive distribution will be around the test set. One way to interpret the size of $\psi_{\delta}$ more effectively is by comparing it to quantities for which our intuitions are better informed. We propose comparing it to $\rho_\delta$, the infimum of $\left\{ \rho :\rho>0  \right\}$ for which
\begin{displaymath}
\Pr\left\{  \mathcal{G} \in \mathbb{B}(\mathcal{G}^m; \rho) \mid \mathcal{G}^m, \gamma \right\}\geq 1-\delta,
\end{displaymath}
holds, \emph{i.e.}, $\rho_\delta$ is the size of the contour set that contains $1-\delta$ of the probability mass under the specified model.

Implementing this simulation in practice is straightforward: we first compute the distances between each sample from the posterior predictive distribution and the element of $\left\{ \mathcal{G}_{n+1},\mathcal{G}_{n+2},\dots,\mathcal{G}_{n_t} \right\}$ closest to it. The estimate of $\psi_{\delta}$ is given by the $1-\delta$ quantile of those distances. Results are summarised in Table~\ref{tab:Concentrate} for the CER\slash CER model and the SN\slash SN model.

We used the random graph models and parameter specifications listed in Table \ref{tab:SimRegimes}. We used the same settings for the MCMC (number of samples from the posterior, burn-in, lag) and choices for the hyperparameters ($\mathcal{G}_0$,$\gamma_0$ and $\alpha_0$) as in Section \ref{ss:samp}. The size of the networks was set to $N=50$.




Results are summarised in Table~\ref{tab:Predict}. Here, larger values of $\psi_\delta$ indicate that a larger open covering of a sample is needed to mimic regions of the posterior predictive distribution with high probability mass. To be able to compare across regimes, we use the quotient of $\psi_\delta$ over $\rho_\delta$, where $\rho_\delta$ serves as a quantile. The results in Table~\ref{tab:Predict} suggest that the size of the neighborhood needed to contain the mass of the predictive decays very slowly with respect to the sample size. We also observed that the results were not very sensitive with respect to the generative model for $\mathcal{G}^m$. Recall that $\rho_\delta$ is model dependent.

\begin{table}[h]
\begin{center}
\begin{tabular}{|l l |c c|}
\hline
$n$ & \text{Generative Model}  & \text{CER}\slash \text{CER} & \text{SN}\slash \text{SN}\\
  & $\text{for }\mathcal{G}^m $&    $ \psi_{\delta}/\rho_\delta$           &     $ \psi_{\delta}/\rho_\delta $         \\   
  
\hline
3 & \text{ ER }      &      1.4447   &      1.0551    \\
5 & \text{ ER }      &      1.3847   &      1.0253     \\
7 & \text{ ER }      &      1.3676   &      1.0072      \\
10 & \text{ ER }      &     1.3612  &       0.9590     \\
3 & \text{ RGG }      &     1.4006    &     1.0516    \\
5 & \text{ RGG }      &     1.3988    &     1.0247    \\
7 & \text{ RGG }      &     1.3953    &     0.9958   \\
10 & \text{ RGG }      &    1.3635    &     0.9366  \\
3 & \text{ SBM }      &     1.4141    &     1.0573   \\
5 & \text{ SBM }      &     1.3824     &    1.0410       \\
7 & \text{ SBM }      &     1.3800   &      0.9898      \\
10 & \text{ SBM }     &     1.3741   &      0.9516    \\
3 & \text{ SW }      &      1.4788    &     1.0697  \\ 
5 & \text{ SW }      &      1.4494    &     1.0419   \\
7 & \text{ SW }      &      1.3953    &     0.9937      \\
10 & \text{ SW }      &     1.3682     &    0.9545     \\
\hline
\end{tabular}
\caption{Average value  $ \psi_{\delta}$ for size of neighborhood needed so $m$ samples from the predictive implied by $n$ data points encloses $1-\delta$ of the predictive distribution associated with the true value of $\mathcal{G}^m$ and $\alpha$. Here we assume a CER\slash CER model (third column, left to right) and a SN\slash SN model (fourth column, left to right). For the CER\slash CER model $\rho_\delta=17$, while for the SN\slash SN model, $\rho_\delta=3642.1$. The size of the network is $50$ and $\alpha=0.01$. We set $\delta=0.1$ and $m=20$ for all regimes. The average is computed over $100$ replications. }\label{tab:Predict}
\end{center}
\end{table} 


\subsection{Robustness}\label{sss:postSamp}
In this section, we evaluate the proposed methodology in terms of robustness regarding model misspecification. This is important, since we are making heavily parametric assumptions about the distribution of the deviations with respect to the Frech\'{e}t mean.  We approach this task in two different ways: (i) by using visual diagnostics based on posterior predictive checks (\cite{GelMeng}), and (ii) by investigating the behaviour of the Bayesian $\chi^2$ (\cite{JohnV}) under different scenarios. These methods are further discussed in  Appendix B.

The types of misspecification we consider in this simulation study are:
\begin{enumerate}
\item Fitting the model when the data was generated by a model based on a different metric on the space of labelled graphs.
\item Fitting the model when the data was generated by a dynamic network model.
\end{enumerate}
For the first type of misspecification, we will fit the SN\slash SN model assuming the diffusion distance~\citep{Hamm} while the generative model is a CER\slash CER model, or vice versa. For the second type of misspecification, we generate data from the dynamic network model implied by making 
$\mathcal{G}_{k+1}(i,j)\mid \mathcal{G}_{k}(i,j)$ the conditional of a bivariate Bernoulli and then, made all entries of $\mathcal{G}_{k+1}$ conditionally independent given $\mathcal{G}_k$, which induces a Markov structure on $\left\{\mathcal{G}_1,\dots,\mathcal{G}_n\right\}$.

To fit the models, we used the same settings for the MCMC (number of samples from the posterior, burn-in, lag) and choices for the hyperparameters ($\mathcal{G}_0$,$\gamma_0$ and $\alpha_0$) as in Sections \ref{ss:samp} and \ref{sss:priorSamp}. The size of the networks was set to $N=50$.

Results are summarised in Table~\ref{tab:RobustPPS}. In this table, we display the proportion of times where each diagnostic provided evidence for lack of fit over 100 simulated data sets. Both types of diagnostic require us to specify a univariate summary of the of the data. We decided to focus on different quantiles of the degree distribution. The results we obtained suggest that is difficult to assess model misspecification in terms of the center of the degree distribution. It was easier to find evidence of model misspecification, via posterior predictive checks or the Bayesian $\chi^2$, when the focus was on the upper tail of the degree distribution.

\begin{table}[ht]
\begin{center}
\begin{tabular}{|l |l  l  l | c c|}
\hline
$n$ & \text{Model used to fit the data }& \text{type of misspecification} & \text{univariate summary} & \text{PPC} & $\text{Bayes }\chi^2$ \\
\hline
3 & \text{ Spherical Network Model}      &       \text{Dependence}     &  \text{10 quantile of degree}    &    0      &     0       \\
3 & \text{ Spherical Network Model}      &       \text{Dependence}     &  \text{50 quantile degree}    &    0      &     0        \\
3 & \text{ Spherical Network Model}      &       \text{Dependence}     &  \text{90 quantile degree}    &    0.08      &     0        \\
10 & \text{ Spherical Network Model}      &       \text{Dependence}     &  \text{10 quantile of degree}    &    0.02      &     0.02        \\
10 & \text{ Spherical Network Model}      &       \text{Dependence}     &  \text{50 quantile degree}    &    0      &     0.01        \\
10 & \text{ Spherical Network Model}      &       \text{Dependence}     &  \text{90 quantile degree}    &    0.09      &     0.11        \\
50 & \text{ Spherical Network Model}      &       \text{Dependence}     &  \text{10 quantile of degree}    &    0.07      &     0.05        \\
50 & \text{ Spherical Network Model}      &       \text{Dependence}     &  \text{50 quantile degree}    &    0.02      &     0.03        \\
50 & \text{ Spherical Network Model}      &       \text{Dependence}     &  \text{90 quantile degree}    &    0.76      &     0.93        \\
3 & \text{ Spherical Network Model}      &       \text{Metric}     &  \text{10 quantile of degree}    &    0.09      &     0        \\
3 & \text{ Spherical Network Model}      &       \text{Metric}     &  \text{50 quantile degree}    &    0      &     0       \\
3 & \text{ Spherical Network Model}      &       \text{Metric}     &  \text{90 quantile degree}    &    0.11      &     0        \\
10 & \text{ Spherical Network Model}      &       \text{Metric}     &  \text{10 quantile of degree}    &    0.03      &     0.07        \\
10 & \text{ Spherical Network Model}      &       \text{Metric}     &  \text{50 quantile degree}    &    0.02      &     0.03        \\
10 & \text{ Spherical Network Model}      &       \text{Metric}     &  \text{90 quantile degree}    &    0.05      &     0.07        \\
50 & \text{ Spherical Network Model}      &       \text{Metric}     &  \text{10 quantile of degree}    &    0.07      &     0.22        \\
50 & \text{ Spherical Network Model}      &       \text{Metric}     &  \text{50 quantile degree}    &    0      &     0.14        \\
50 & \text{ Spherical Network Model}      &       \text{Metric}     &  \text{90 quantile degree}    &    0.09      &     0.97        \\
3 & \text{ Centred Erd\"{o}s--R\'{e}nyi Model}      &       \text{Dependence}     &  \text{10 quantile of degree}    &    0      &     0       \\
3 & \text{ Centred Erd\"{o}s--R\'{e}nyi Model}      &       \text{Dependence}     &  \text{50 quantile degree}    &    0      &     0        \\
3 & \text{ Centred Erd\"{o}s--R\'{e}nyi Model}      &       \text{Dependence}     &  \text{90 quantile degree}    &    0.03      &     0.04        \\
10 & \text{ Centred Erd\"{o}s--R\'{e}nyi Model}      &       \text{Dependence}     &  \text{10 quantile of degree}    &    0.02      &     0.01        \\
10 & \text{ Centred Erd\"{o}s--R\'{e}nyi Model}      &       \text{Dependence}     &  \text{50 quantile degree}    &    0      &     0.00        \\
10 & \text{ Centred Erd\"{o}s--R\'{e}nyi Model}      &       \text{Dependence}     &  \text{90 quantile degree}    &    0.07      &     0.14        \\
50 & \text{ Centred Erd\"{o}s--R\'{e}nyi Model}      &       \text{Dependence}     &  \text{10 quantile of degree}    &    0.07      &     0.05        \\
50 & \text{ Centred Erd\"{o}s--R\'{e}nyi Model}      &       \text{Dependence}     &  \text{50 quantile degree}    &    0.03      &     0.02        \\
50 & \text{ Centred Erd\"{o}s--R\'{e}nyi Model}      &       \text{Dependence}     &  \text{90 quantile degree}    &    0.74      &     0.89        \\
3 & \text{ Centred Erd\"{o}s--R\'{e}nyi Model}      &       \text{Metric}     &  \text{10 quantile of degree}    &    0.03      &     0        \\
3 & \text{ Centred Erd\"{o}s--R\'{e}nyi Model}      &       \text{Metric}     &  \text{50 quantile degree}    &    0      &     0       \\
3 & \text{ Centred Erd\"{o}s--R\'{e}nyi Model}      &       \text{Metric}     &  \text{90 quantile degree}    &    0.12      &     0        \\
10 & \text{ Centred Erd\"{o}s--R\'{e}nyi Model}      &       \text{Metric}     &  \text{10 quantile of degree}    &    0.06      &     0.05        \\
10 & \text{ Centred Erd\"{o}s--R\'{e}nyi Model}      &       \text{Metric}     &  \text{50 quantile degree}    &    0.03      &     0.01        \\
10 & \text{ Centred Erd\"{o}s--R\'{e}nyi Model}      &       \text{Metric}     &  \text{90 quantile degree}    &    0.07      &     0.11        \\
50 & \text{ Centred Erd\"{o}s--R\'{e}nyi Model}      &       \text{Metric}     &  \text{10 quantile of degree}    &    0.07      &     0.16        \\
50 & \text{ Centred Erd\"{o}s--R\'{e}nyi Model}      &       \text{Metric}     &  \text{50 quantile degree}    &    0      &     0.12        \\
50 & \text{ Centred Erd\"{o}s--R\'{e}nyi Model}      &       \text{Metric}     &  \text{90 quantile degree}    &    0.11      &     0.93        \\
\hline
\end{tabular}
\caption{Proportion of times where each diagnostic provided evidence for lack of fit over 100 simulated data sets. The regimes are given by the generative model, the type of misspecification and the univariate summary used for the diagnostics. }\label{tab:RobustPPS}
\end{center}
\end{table}

\section{Data Analysis}\label{Sec:Data}
\subsection{Gene Interaction Data}\label{Sec:SysBio}
It has become common practice in systems biology to estimate networks that have either genes or proteins as nodes and where the edges represent, either a potential flow of information (protein signalling) or other evidence of association. Estimating the network is often an intermediate step within a series of inferences and\slash or decisions; this is for example the case for the research aimed for the development of new treatments and vaccines. In this context, having an appropriate characterisation of the variability across different estimated networks can prove key when trying to assess the uncertainty to be associated to the final inferences\slash decisions. The variability of the inference of such networks can be due to: i) use of different data bases, ii) use of different technologies to pre-process the data, iii) use of different criteria to decide what constitutes and edge.

\begin{figure}[!t]
\begin{center}
  \includegraphics[height=40mm]{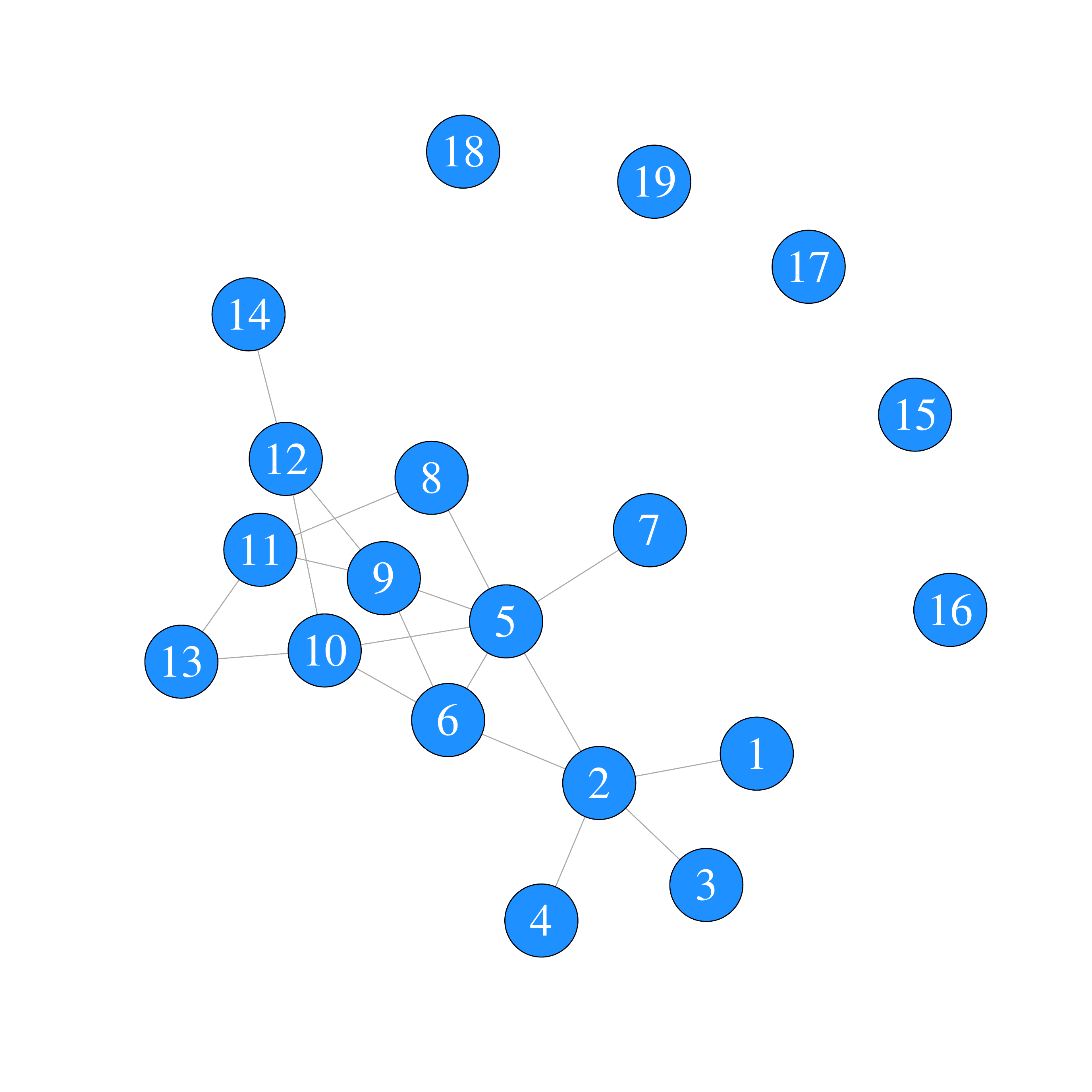}
  \includegraphics[height=40mm]{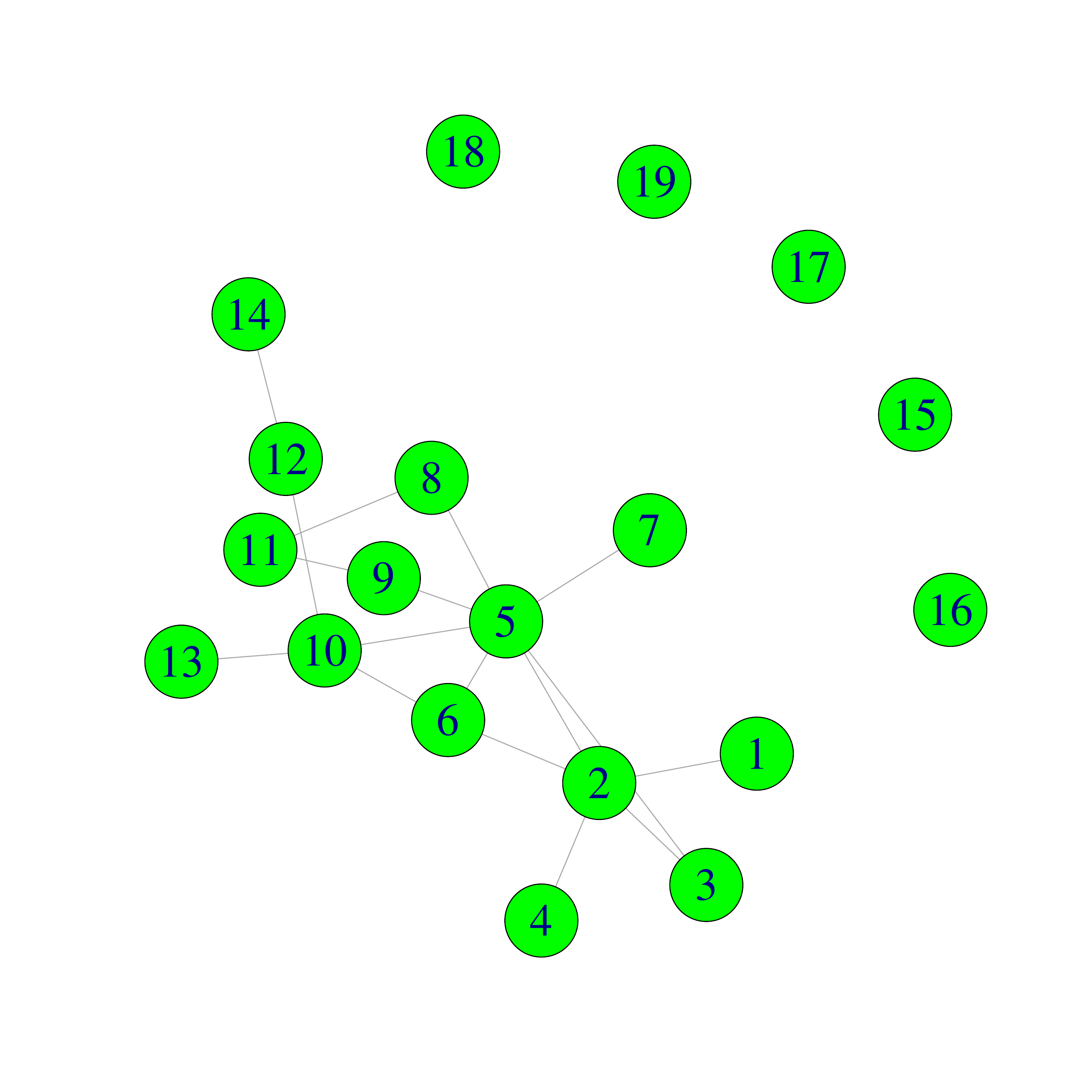}
  \includegraphics[height=40mm]{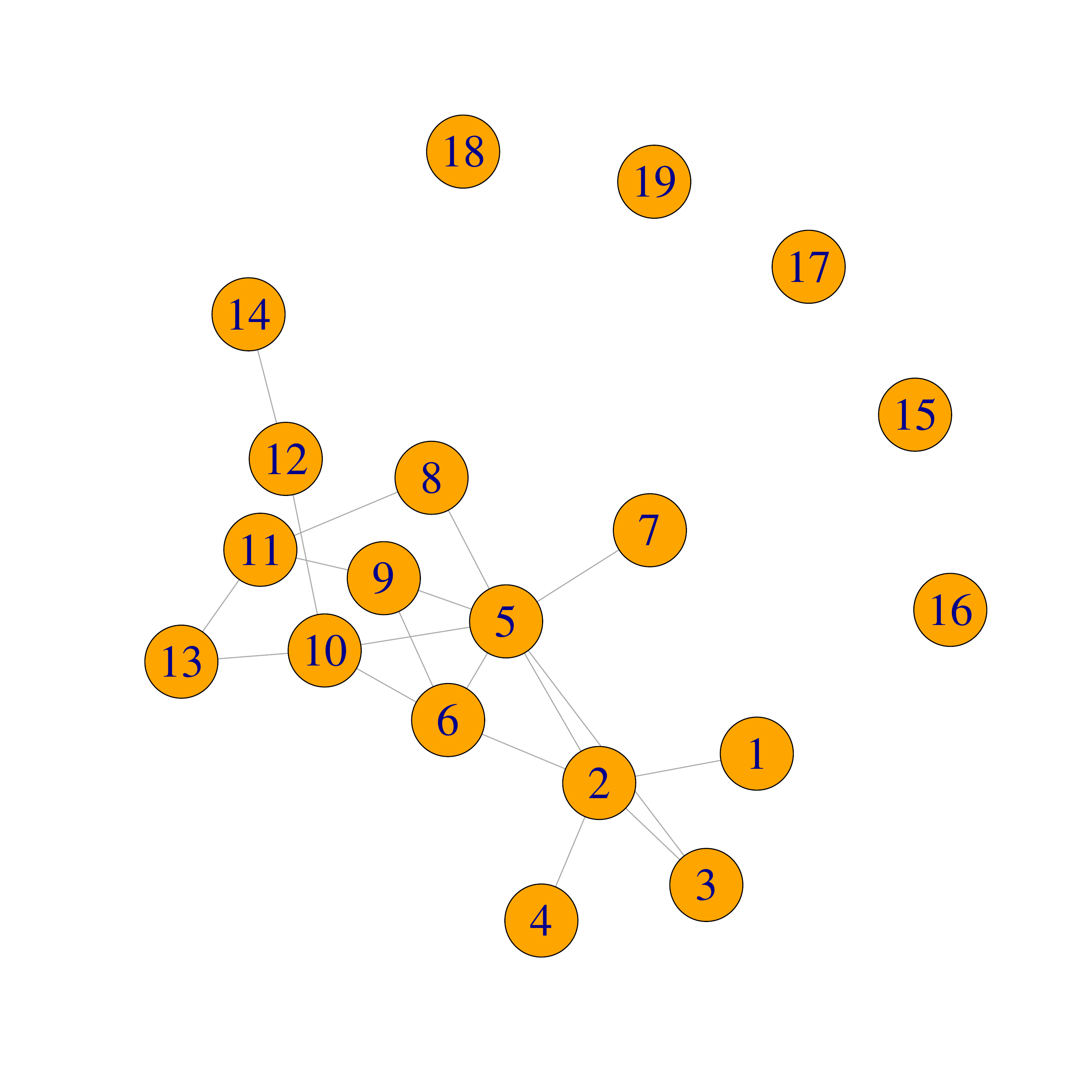}
  \includegraphics[height=40mm]{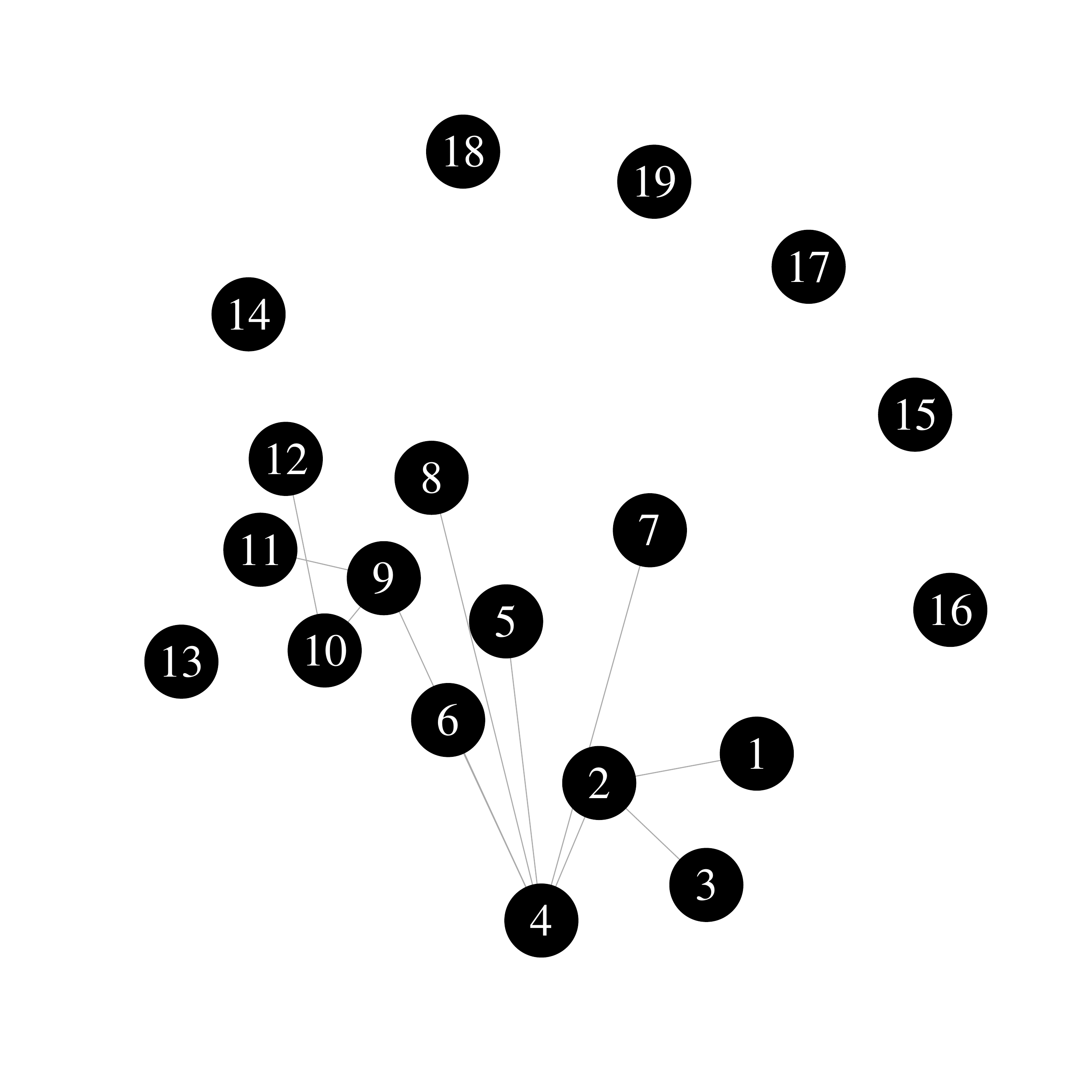}
  \caption{Example of multiple network data in the context of cancer genomics, with each node one of the 19 most frequently mutated human cancer genes (see Section~\ref{Sec:Data}). Top Left: Network $\text{N1}$ inferred from curated databases; Top Right: Network $\text{N2}$ determined by a series of individual experiments; Bottom Left: Network $\text{N3}$ inferred via text mining; Bottom Right: Network $\text{N4}]$ inferred via co-expression. The set of nodes of this network is formed by the 19 most frequently mutated human cancer genes .}
 \label{fig:GData1}
\end{center}
\end{figure}


\begin{figure}[htbp]
\begin{center}
\includegraphics[height=62mm]{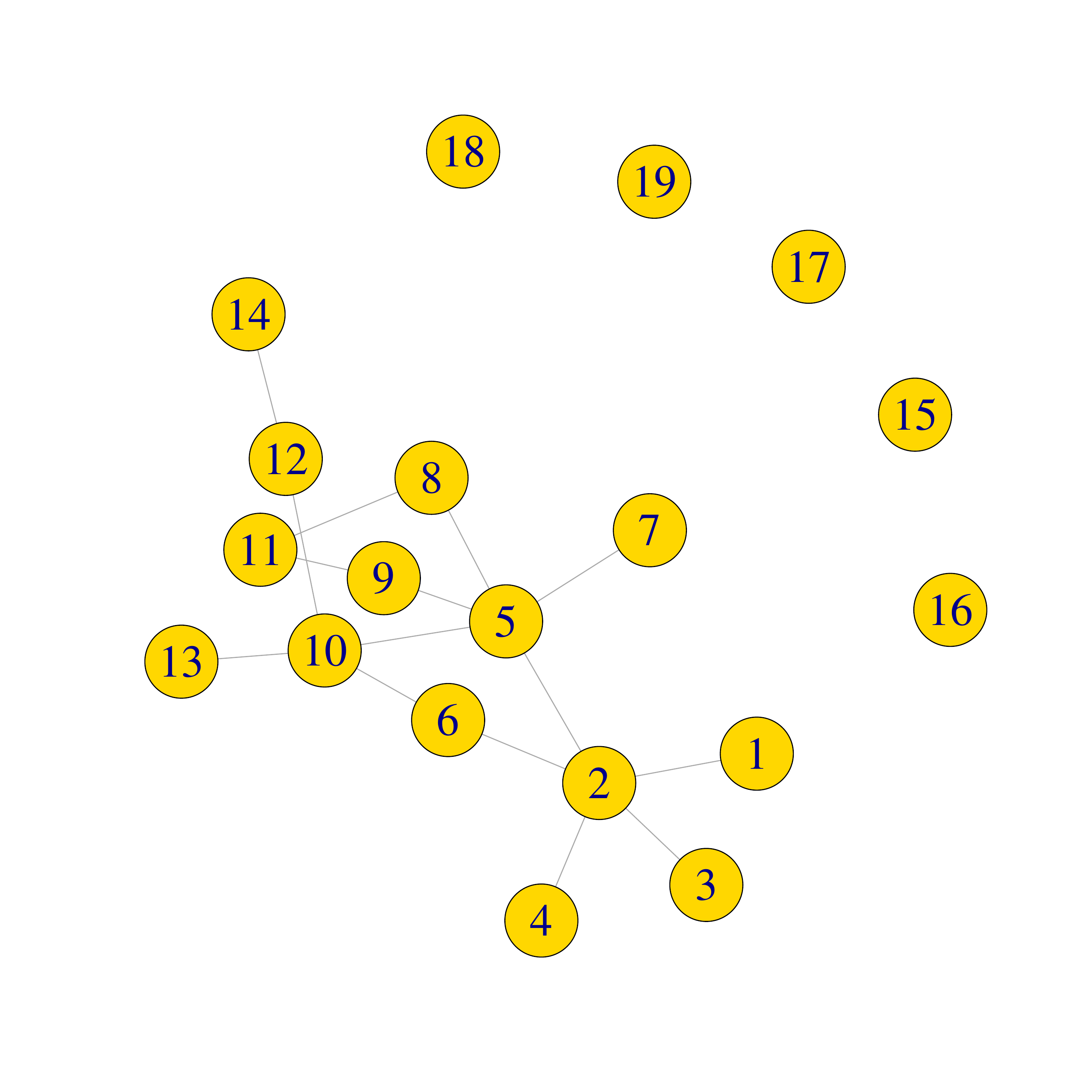}
\includegraphics[height=62mm]{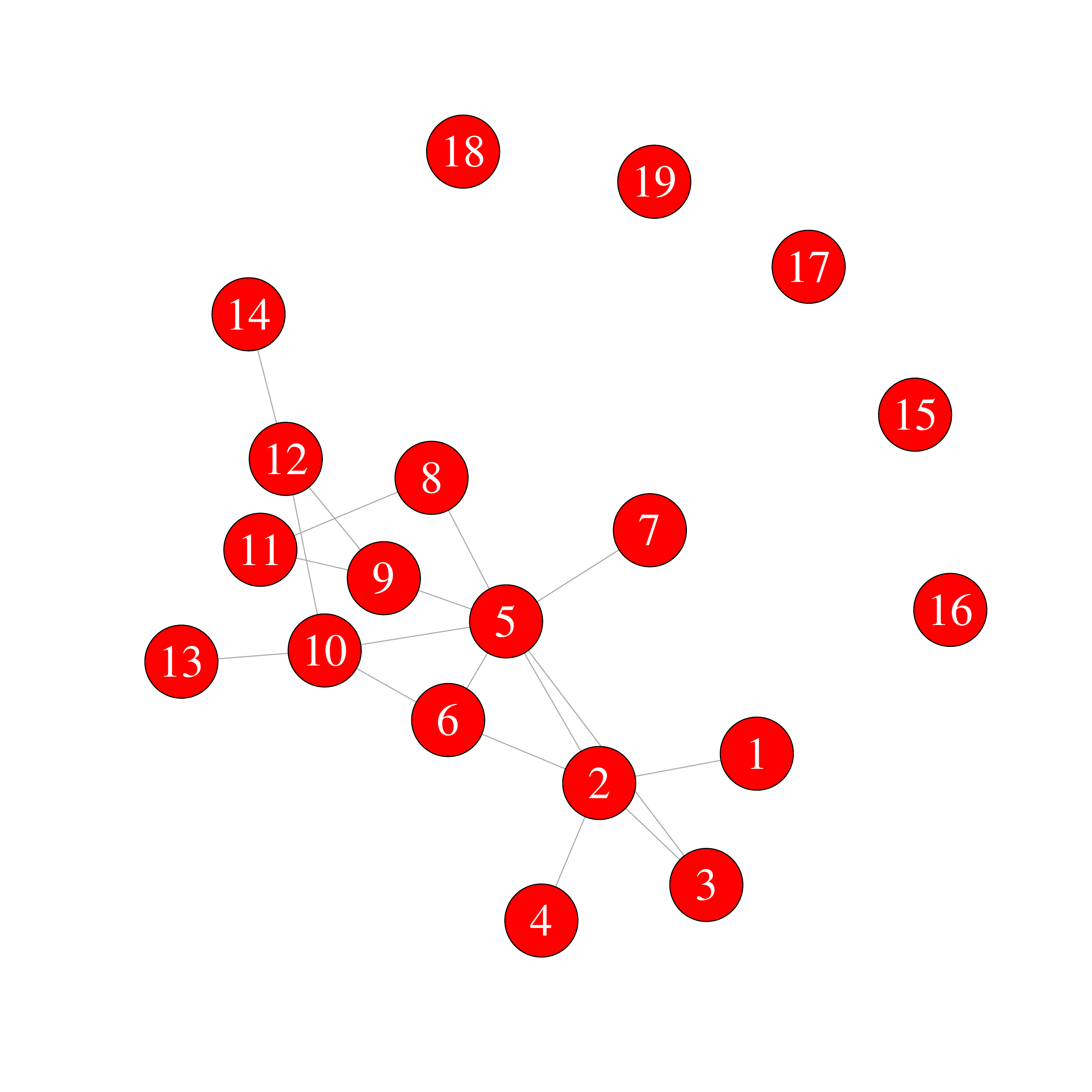}\\
\includegraphics[height=55mm]{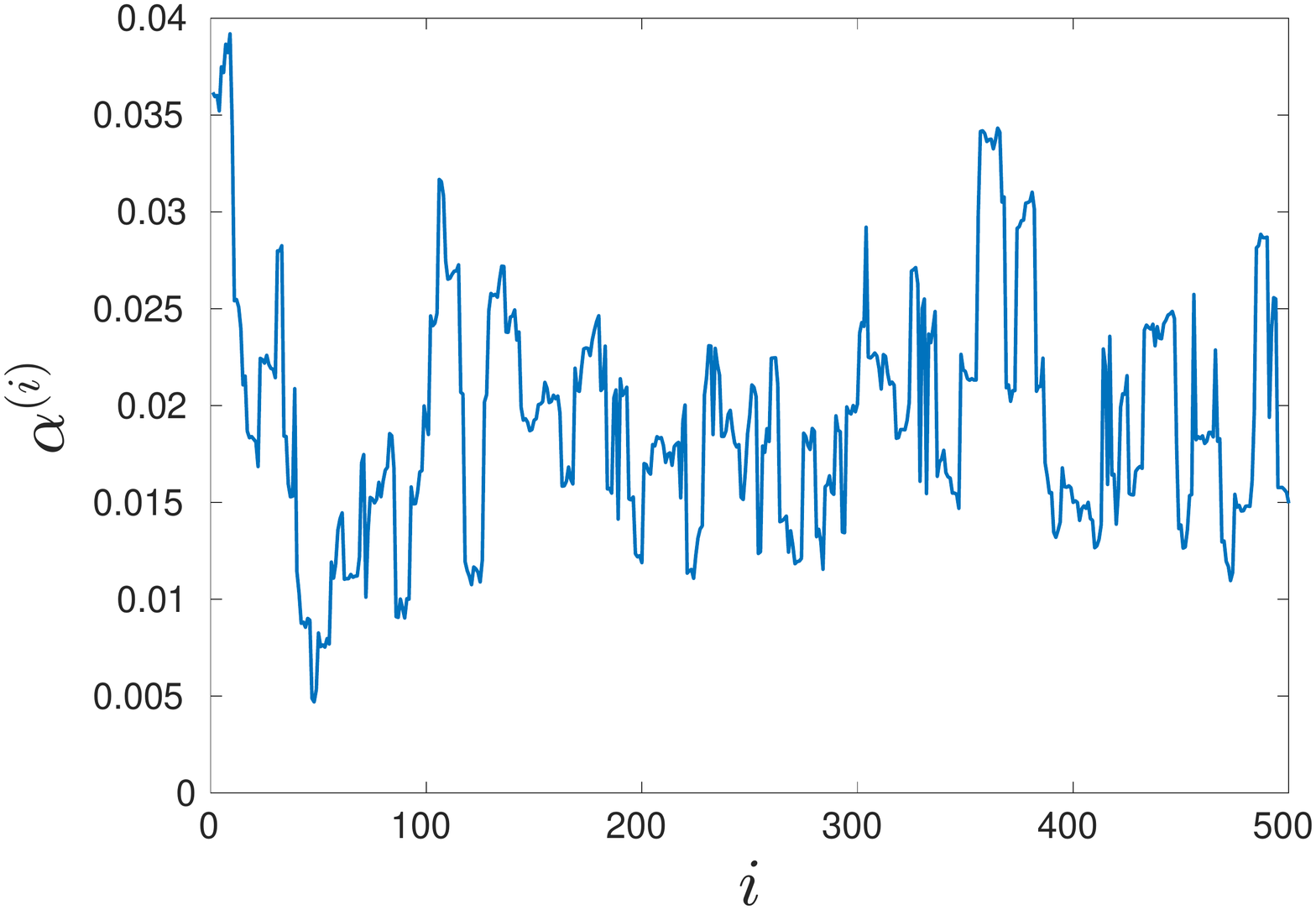}
 \includegraphics[height=55mm]{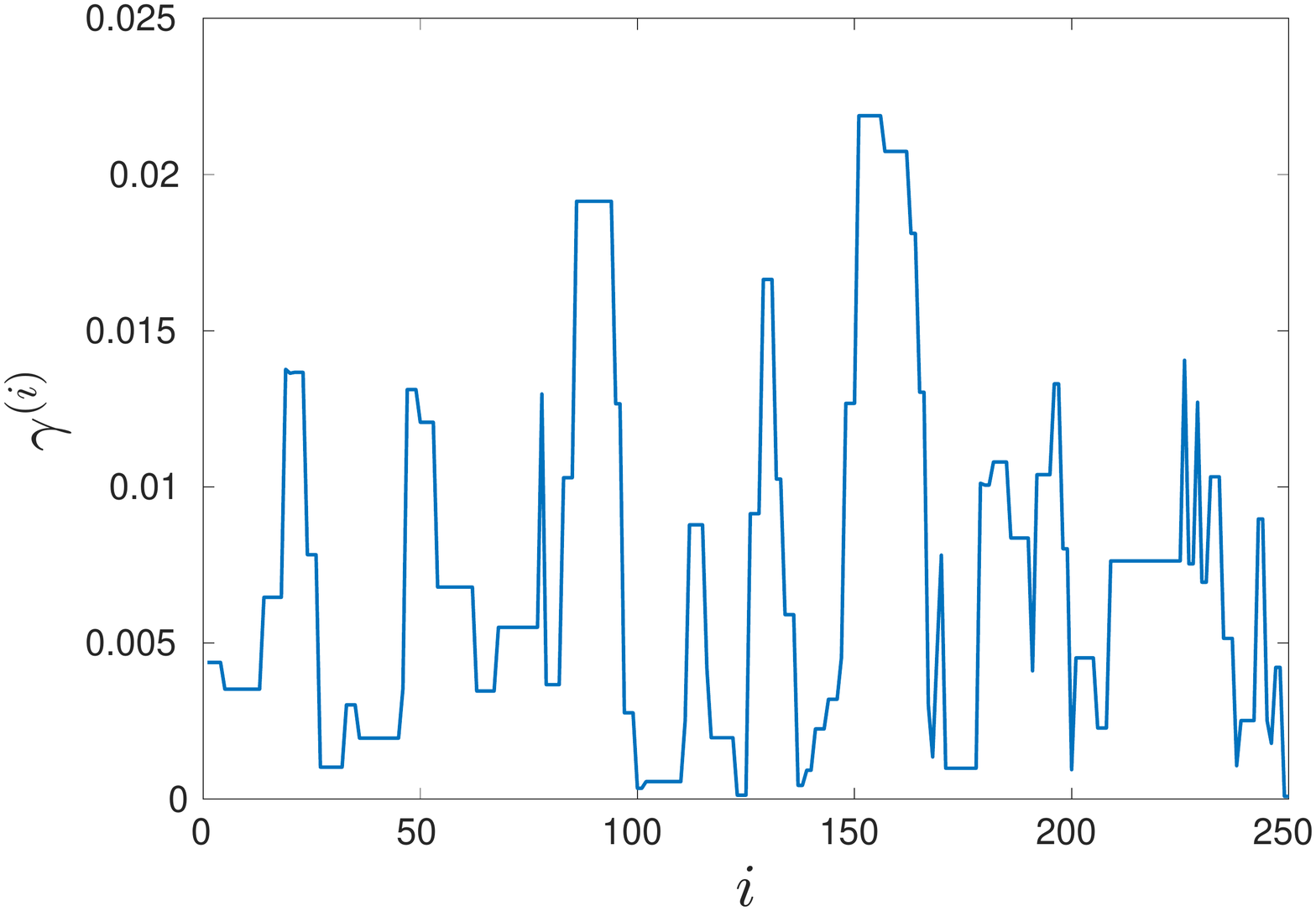}\\
 \includegraphics[height=55mm]{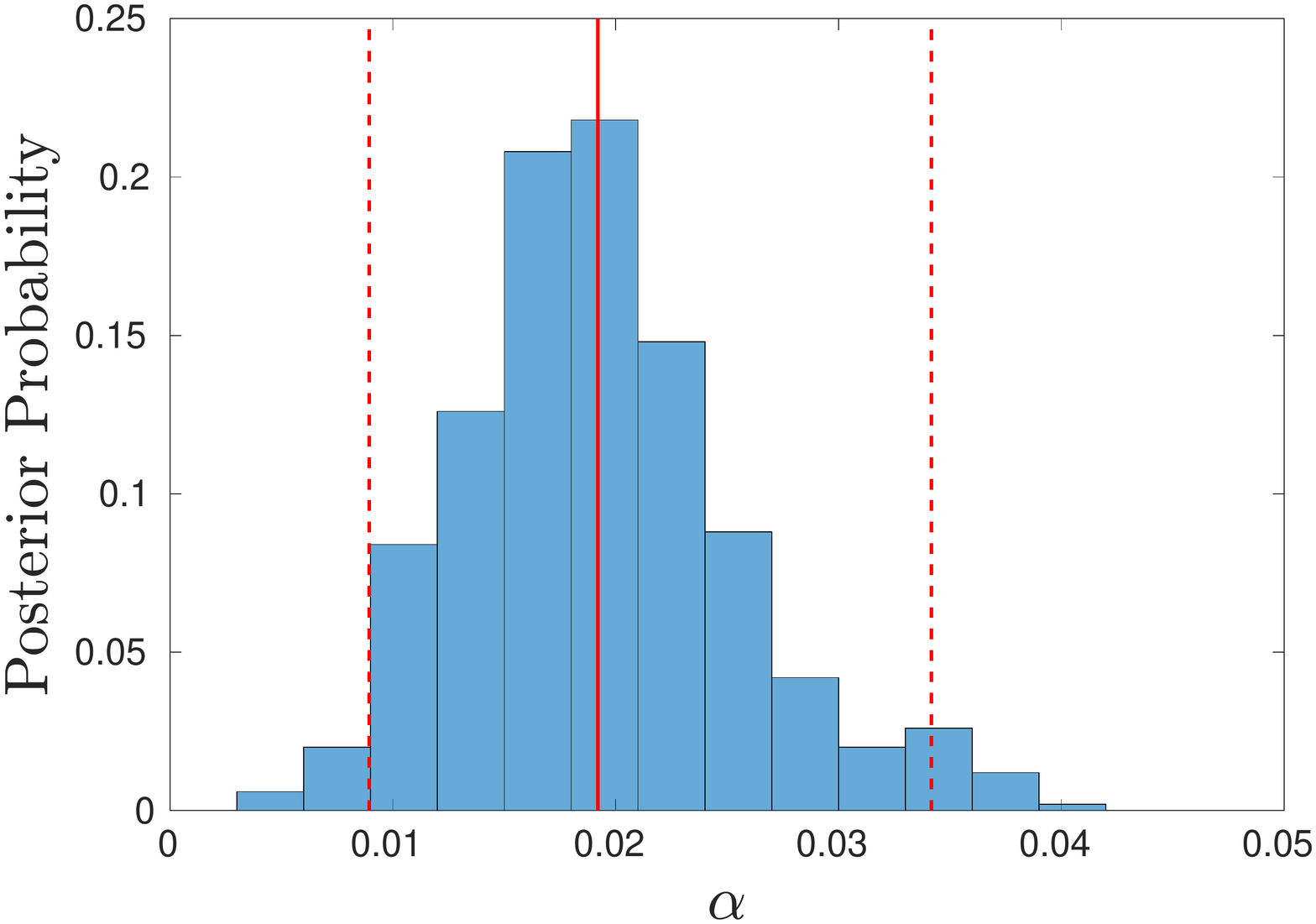}
 \includegraphics[height=55mm]{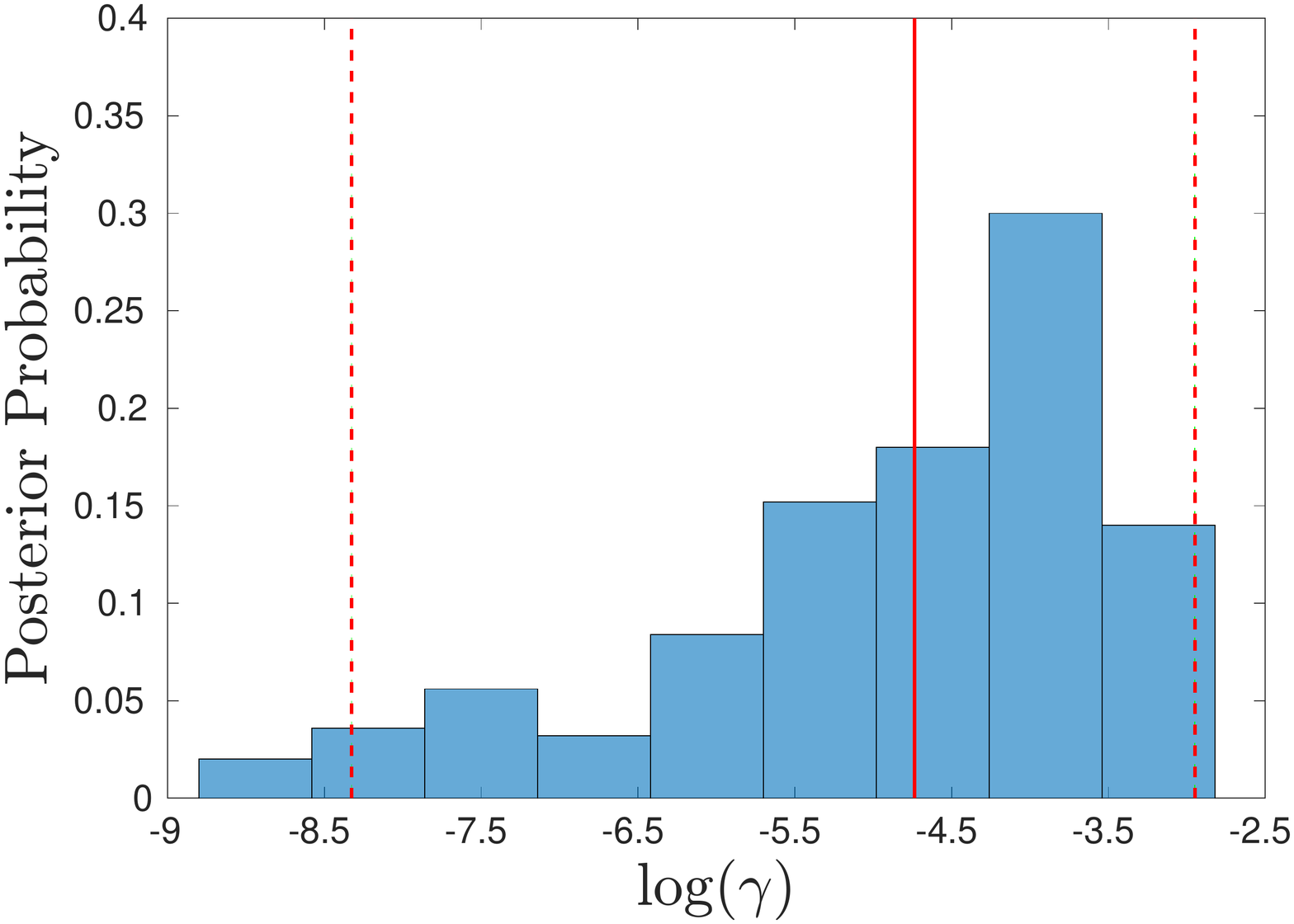}\\
 \caption{Upper Left: Posterior mode from the SER\slash SER model applied to the data set $\left\{\text{N1},\text{N2},\text{N3} \right\}$ and prior for the centroid centered at $\text{N4}$. Centre Left: Traceplot for 500 posterior samples for $\alpha$ after a burn-in of 150,000 and a lag of 50. Lower Left: Histogram for $\alpha$. The posterior mean (highlighted by the red solid line) is equal to 0.0192. The $95$\% credible interval for $\alpha$ (delimeted by the dotted lines) is (0.0089,0.0342). Upper Right: Posterior mode obtained from fitting the SN\slash SN model to the data set $\left\{\text{N1},\text{N2},\text{N3} \right\}$. This graph concentrates 0.544 of the posterior mass. Centre Right: Traceplot for 250 posterior samples for $\gamma$ after a burn-in of 100,000 and a lag of 50. Lower Right: Histogram for $\log(\gamma)$. The posterior mean is equal to -4.6177. The $95$\% credible interval for $\log(\gamma)$ is (-8.4130,-2.9866).}\label{fig:Summaries_Bayes}
\end{center} 
\end{figure}

An example of a set of networks where the variability is can be attributed to the use of different data bases and\slash or technologies is displayed in Figure \ref{fig:GData1}. Here, the nodes stand for the 19 most frequently mutated human cancer genes (the key is provided in Table~\ref{tab:Key}). These genes have a higher-than-expected degree of interconnectivity, this is with respect to sets of genes of similar size selected at random. We consider four types of inferred edges: $\text{N1}$ Inferred from expert opinion using curated databases, $\text{N2}$ Experimentally determined, $\text{N3}$ Obtained via textmining, and
$\text{N4}$ Obtained via co-expression.

These genes have been widely studied in both the systems biology and cancer research literature. Figure~\ref{fig:GData1} suggests that the set composed by $\left\{\text{N1},\text{N2},\text{N3} \right\}$ reasonably fulfils the assumptions of our methodology. The edges of $\text{N4}$ have a different interpretation, since that graph was obtained via a graphical model. Still, $\text{N4}$ can be interpreted as a rough approximation of each element of $\left\{\text{N1},\text{N2},\text{N3} \right\}$. This data is publicly available from \begin{verbatim}
://string-db.org/cgi/network.pl?taskId=PjAoqaYLxdta.
\end{verbatim}
Note that nodes 15-19 are isolated. This presents no additional challenge to our methodology since we make no assumptions regarding the connectivity of the observed networks.

We fit the CER\slash CER model to the networks $\left\{\text{N1},\text{N2},\text{N3} \right\}$ and centered the prior for the centroid at $\text{N4}$. Results are summarized in Table~\ref{tab:SummarySER} and Fig.~\ref{fig:Summaries_Bayes}. The edge sets corresponding to four networks with highest posterior probability are displayed in Table~\ref{tab:SummarySER}. The posterior mode is displayed in Fig.~\ref{fig:Summaries_Bayes} (Upper Left), along with summaries for $\alpha$. We observed that these four networks concentrate more than half of the posterior mass and that the posterior mode concentrates almost $0.25$ of the posterior mass. We also observed that nearly 35\% of the posterior probability was spread between models (centroids) that were visited by the MCMC only once or twice.

We also fit the SN\slash SN model to the data set formed by $\left\{\text{N1},\text{N2},\text{N3} \right\}$ and centered the prior for the centroid at the minimum spanning tree obtained from assigning random weights to the edges of the graph displayed in Fig.~\ref{fig:Summaries_Bayes} (Upper Left). We centered the prior at this graph instead of using $\text{N4}$ because that graph is too far with respect to the data in terms of the graph diffusion distance (\cite{Hamm}), for which the creation\slash merging of connected components is expensive. In Table~\ref{tab:SummarySN}, we display the three networks with highest posterior probability. We display the posterior mode in Fig.~\ref{fig:Summaries_Bayes} (Upper Right), along with summaries for $\gamma$. We observed that these three networks concentrate almost all of the posterior mass and that the posterior mode concentrates more than half of the posterior mass.

The presence of singletons (nodes 15-19) manifests differently in the results, depending on the metric: for the Hamming distance, we observed that the singletons merged to the connected component formed by nodes 1-14 for some of the posterior samples, producing a set of graphs that were visited once or twice by the MCMC, in contrast, when we specified the model in terms the diffusion distance, connected components do not tend to merge or split, which made the set of singletons (nodes 15-19) to remain constant across the MCMC samples.

By fitting both models, we learned that the posterior for the Fr\'{e}chet mean is sensitive with respect to the metric the model assumes for $\left\{\mathcal{G}_{[N]} \right\}$; this becomes evident from comparing Tables \ref{tab:SummarySER} and \ref{tab:SummarySN} and the two panels at the top of Fig.~\ref{fig:Summaries_Bayes}. The choice of the metric penalises discrepancies between the posterior mode and the Fr\'{e}chet mean. One way of looking at this, is that, by choosing the metric, the statistician is making decisions regarding which features of the Fr\'{e}chet mean should be retrieved when computing the posterior. This is a consequence of Proposition \ref{Prop:Frechet}.  For this data, we observed an instance of a situation where there are clear differences between  choosing $d_G(\cdot,\cdot)$ with input from the practitioner and\slash or considerations from the application (SN\slash SN model), and choosing the metric based on computational or mathemathical convenience (SER \slash SER model).

\begin{table}[h]
\begin{center}
\begin{tabular}{|l l| l l | l l |}
\hline
\text{Index} & \text{Gene} & \text{Index} & \text{Gene} & \text{Index} & \text{Gene}\\
\hline
1            & \text{BRAF}          & 8            & \text{PTEN}           & 15    & \text{CIC} \\
2            & \text{NRAS}          & 9            & \text{CDKN2A}         & 16    &\text{DNMT3A} \\
3            & \text{ERBB3}         & 10           & \text{CTNNB1}         & 17    &\text{BFXW7} \\
4            & \text{NF1}           & 11           & \text{TP53}           & 18    &\text{SF3B1} \\
5            & \text{PIK3CA}        & 12           & \text{SMAD4}          & 19    &\text{LPHN2} \\
6            & \text{PIK3R1}        & 13           & \text{APC }           &              &  \\
7            & \text{FLT3}          & 14           & \text{NCOR1}          &              &  \\
\hline
\end{tabular}
\caption{Key for indices assigned to the 19 genes related to cancer. }\label{tab:Key}
\end{center}
\end{table}

\begin{table}[h]
\begin{center}
\begin{tabular}{|c | l|}
\hline
\text{Posterior Probability} & \text{Edge Set} \\
\hline
  0.246 & \text{1-2, 2-3, 2-4, 2-5, 2-6, 5-6, 5-7,}\\
        & \text{5-8, 5-9, 5-10, 6-10, 8-11, 9-11,}\\
        & \text{10-12, 10-13, 12-14} \\
  0.168 & $\mathcal{E}_{\text{mode}} + (\text{3-5})$ \\
  0.140 & $\mathcal{E}_{\text{mode}} + (\text{11-13})$ \\
  0.114 & $\mathcal{E}_{\text{mode}} + (\text{6-9}) $\\        
  \hline
\end{tabular}
\caption{The four networks with highest posterior mass obtained by fitting the CER\slash CER model to the data set $\left\{\text{N1},\text{N2},\text{N3} \right\}$. Here $\mathcal{E}_{\text{mode}}$ denotes the edge set for the posterior mode. }\label{tab:SummarySER}
\end{center}
\end{table}



\begin{table}[h]
\begin{center}
\begin{tabular}{|c| l|}
\hline
\text{Posterior Probability} & \text{Edge Set} \\
\hline
  0.544 & \text{1-2, 2-3, 2-4, 2-5, 2-6, 3-5, 5-6, 5-7,}\\
        & \text{5-8, 5-9, 5-10, 6-10, 8-11, 9-11, 9-12,}\\
        & \text{10-12, 10-13, 12-14} \\
  0.216 & $\mathcal{E}_{\text{mode}}+ (\text{3,12})+(\text{4,6}) +(\text{6,9})-(\text{9,12})$\\
        &$ -(\text{5,10})$\\
  0.188 &$ \mathcal{E}_{\text{mode}}+ (\text{3,12})+(\text{4,6})- (\text{5,10})$\\
\hline
\end{tabular}
\caption{The three networks with highest posterior mass obtained by fitting the SN\slash SN model to the data set $\left\{\text{N1},\text{N2},\text{N3} \right\}$. Here $\mathcal{E}_{\text{mode}}$ denotes the edge set for the posterior mode. }\label{tab:SummarySN}
\end{center}
\end{table}


\subsection{Connectome Data}\label{Sec:GeneInter}

Connectome data is an instance of measurements of brain activity that are collected, among other purposes: to describe brain structure, to find associations between brain structure and function and to correlate brain structure to covariate information. Among the questions that can be posed given the availability of this type of data, we focus on the following: which is an appropriate representative for either the population or a subpopulation of individuals? One key aspect of this problem consists on making decisions regarding what does it mean for connectomes to be similar. As discussed in \cite{DonnHolm}, for different metrics in graph space, different representatives and different groupings of the data points may seem appropriate.



We analyzed the dataset discussed in \cite{ArroAthr} and \cite{ZuoZhou}. The data consists on 300 instances of connectome data. The connectomes are graphs constructed via diffusion magnetic resonance imaging (dMRI). These measurements were obtained from 30 healthy individuals; 10 measurements were obtained during the curse of a month for each individual. Each of these networks has $200$ nodes over the same regions of the brain. The vertices are registered according to the CC200 atlas \cite{CradJame}. The goal of the analysis performed by \cite{ArroAthr} was to cluster the graphs according to their community structure (at node level) to see if they could find differences between individuals. We approach this dataset from a different perspective: we assume the metric based on diffusion and based on that, estimate a representative of the population. We also explore to what extend there is evidence for clusters in the data. We also perform these inferences assuming a Hamming distance.

\begin{figure}[!t]
\begin{center}
  \includegraphics[height=70mm]{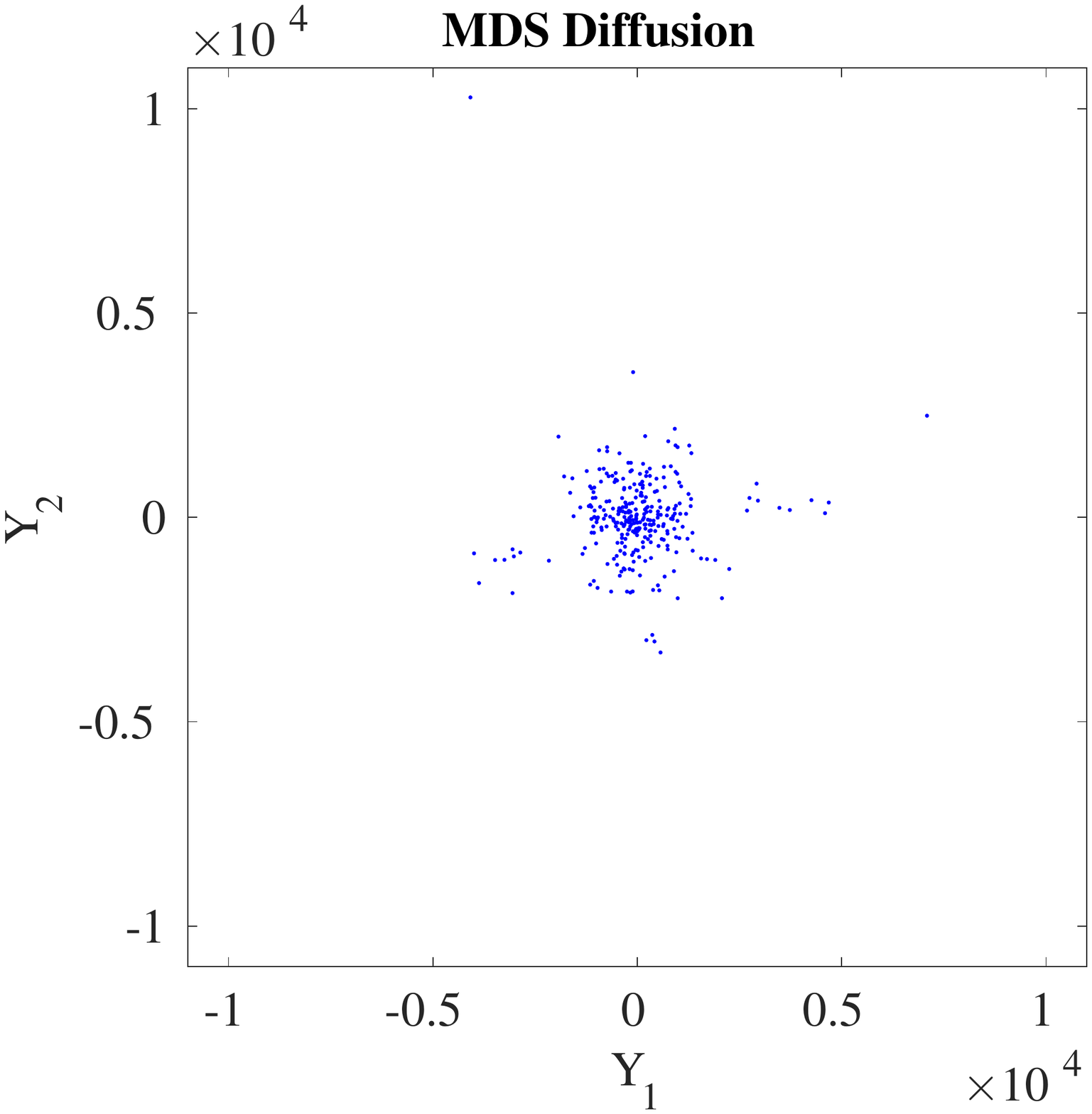}
  \caption{The two-dimensional map obtained from applying multidimensional scaling on the $300$ connectomes. The similarity is given by the diffusion distance. }
  \label{fig:MDS_Diff}
\end{center}
\end{figure}

One of the key assumptions of our methodology is that the data was generated from a unimodal distribution over the space of labelled graphs defined over the same vertex set. The validity of such an assumption depends on the metric. In practice, this assumption can be verified by using a reasoning similar to the one deployed by \cite{DonnHolm} when studying the different metrics. We applied multidimensional scaling (MDS) on the data to assess if there is more than one cluster, where each cluster suggests the existence of a different mode. The two-dimensional map for the $300$ networks implied by the diffusion distance is displayed in Figure \ref{fig:MDS_Diff}. It suggests that modelling the data as unimodal is a reasonable first approximation.

Fitting the CER\slash CER model is not a major challenge when analyzing this dataset; the same MCMC scheme as the one used in Section \ref{Sec:SER} can be implemented. For this data set, we used 2,000,000 iterations for burn-in and obtained 5,000 samples with a lag of 1,000. In contrast, fitting the SN\slash SN model for a data set is not straightforward. We followed the divide-and-conquer strategy proposed by \cite{WuRobe}. We divided the data into ten subsets of the same size, where each subset preserves the pattern suggested in Figure \ref{fig:MDS_Diff} as much as possible. Each subset is constituted by 30 of the observed networks. We parametrized the model so the posteriors for $\gamma$ implied by each subset of the data can be transformed into a distribution with roughly same dispersion as the posterior we would obtain by using the whole data set. To compute summaries from the posterior, we proceed as follows:
\begin{itemize}
    \item For $\mathcal{G}^{m}$: We let $\mathcal{G}^{m,(i)}$ be the centroid of 
    \begin{displaymath}
    (\mathcal{G}^{m,(i)}_1,\mathcal{G}^{m,(i)}_2,\dots,\mathcal{G}^{m,(i)}_{10})
    \end{displaymath}
    with respect to $d(\cdot,\cdot)$. The point estimator for $\mathcal{G}^{m}$ was obtained by computing the centroid of the posterior modes associated to each subset (as in Section \ref{Sec:GeneInter}).
    \item For $\gamma$: since the model is parametrized so $\gamma$ is on the same scale across the ten subsets. We only need to: i) re-center the all samples with respect to the sample mean of the corresponding subset; ii) re-scale so each posterior has roughly the same dispersion as the full posterior; iii) re-center again using the global sample mean. 
\end{itemize}
For each subset, we ran the MCMC  described in Section \label{Sec:ENF}. We used 500,000 iterations for burn-in and obtained 1,000 samples with a lag of 500.



\begin{figure}[htbp]
\begin{center}
\includegraphics[height=65mm]{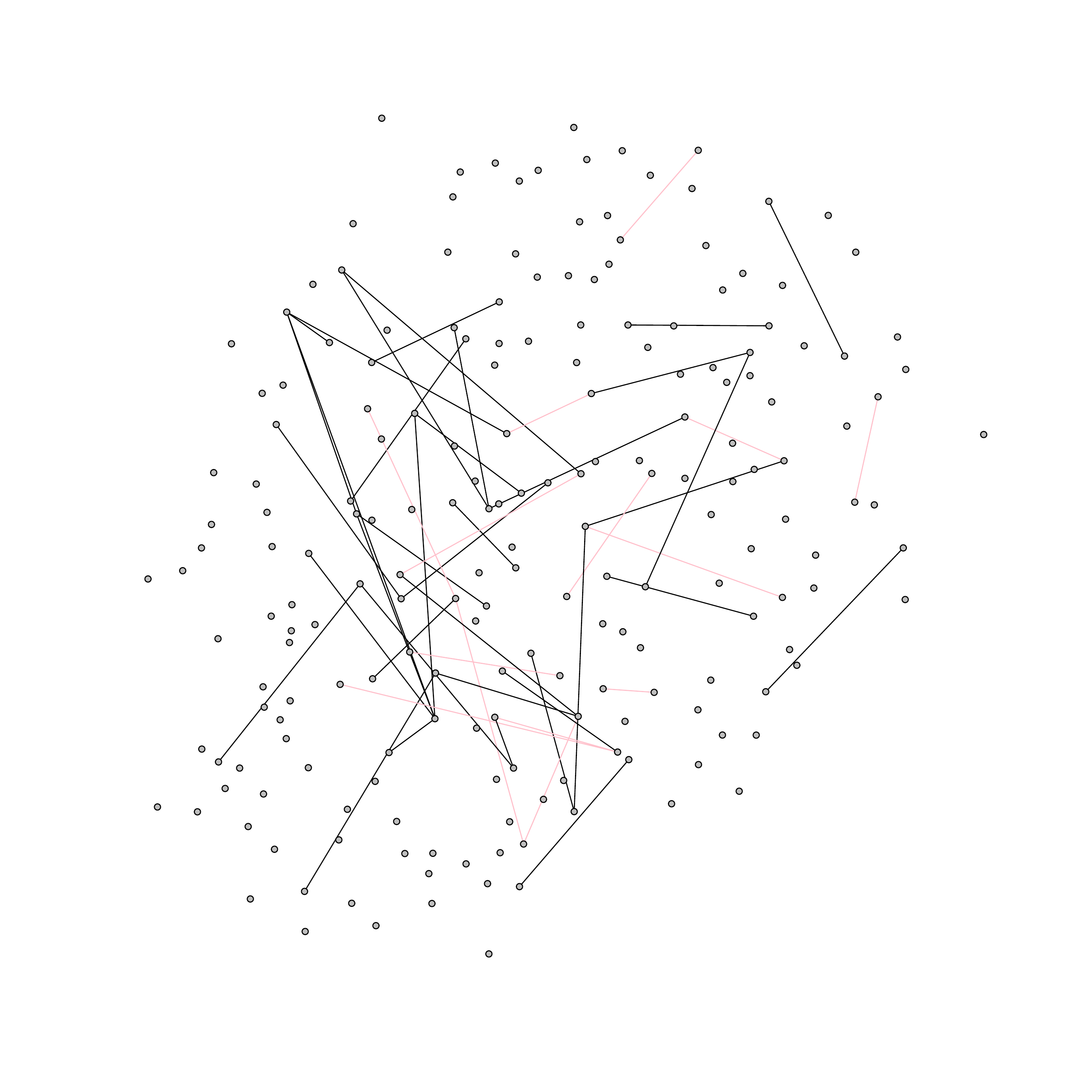}
\includegraphics[height=65mm]{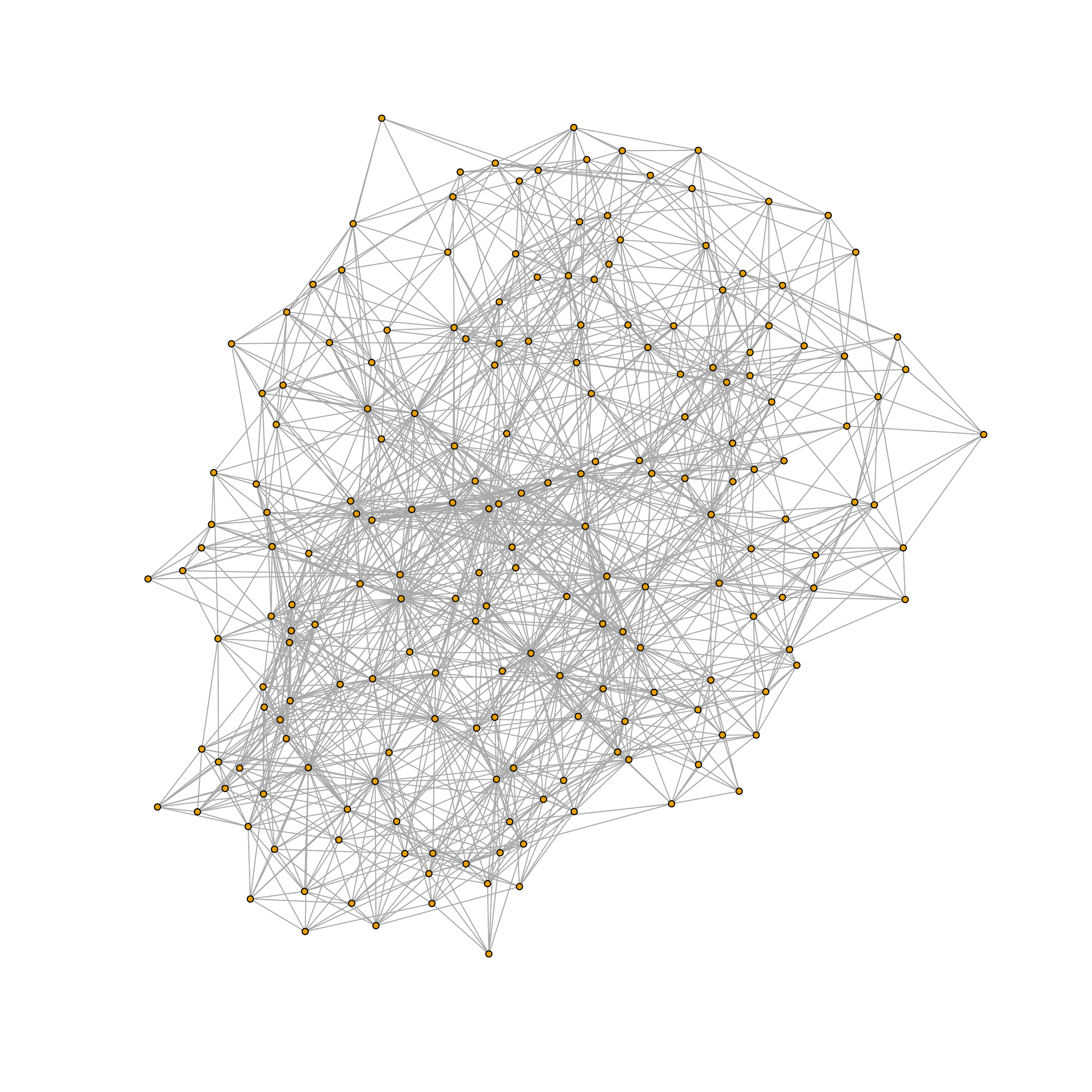}\\
\includegraphics[height=50mm]{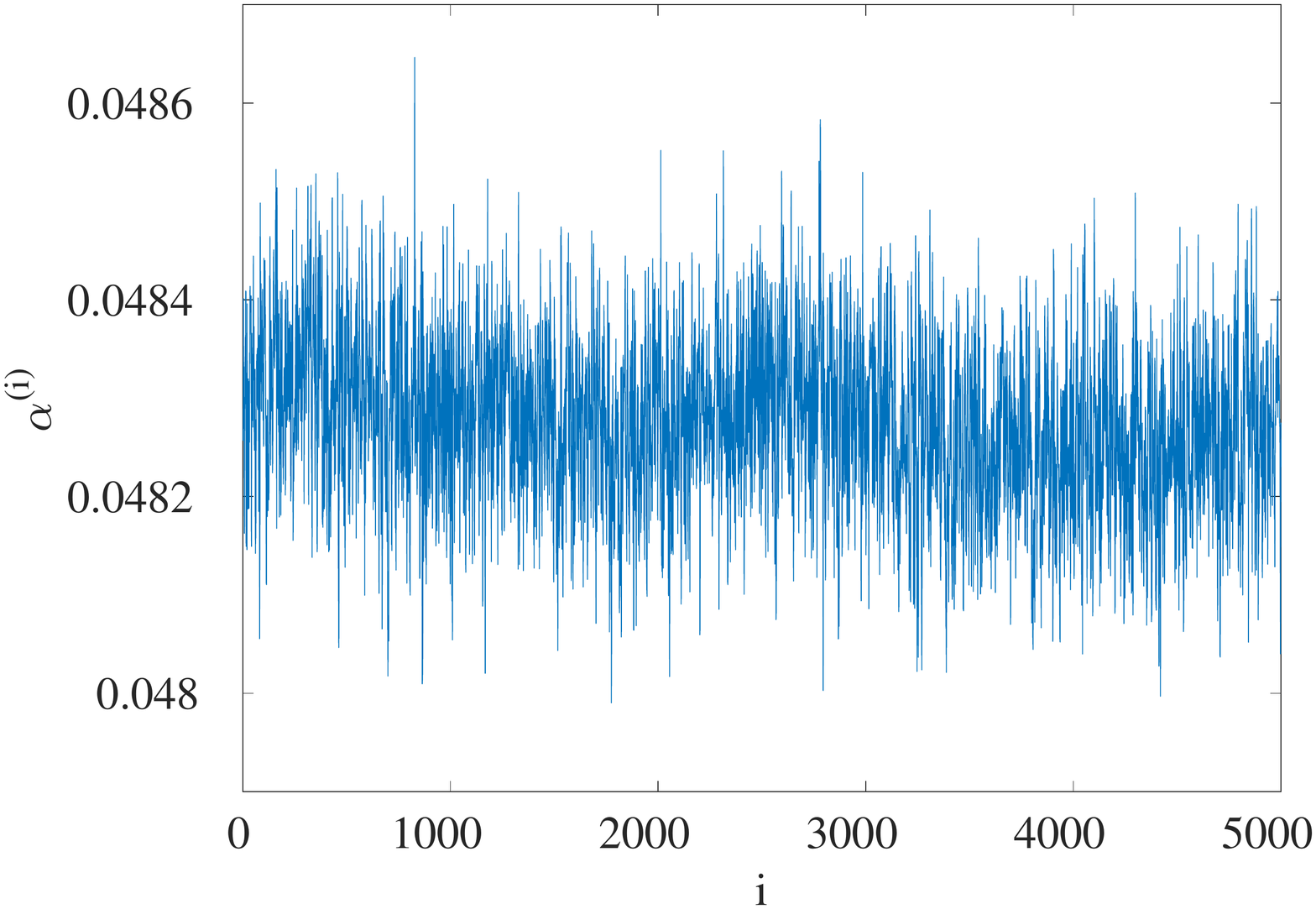}
 \includegraphics[height=50mm]{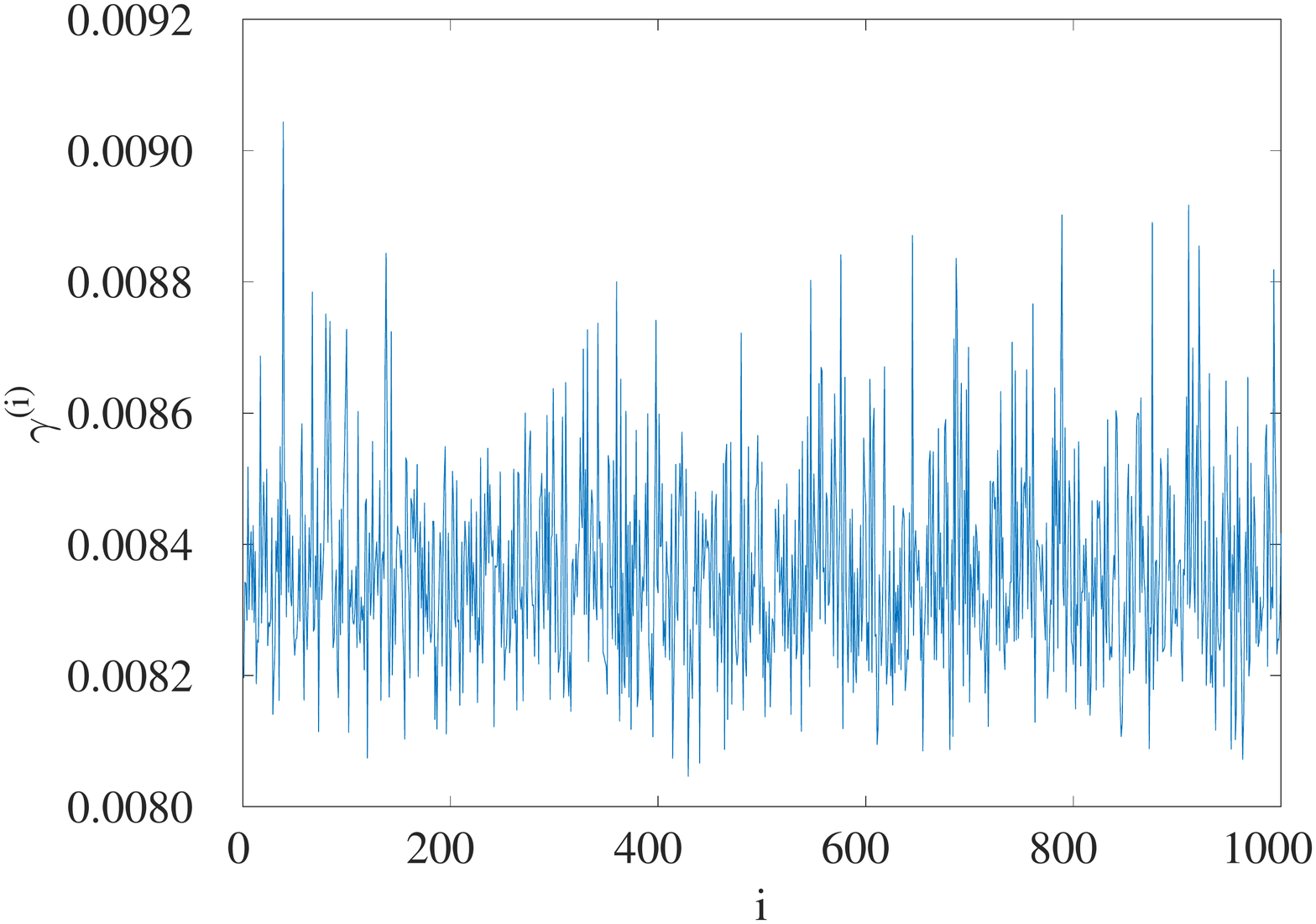}\\
 \includegraphics[height=50mm]{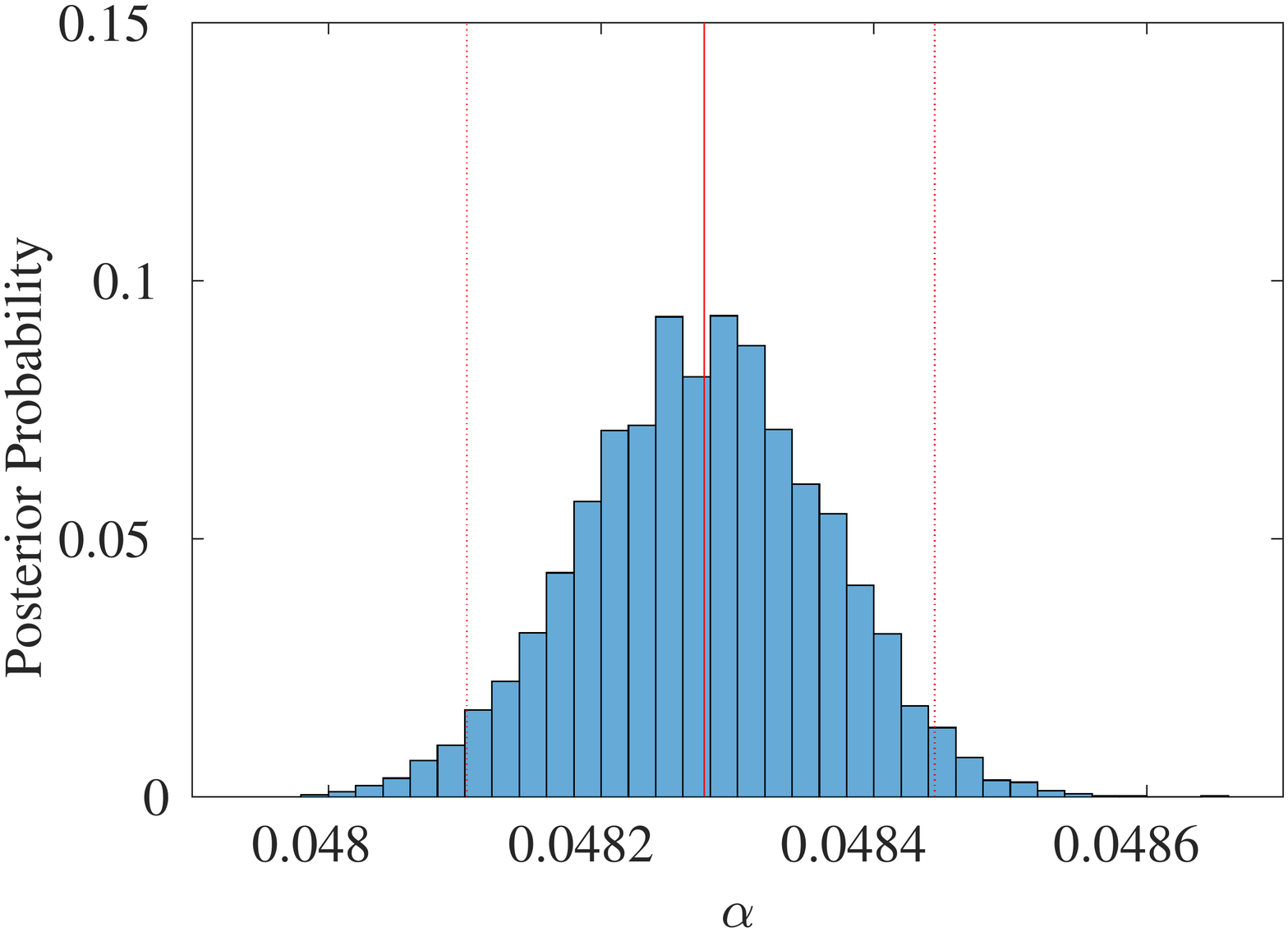}
 \includegraphics[height=50mm]{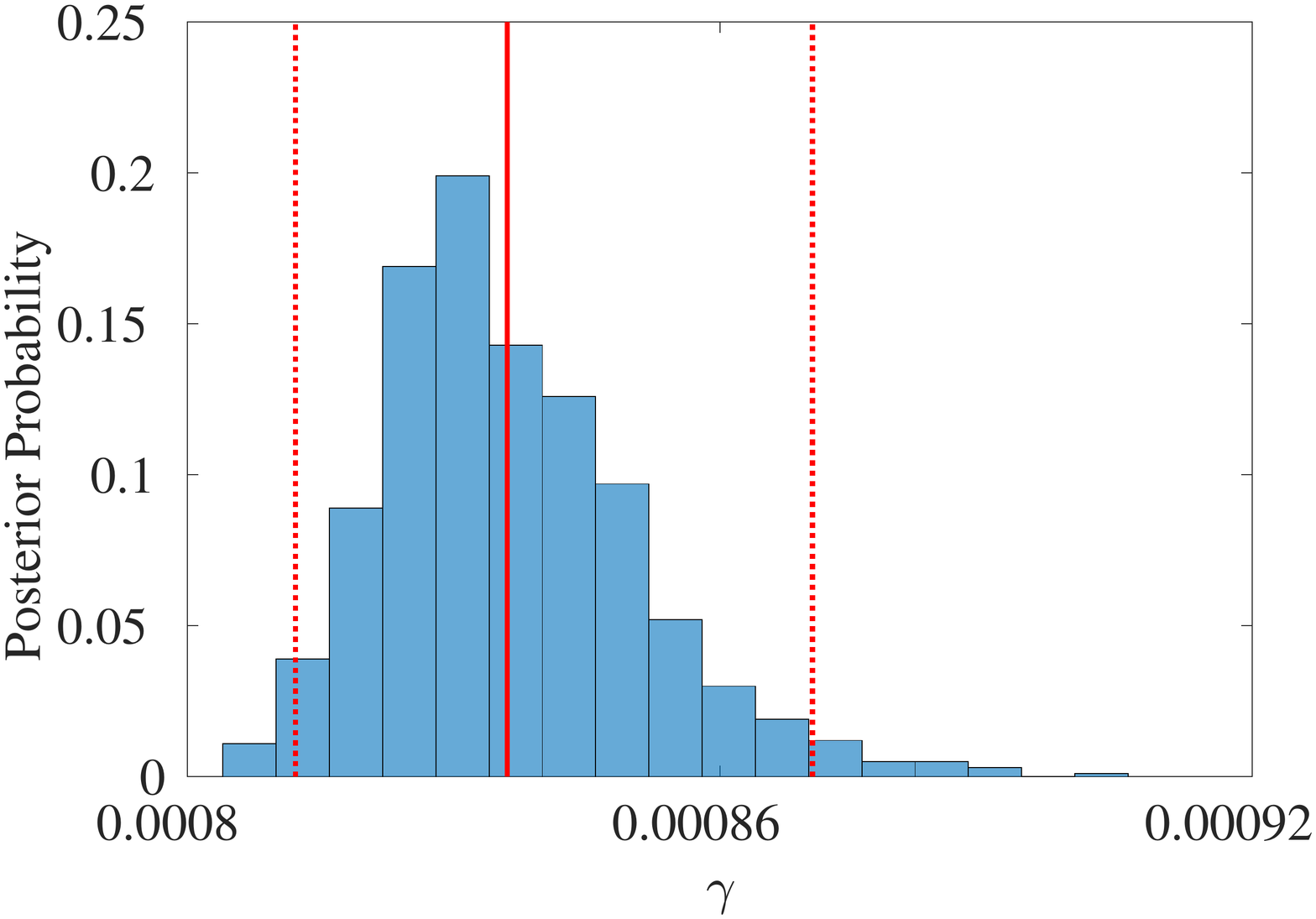}\\
 \caption{Upper Left: Posterior mode from the SER\slash SER model applied to the whole connectome data set and prior for $\mathcal{G}^{m}$ centered at the centroid of one of the subsets of the data. The posterior mode is presented in terms of its discrepancies with respect to the point estimate from the SN\slash SN model. Edges only present in the posterior mode from SER\slash SER are colored in blue, while edges present in the point estimate from  the SN\slash SN model but not the posterior mode are colored in pink. Centre Left: Traceplot for 5000 posterior samples for $\alpha$ after a burn-in of 150,000 and a lag of 50. Lower Left: Histogram for $\alpha$. The posterior mean (highlighted by the red solid line) is equal to 0.0483. The $95$\% credible interval for $\alpha$ (delimeted by the dotted lines) is (0.0481,0.0485). Upper Right: Point estimate obtained from fitting the SN\slash SN model to the full data set. Centre Right: Traceplot for 250 posterior samples for $\gamma$ after a burn-in of 100,000 and a lag of 50. Lower Right: Histogram for $\log(\gamma)$. The posterior mean is equal to -4.6177. The $95$\% credible interval for $\log(\gamma)$ is (-8.4130,-2.9866).}\label{fig:Summaries_JHU}
\end{center} 
\end{figure}

\begin{figure}[htbp]
\begin{center}
\includegraphics[height=65mm]{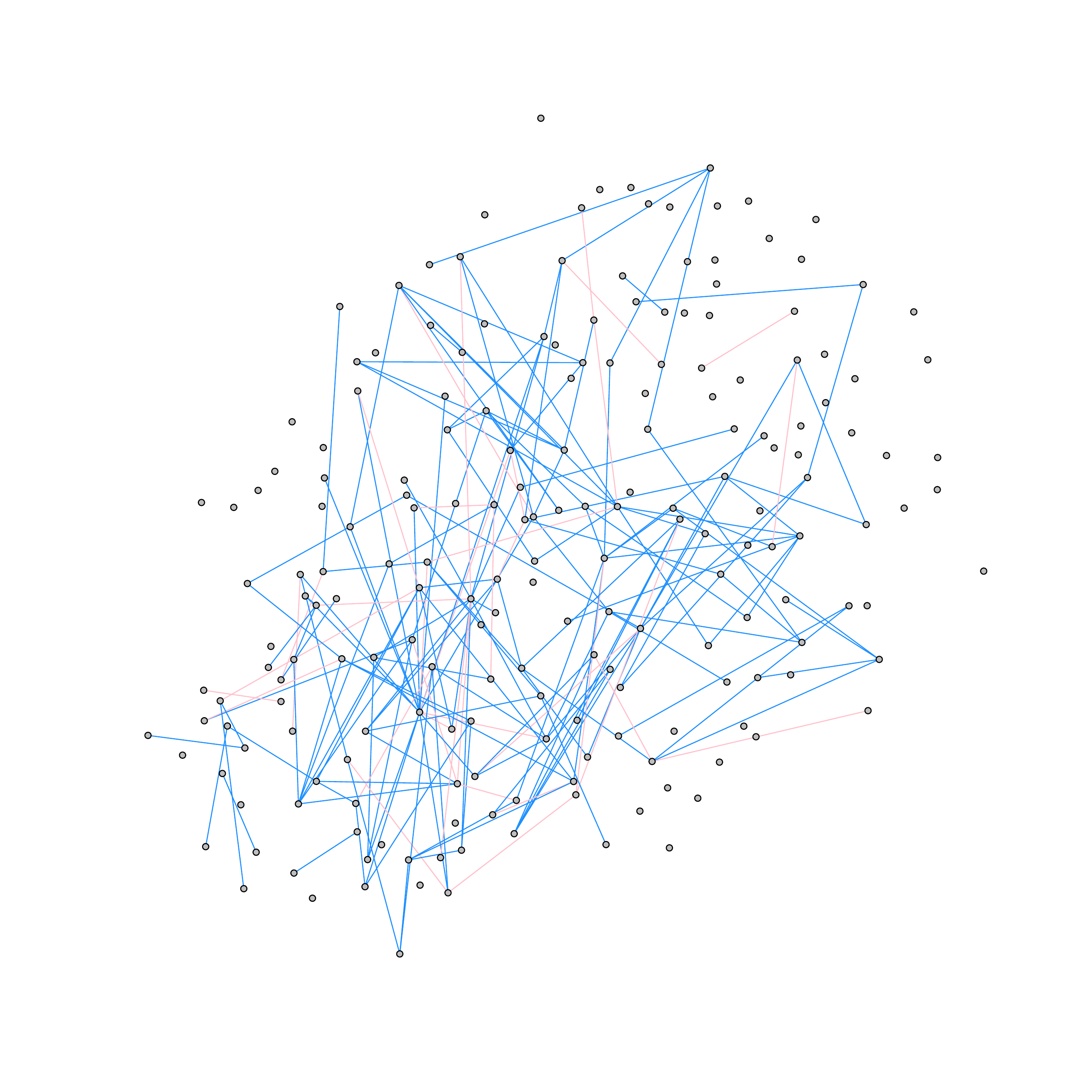}
\includegraphics[height=65mm]{ModeDiffusion_JHU_V4.pdf}\\
 \caption{Right: Point estimate for $\mathcal{G}^{m}$ obtained from computing the centroid of the posterior modes associated to each of the 10 subsets of the data . Each of these posterior modes was obtained from fitting the SN\slash SN model. Left: Discrepancies of the posterior mode corresponding to one of the subsets of the data. Edges only present in the centroid are colored in blue, while edges present in the mode but not the centroid are colored in pink.  }\label{fig:Summaries_JHU_Centroids}
\end{center} 
\end{figure}

Results for the Hamming distance are summarized in Figure \ref{fig:Summaries_JHU} (Left). Results for the diffusion distance are summarized in Figure \ref{fig:Summaries_JHU} (Right). A traceplot of posterior samples for $\gamma$ corresponding to one of the subsets of the data is displayed in Figure \ref{fig:Summaries_JHU}. Summaries for $\gamma$ obtained from combining the samples from the different data sets are also displayed in Figure \ref{fig:Summaries_JHU}. As a complementary summary, we also show the point estimate for $\mathcal{G}^{m,(i)}$ for one of the subsets of the data in terms of its discrepancies to the point estimate for the whole data set (Figure \ref{fig:Summaries_JHU_Centroids}).

\section{Discussion}\label{Sec:Discussion}

Network data has caught the imagination of statistical researchers and data analysis practitioners. Despite this interest a number of very fundamental questions lie unresolved in pursuing {\em multiple} network data analysis. To be able to understand not one network but multiple networks collected simultaneously one has to ask questions like: a) what is the ``mean'' network (rather than how do we estimate the success-probabilities of an inhomogeneous random graph), and do we want the ``mean'' itself to be a network? b) what is the degree of variation in realizations away from that ``mean'', and how can we make statistical inference in such scenarios? This requires a number of modeling choices, that need to be made for us to make inferences. We in this paper have designed a modular framework that allows us to specify each component, and thus to model.

This modular framework can be compared to the modelling framework of others, such as~\citep{Newm,LeLevLevi2018,ChanKola,DurDunVog,Peix}.
In comparison to~\citet{DurDunVog}, for example, we adopt a less flexibly nonparametric approach but allow for our notion of an average or typical network to have complex structure; relative to the approach of \citep{Newm,LeLevLevi2018,ChanKola,Peix}, by contrast, our parametrisations are more complex while we adopt a similarly simple characterisation of perturbations from the typical network.


The use of the Fr\'{e}chet mean as a parameter that encodes what the centre of the distribution is supposed to be, as well as the use of the entropy to encode the notion of dispersion, are insights that we borrow exactly from shape theory~\citep{DrydMard}. Even more, the problem of finding a representative for a population of shapes and the problem of modelling the variability of a homogeneous population of shapes are listed as two of the main challenges in that area in \cite{SrivKlas} (Section 1.3). We pose these challenges in the context of network data and offer solutions for the implied inference problems via Bayesian modelling. Some of our theoretical results (Propositions 3 and 4) borrow heavily from shape analysis ideas. From functional data analysis, we adopt the rationale of using a complicated object (a network without a pre-specified structure) to model the trend, while using a simple model to account for the error. The trade-off between the complexity of the trend and the complexity of the error distribution has been widely studied in the functional data analysis literature; a similar tension will arise in our context.  In this setting the mean function is often left mainly unspecified (or even just restricted to a form of regularity such as Besov regularity), but the noise is not permitted much structure. The noise or perturbation from that network we chose to be very simple, normally just uncorrelated white noise. This could be construed as the Goldilocks principle at work, where things are made complex, but not too complex, rather just right in their complexity to capture realistic features. This remains a topic for exploration and\slash or future developments. One interesting challenge that arises in the context of network data is that there is a lot to be learned regarding which metric in graph space should be adopted for a given problem. This is an interesting contrast to functional data analysis, since in that context, practitioners are more familiar with the idea of pairing a specific metric to a given application (such as the $l^2$ norm for signal processing).

There are also inevitably limits of resolvability to this problem, linked to 
being able to resolve the blocks of the stochastic block model~\citep{hajek2017information}.
Here we see identifiability starts to depend on the number of nodes, and the observed number of networks, as well as the level of variability of each individual network. Our study of small world networks, show that if the number of observed networks are sufficiently few, then the regularizing effect of the prior can indeed be too strong.




In Section \ref{s:intro} we mentioned that it would be challenging to extend the proposed methods to a setting where the vertex set is allowed to vary in an unconstrained manner. However, if the vertex set varies so it is always a subset of a maximal finite collection of vertices, then the methods proposed in this paper are still valid (provided an appropriate metric is provided) and the the theory results will hold. In terms of computation, an MCMC based on a saturated model approach (as the one proposed in Section 5 of \cite{BrookGiud}) can be used to obtain samples from the posterior. The key aspect here is that, for this setting, we still have a finite discrete space endowed with a metric, which is the core assumption for our method.

In contrast to the methodology proposed by \cite{DurDunVog}, which focuses on clustering, our methodology in turn is designed for providing summaries that are easy to interpret in the context of replication and on prior elicitation, in addition, our methodology makes explicit what the estimand for a central network is, instead of just providing an estimator with no obvious estimand associated to it. The main advantage of our method with respect to approaches that use the idea of a Fr\'{e}chet mean as a centre, but derive the uncertainty around that centre via asymptotics (\cite{GineBalaKola2017}) are: i) that our method enables the statistician to propagate uncertainty to subsequent inferences, since we are able to sample from a posterior, and ii) our method is not constrained to use of a single metric, in contrast to \cite{GineBalaKola2017}, which relies on a specific metric to derive the asymptotic results they need. In a broad sense, this last point also applies to the approach proposed by \cite{DurDunVog}, since their MCMC scheme relies heavily on the metric induced by a random dot product model to take advantage of conjugacy.

The proposed methodology can help in the development of informative priors for graphical models. The example discussed in Section \ref{Sec:SysBio} suggests how to proceed: i) obtain the posterior mode from previous\slash similar studies; ii) apply the proposed methodology with a metric that can be related to a measure of similarity in distribution space (such as the Kullback-Leibler divergence); and iii) use the posterior produced this way as the prior for the data associated to the graphical model we want to infer.

Future work includes: i) to develop methodology that enables the use of mixture distributions at the level of the centroid network. There are two possibilities for achieving this: to specify the number of elements in the mixture (hierarchical model approach) or to leave the number of elements unspecified (the Bayesian nonparametric approach); ii) to extend the current methodology to allow for missing data and/or partial observation of the network due to sampling. This would raise interesting challenges, since in our approach the network is treated as the observational unit; iii) to constrain the structure of the centroid by using a parametric model (such as~\cite{Peix2018,Newm}), or to impose specific constrains on graph features of the centroid. Such an extension demands a formulation in terms of hierarchical models. By constraining the possible values for the centroid, we should be able to propose richer models for the error distribution.

From our perspective, to get a better understanding of the trade-offs between imposing structure on the centroid versus imposing structure for the error distribution is a promising area for future research. It is not straightforward to anticipate which combinations of assumptions for the centroid and the error distribution will lead to useful models, since, both, the use of metrics on a graph space and the use of random graphs as error models have not been explored from a statistician's perspective. 

\bibliographystyle{plainnat}
\bibliography{mybib}

\appendix

\section{Proofs}

Proof of Proposition \ref{Prop:CER}\\
Let $\mathcal{G}_1$ and $\mathcal{G}_2$ in $\left\{ \mathcal{G}_{[N]} \right\}$. Here $\mathcal{G}^m\in \left\{ \mathcal{G}_{[N]} \right\}$ is fixed and $d_H(\cdot,\cdot)$ denotes the Hamming distance. Let $N_e=\binom{N}{2}$ be the total number of edges possible in the graph.  It follows:
\begin{eqnarray}
\frac{p\left({\cal G}_1|\mathcal{G}^m,\alpha \right)}{p\left({\cal G}_2|\mathcal{G}^m,\alpha \right)}& = & \frac{\alpha^{d_H(\mathcal{G}_1,\mathcal{G}^m)}(1-\alpha)^{N_e-d_H(\mathcal{G}_1,\mathcal{G}^m)}}{\alpha^{d_H(\mathcal{G}_2,\mathcal{G}^m)}(1-\alpha)^{N_e-d_H(\mathcal{G}_2,\mathcal{G}^m)}}\nonumber \\
 &  =  & \left( \frac{\alpha}{1-\alpha}\right)^{d_H(\mathcal{G}_1,\mathcal{G}^m)-d_H(\mathcal{G}_2,\mathcal{G}^m)},\nonumber \\
 &  =  & \left(\frac{1-\alpha}{\alpha} \right)^{d_H(\mathcal{G}_2,\mathcal{G}^m)-d_H(\mathcal{G}_1,\mathcal{G}^m)}.\nonumber
\end{eqnarray}
If $0.5>\alpha>0$, then $\frac{1-\alpha}{\alpha}>1$. Under this condition, $d_H(\mathcal{G}_2,\mathcal{G}^m)>d_H(\mathcal{G}_1,\mathcal{G}^m)$ if and only if:
\begin{equation}
\frac{p\left({\cal G}_1|\mathcal{G}^m,\alpha \right)}{p\left({\cal G}_2|\mathcal{G}^m,\alpha \right)}>1.
\end{equation}
Since $d_H(\cdot, \cdot)$ is a metric, this reasoning implies that the distribution $p(\cdot | \mathcal{G}^m , α)$ is unimodal. The proof for the case $d_H(\mathcal{G}_1,\mathcal{G}^m)=d_H(\mathcal{G}_2,\mathcal{G}^m)$ follows \emph{mutatis mutandis}.

Our second task is to show that the  Centred Erd\"{o}s--R\'{e}nyi graph defined on $\left\{ \mathcal{G}_{[N]} \right\}$ fulfills
Definition \ref{def:RGDisEnt}. For this, we need to investigate how the entropy $H_{\text{CER}(\mathcal{G}^m,\alpha)}$ of the distribution
relates to $\alpha$. Remember that, if $X$ and $Y$ are independent random variables, then the entropy of their joint distribution $H_{X,Y}$ and the entropy of the individual variables ($H_X$ and $H_Y$) relate as follows:
\begin{displaymath}
H_{X,Y}=H_X+H_Y.
\end{displaymath}
as explained in Following \cite{MezaMont}, Section 1.2. The CER model with parameters $(\mathcal{G}^m,\alpha)$ can be represented as a random vector of size $N_e=\frac{(N-1)N}{2}$, where entries iid $\text{Ber}(\alpha)$. Therefore, the entropy of the CER with parameters $(\mathcal{G}^m,\alpha)$ is given by:
\begin{displaymath}
H_{\text{CER}(\mathcal{G}^m,\alpha)}=-N_e\times\left[ (1-\alpha)\log (1-\alpha) + \alpha \log (\alpha) \right].
\end{displaymath}
Since
\begin{displaymath}
\frac{\partial}{\partial\gamma}H_{\text{CER}(\mathcal{G}^m,\alpha)}=-N_e\times\left[ \log \left(\frac{\alpha}{1-\alpha} \right) \right],
\end{displaymath}
we conclude that the entropy of the distribution is a strictly increasing function of $\alpha$ in $[0, 0.5]$, with
$H_{\text{CER}(\mathcal{G}^m,\alpha)} = 0$ and such that the maximum entropy is reached at $\alpha = 0.5$, for which all elements in $\left\{ \mathcal{G}_{[N]} \right\}$ are assigned equal mass (See Example 1.6 from \cite{MezaMont} (2009)). This computation implies that part (2) of Definition \ref{def:RGDisEnt} is fulfilled. Part (1) of Definition \ref{def:RGDisEnt} is fulfilled since the proof that CER model satisfies Definition \ref{def:RGDis} was carried out for $\alpha$ specified and no
property of $\mathcal{G}^m$ was invoked or constrain on it was imposed.

Proof of Proposition \ref{CERSNF}
\begin{proof}
Let us start from the PMF of the graph. Let $N$ be the number of nodes in the graph, and $N_e$ the total number of possible edges.
Let $n_j$ be the number of switches of ${\cal G}_j$ away from 
${\cal G}^m$. We can then write
\begin{align}
\nonumber
p\left({\cal G}_j|{\cal G}^m,\alpha\right)&=\alpha^{n_j}\left( 1-\alpha\right)^{N_e-n_j}\\
\nonumber
&=\exp\left\{n_j\log(\alpha)+(N_e-n_j)\log(1-\alpha) \right\}\\
\label{ER}
&=\exp\{N_e\log\left(1-\alpha\right) \}\exp\left\{n_j\log\left(\frac{\alpha}{1-\alpha}\right)\right\}.
\end{align}
We note directly that
\[ n_j=\|A_{{\cal G}_j}-A_{{\cal G}^m}\|_H=d_H({\cal G}_j,{\cal G}^m).\]
We note that
\[p\left({\cal G}_j|{\cal G}^m,\alpha\right)=\exp\{N_e\log\left(1-\alpha\right) \}\exp\left\{d_H({\cal G}_j,{\cal G}^m)\log\left(\frac{\alpha}{1-\alpha}\right)\right\}.\]
We therefore see that we have
\[\phi(x)=x,\]
with $\gamma=\log(\frac{1-\alpha}{\alpha})$, or $e^{\gamma}=\frac{1-\alpha}{\alpha}$, or $1/\alpha=e^{\gamma}+1$. Thus
\[\alpha=\frac{1}{1+e^\gamma}\Rightarrow 1-\alpha=\frac{e^{\gamma}}{1+e^\gamma}\Rightarrow
\log(1-\alpha)=\gamma-\log(1+e^\gamma). \]
and so
\begin{align}
\nonumber
p\left({\cal G}_j|{\cal G}^m,\gamma\right)&=\exp\{N_e\log\left(1-\alpha\right) \}\exp(-\gamma\phi(d_H({\cal G}_j,{\cal G}^m)))\\
\nonumber
&=\exp\{N_e\left[\gamma-\log(1+e^\gamma) \right] \}\exp(-\gamma\phi(d_H({\cal G}_j,{\cal G}^m)))\\
&=\frac{\exp\{N_e\gamma\}}{(1+e^\gamma)^{N_e}}\exp(-\gamma\phi(d_H({\cal G}_j,{\cal G}^m)))
\end{align}
Thus the  CER is a member of the spherical network family (Definition 3.4). We find with the abbreviations Centered Erd\H{o}s--Renyi (CER), Spherical Network Family (SNF), Unimodal network Distribution based on location (UDL), as well as unimodal network distribution based on location and scale (UDLS), that there is a natural nestedness
\begin{equation}
{\mathrm{CER}}\in {\mathrm{SNF}} \subset {\mathrm{UDLS}}\subset {\mathrm{UDL}}.
\end{equation}
From these, the only inclusion that requires some qualification is ${\mathrm{SNF}} \subset {\mathrm{UDLS}}$. This point is addressed in Proposition \ref{Prop:Spher}.
\end{proof}

Proof of Proposition \ref{Prop:Spher}
\begin{proof}
First, we will prove that Spherical Network Family defined on $\left\{ \mathcal{G}_{[N]} \right\}$  fulfills Definition \ref{def:RGDis}. We start with property (2). Let $(\mathcal{G}^m ,\gamma)$ be pre-specified. Let $\mathcal{G}_1$ and $\mathcal{G}_2$ be such that:
\begin{equation}
\label{DG1}
d_G(\mathcal{G}_1,\mathcal{G}^m)>d_G(\mathcal{G}_2,\mathcal{G}^m),
\end{equation}
or $\mathcal{G}_1$ is further from $\mathcal{G}^m$ than
$\mathcal{G}_2$. 

The Spherical Network Family is a Boltzmann distribution on the space of graphs. Boltzmann (or Gibbs) distributions take the form of
\begin{equation}
\label{sphere}
p(\mathcal{G}\mid \mathcal{G}^m,\gamma)=Z^{-1}( \mathcal{G}^m,\gamma)
\exp\left\{-\gamma\phi(d_G(\mathcal{G},\mathcal{G}^m))\right\},
\end{equation}
where $Z^{-1}( \mathcal{G}^m,\gamma)$ is a normalizing constant ($Z( \mathcal{G}^m,\gamma)$ is the partition function) and $\gamma>0$. We note that
maximum entropy is approached as $\gamma\to 0$ (when the distribution becomes uniform) and the distribution degenerates to a point mass at $\mathcal{G}^m$ as $\gamma \to \infty$. This means that, in the limit cases, the model has the desired behaviour. 

The relationship of~\eqref{DG1} occurs if and only if
\begin{eqnarray}
  &   &\phi(d_G(\mathcal{G}_1,\mathcal{G}^m))  >   \phi(d_G(\mathcal{G}_2,\mathcal{G}^m) \nonumber\\
  & \iff  &-\gamma\phi(d_G(\mathcal{G}_1,\mathcal{G}^m)) <   -\gamma\phi(d_G(\mathcal{G}_2,\mathcal{G}^m)) \nonumber \\
  & \iff  &\exp\left\{-\gamma\phi(d_G(\mathcal{G}_1,\mathcal{G}^m))\right\} <  \exp\left\{-\gamma\phi(d_G(\mathcal{G}_2,\mathcal{G}^m))\right\} \nonumber\\
  & \iff  &p(\mathcal{G}_1\mid \mathcal{G}^m,\gamma) < p(\mathcal{G}_2\mid \mathcal{G}^m,\gamma),\nonumber
\end{eqnarray}
as $\exp\left\{ \cdot \right\}$ is a strictly increasing function and the constants of proportionality cancel. The argument for the equality case follows \emph{mutatis mutandis}. 

We now proceed to prove Definition \ref{def:RGDis}(1) holds: Let $\mathcal{G}_1 \neq \mathcal{G}^m$, then $d_G(\mathcal{G}_1,\mathcal{G}^m)  > 0$ since $d_G(\cdot,\cdot)$ is a metric. It follows that
$d_G(\mathcal{G}_1,\mathcal{G}^m)  >  d_G(\mathcal{G}^m,\mathcal{G}^m)$, since $d_G(\mathcal{G}^m,\mathcal{G}^m)=0$. We conclude that:
\begin{displaymath}
p(\mathcal{G}_1\mid \mathcal{G}^m,\gamma)< p(\mathcal{G}^m\mid \mathcal{G}^m,\gamma),
\end{displaymath}
which shows that $\mathcal{G}^m$ is a mode and that the mode is unique.

Second, we will prove that Spherical Network Family defined on $\left\{ \mathcal{G}_{[N]} \right\}$  fulfills Definition \ref{def:RGDisEnt}, that is, that the parameter $\gamma$ controls the entropy of the distribution. This is to achieve the analogy of a Gaussian distribution. The parameter $\gamma$ is indexing the family of distributions that we study. Any member of the family is characterised by its entropy. We want the indexing to be such that if  $\gamma$ increases, the entropy decreases, and the distribution becomes better concentrated. 

To be able to understand the spherical network family, we shall study the so-called Boltzmann or Gibbs distributions. Boltzmann distributions are common in statistical mechanics, and further discussed in \citep{MezaMont}, Section 2.2. The SNF falls in this class, as is directly apparent from Definition~\ref{def:ENF}.

Our next objective is to investigate the entropy of the spherical network family. We first introduce some notation. Let:
\begin{displaymath}
K(\mathcal{G},\mathcal{G}^m,\gamma)=\exp\left\{ -\gamma \phi\left[d(\mathcal{G},\mathcal{G}^m)\right] \right\};
\end{displaymath}
this function is decreasing in $\gamma$. Let:
\begin{displaymath}
Z(\mathcal{G}^m,\gamma)=\sum_{\mathcal{G}\in \left\{\mathcal{G}_{[N]}\right\}} \exp\left\{ -\gamma \phi\left[d(\mathcal{G},\mathcal{G}^m)\right] \right\}.
\end{displaymath}
denote the partition function. It follows that both $Z(\mathcal{G}^m,\gamma)$ and $\log\left\{  Z(\mathcal{G}^m,\gamma) \right\}$ are decreasing in $\gamma$, as $Z(\mathcal{G}^m,\gamma)$ is the sum of decreasing functions.\\

The entropy of a member of the spherical network family is given by:
\begin{eqnarray}\label{Eq:CompEntropy}
H_{\text{SNF}} & = & -\sum_{\mathcal{G}\in \left\{\mathcal{G}_{[N]}\right\}} p(\mathcal{G}\mid \mathcal{G}^m,\gamma) \log\left\{  p(\mathcal{G}\mid \mathcal{G}^m,\gamma)  \right\}\nonumber \\
& = & -\sum_{\mathcal{G}\in \left\{\mathcal{G}_{[N]}\right\}} p(\mathcal{G}\mid \mathcal{G}^m,\gamma) \log\left\{  \frac{1}{Z(\mathcal{G}^m,\gamma)}\exp\left\{ -\gamma \phi\left[d(\mathcal{G},\mathcal{G}^m) \right]\right\}  \right\}\nonumber \\
 & = & -\sum_{\mathcal{G}\in \left\{\mathcal{G}_{[N]}\right\}} p(\mathcal{G}\mid \mathcal{G}^m,\gamma)\left[ \log\left\{ K(\mathcal{G},\mathcal{G}^m,\gamma)\right\} - \log\left\{ Z(\mathcal{G}^m,\gamma)\right\}\right]\nonumber \\
 & = & \log\left\{ Z(\mathcal{G}^m,\gamma)\right\} + \gamma \sum_{\mathcal{G}\in \left\{\mathcal{G}_{[N]}\right\}} \phi\left[d(\mathcal{G},\mathcal{G}^m)\right] p(\mathcal{G}\mid \mathcal{G}^m,\gamma) \nonumber\\
  & = &  \log\left\{ Z(\mathcal{G}^m,\gamma)\right\} +\gamma \times \mathbb{E}\left\{\phi\left[ d(\mathcal{G},\mathcal{G}^m) \right] \right\}.
\end{eqnarray}
The next task is to determine under which conditions, the entropy of the spherical network family is decreasing in $\gamma$. We introduce some additional notation. Let $F(\gamma)$ denote the free energy:
\begin{displaymath}
F(\gamma)=-\frac{1}{\gamma}\log\left\{  Z(\mathcal{G}^m,\gamma) \right\}.
\end{displaymath}
The following identity is a standard result for the Boltzmann distribution \cite{MezaMont}, p. 25-29.
\begin{equation}
\mathbb{E}\left\{\phi \left[d(\mathcal{G},\mathcal{G}^m)\right]  \right\}  =  \frac{\partial}{\partial \gamma}\left[\gamma F(\gamma)    \right],\nonumber
\end{equation}
which implies
\begin{equation}\label{Eq:FreeEnt}
\mathbb{E}\left\{\phi\left[ d(\mathcal{G},\mathcal{G}^m)\right]  \right\} =  -\frac{\partial}{\partial \gamma}\log\left\{  Z(\mathcal{G}^m,\gamma) \right\}. 
\end{equation}
We now compute the derivative of the entropy of the spherical network family with respect to $\gamma$. From Equation \ref{Eq:CompEntropy}, we have:
\begin{eqnarray}
\frac{\partial}{\partial \gamma} H_{\text{SNF}} & = & \frac{\partial}{\partial \gamma}\log\left\{ Z(\mathcal{G}^m,\gamma)\right\} +\frac{\partial}{\partial \gamma}\left[ \gamma \times \mathbb{E}\left\{ \phi\left[d(\mathcal{G},\mathcal{G}^m) \right] \right\} \right] \nonumber\\
 & = & -\mathbb{E}\left\{ \phi\left[d(\mathcal{G},\mathcal{G}^m) \right] \right\} + \mathbb{E}\left\{\phi \left[ d(\mathcal{G},\mathcal{G}^m) \right] \right\} +\gamma \times \frac{\partial}{\partial \gamma}\mathbb{E}\left\{ \phi\left[d(\mathcal{G},\mathcal{G}^m)\right]\right\}\label{Eq:DerEntropy}\\
 & = & \gamma \times \frac{\partial}{\partial \gamma}\mathbb{E}\left\{ \phi\left[d(\mathcal{G},\mathcal{G}^m)\right]\right\}, \label{Eq:DerEntropyExpect}
\end{eqnarray}
where the equality in Expression \ref{Eq:DerEntropy} follows from applying Equation \ref{Eq:FreeEnt}. By definition, 
\begin{equation}\label{Eq:ExpectEnergyF}
\mathbb{E}\left\{ \phi\left[d(\mathcal{G},\mathcal{G}^m)\right]\right\}=\frac{1}{Z(\mathcal{G}^m,\gamma)} \sum_{\mathcal{G}\in \left\{\mathcal{G}_{[N]}\right\}} \phi\left[d(\mathcal{G},\mathcal{G}^m)\right]\exp\left\{  -\gamma \phi\left[ d(\mathcal{G},\mathcal{G}^m) \right]\right\}.
\end{equation}
From Equation \ref{Eq:ExpectEnergyF}, we obtain that $\frac{\partial\mathbb{E}\left\{\phi\left[ d(\mathcal{G},\mathcal{G}^m)\right]\right\}}{\partial \gamma}$ is equal to
\begin{displaymath}
    \frac{-\frac{\partial Z(\mathcal{G}^m,\gamma)}{\partial \gamma}}{Z(\mathcal{G}^m,\gamma)^2}\sum_{\mathcal{G}\in \left\{\mathcal{G}_{[N]}\right\}} \phi\left[d(\mathcal{G},\mathcal{G}^m)\right]K(\mathcal{G},\mathcal{G}^m,\gamma)+\frac{1}{Z(\mathcal{G}^m,\gamma)}\sum_{\mathcal{G}\in \left\{\mathcal{G}_{[N]}\right\}} (-1)\phi^2\left[d(\mathcal{G},\mathcal{G}^m)\right]K(\mathcal{G},\mathcal{G}^m,\gamma).
\end{displaymath}
Therefore
\begin{eqnarray}
\frac{\partial\mathbb{E}\left\{ \phi\left[d(\mathcal{G},\mathcal{G}^m)\right]\right\}}{\partial \gamma} & = &
\frac{-1}{Z(\mathcal{G}^m,\gamma)}\sum_{\mathcal{G}\in \left\{\mathcal{G}_{[N]}\right\}} \left( \phi\left[d(\mathcal{G},\mathcal{G}^m)\right]+ \frac{\frac{\partial Z(\mathcal{G}^m,\gamma)}{\partial \gamma}}{Z(\mathcal{G}^m,\gamma)}   \right)\phi\left[d(\mathcal{G},\mathcal{G}^m)\right]K(\mathcal{G},\mathcal{G}^m,\gamma)\nonumber\\
& = & - \mathbb{E}\left\{ \left(\phi\left[ d(\mathcal{G},\mathcal{G}^m)\right]-\mathbb{E}\left\{\phi\left[ d(\mathcal{G},\mathcal{G}^m)\right] \right\} \right) \phi\left[d(\mathcal{G},\mathcal{G}^m)\right]\right\}\label{Eq:AppBoltz}\\
 & = & -\mathbb{V}\text{ar}\left\{\phi\left[d(\mathcal{G},\mathcal{G}^m)\right]\right\} <0. \label{Eq:VarEner}
\end{eqnarray}
Here, Equation \ref{Eq:AppBoltz} follows from Equation \ref{Eq:FreeEnt}. By applying Equation \ref{Eq:VarEner} to Equation \ref{Eq:DerEntropyExpect}, we obtain
\begin{equation}
\frac{\partial}{\partial \gamma} H_{\text{SNF}} =  -\gamma \times\mathbb{V}\text{ar}\left\{\phi\left[d(\mathcal{G},\mathcal{G}^m)\right]\right\} <0. 
\end{equation}
It follows that the spherical network family is parametrized in terms of $\mathcal{G}^m$, which is the mode of the distribution, and $\gamma$, which is a monotone function of the entropy as long as $p(\cdot\mid \mathcal{G}^m,\gamma)$ is not a point mass. We also have that each of these parameters can be specified without constrains imposed by the other, therefore Definition \ref{def:RGDisEnt}  is satisfied. 
\end{proof}
For further understanding of the behaviour of the entropy as a function of $\alpha$ we refer to the Proof of Proposition 3.1, earlier in this appendix.

Proof of Proposition \ref{Prop:CheckPropp}
\begin{proof}
Let $N\geq 2$. This means that $A_{\mathcal{G}^m}$ has at least one entry in its upper-triangular section, outside of the diagonal; this also means that there is at least one graph $\mathcal{G}_1\in $, such that $\mathcal{G}_1\neq \mathcal{G}^m$. For both, the Hamming distance and the diffusion distance, we have:
\begin{displaymath}
d(\mathcal{G}^m,\mathcal{G}^m)=0\qquad \text{and} \qquad 0<d(\mathcal{G}^m,\mathcal{G}_1)<\infty,
\end{displaymath}
and
\begin{displaymath}
p(\mathcal{G}^m\mid \mathcal{G}^m,\gamma)>p(\mathcal{G}_1\mid \mathcal{G}^m,\gamma)>0,
\end{displaymath}
for both models. Remember that, for the CER, $\gamma$ is a function of $\alpha\in (0,1)$. It follows, that $\mathbb{E}\left\{ d(\mathcal{G},\mathcal{G}^m)\right\}>0$ and therefore $\mathbb{V}\text{ar}\left\{d(\mathcal{G},\mathcal{G}^m)\right\} >0$.
\end{proof}

Proof of Proposition \ref{Prop:Frechet}
\begin{proof}
For $n$ observations $\left\{y_1, \dots y_n \right\}$ from $\left\{ \mathcal{G}_{[N]} \right\}$  The sample Frech\'{e}t mean is given by:
\begin{equation}\label{Eq:SampleFrech}
\hat{\psi}_n=\arg \min_{\psi \in \mathcal{Y}} \frac{1}{n} \sum_{i=1}^n d(y_i,\psi)^2;
\end{equation} 
see Eqn~\ref{Eq:FrechMean}.  By sampling elements of $\left\{ \mathcal{G}_{[N]} \right\}$ via a distribution with full support, each individual expectation in Eqn~\ref{Eq:FrechMean} is the limit of the corresponding sample mean. These sample means are part of the computation in Equation \ref{Eq:SampleFrech}.

Let $\psi$ be an element of $\left\{ \mathcal{G}_{[N]} \right\}$ such that $\mathbb{E}(d^2(Y,\psi))$ is finite. We only need to considers those $\psi\in \left\{ \mathcal{G}_{[N]} \right\}$ for which the expectation is finite, since, both, the Frech\'{e}t mean and the sample Frech\'{e}t mean are obtained by computing the minimum. Note that, if $\mathbb{E}(d^2(Y,\psi))=\infty$  for all $\psi \in \left\{ \mathcal{G}_{[N]} \right\}$, the assumption of a unique Frech\'{e}t mean would not be fullfilled. Let $n$ be the number of observed networks, note that, as $n \to \infty$:
\begin{displaymath}
\frac{1}{n} \sum_{i=1}^n d(y_i,\psi)^2 \to \mathbb{E}(d^2(Y,\psi)), \quad \text{a.s. for all}\quad \psi \in \left\{ \mathcal{G}_{[N]} \right\} \quad \text{such that} \quad\mathbb{E}(d^2(Y,\psi))<\infty,
\end{displaymath}
by the Strong Law of Large Numbers, as $d^2(Y,\psi)$ is a scalar.  Here, $\psi$ is fixed and the $y_i$'s are random. This argument tells us that each individual expectation is the limit of the corresponding sample mean, the next part of the argument is to prove that all the expectations in Equation Eqn~\ref{Eq:FrechMean} can be estimated simultaneously with enough accuracy (encoded by $\epsilon$), so the minimization entailed by \ref{Eq:SampleFrech} can be carried out without errors with high probability.

For every $\psi \in \left\{ \mathcal{G}_{[N]} \right\}$ and all $\epsilon >0$ and $\delta \in (0,1)$, there exists $N(\epsilon, \delta) \in \mathbb{N}$ such that:
\begin{displaymath}
\Pr \left\{ \left\|\frac{1}{n} \sum_{i=1}^n d(y_i,\psi)^2 - \mathbb{E}(d^2(Y,\psi)) \right\| <\epsilon \right\}> 1-\delta,
\end{displaymath}
for all $n>N(\epsilon,\delta)$. Since $\left\{ \mathcal{G}_{[N]} \right\}$ is finite, this is true for 
\begin{displaymath}
\epsilon < \frac{1}{2}\min_{(\psi_1,\psi_2)\in \left\{ \mathcal{G}_{[N]} \right\}\times \left\{ \mathcal{G}_{[N]} \right\} }\left\|  \mathbb{E}(d^2(Y,\psi_1)) - \mathbb{E}(d^2(Y,\psi_2)) \right\|,
\end{displaymath}
and $0<\delta<1$ pre-specified, for all $\psi \in \left\{ \mathcal{G}_{[N]} \right\}$. This is, we can make the noise of the sampled means smaller than any pairwise difference of the expectations, for $\psi_1,\psi_2$ in $ \left\{ \mathcal{G}_{[N]} \right\}$ .  Since $\left\{ \mathcal{G}_{[N]} \right\}$ is finite, we can make $N(\epsilon,\delta)$ constant with respect to $\psi \in \left\{ \mathcal{G}_{[N]} \right\}$ by taking the maximum. This means, that for all $n>N(\epsilon,\delta)$, 
\begin{displaymath}
\Pr \left\{\hat{\psi}_n=\psi^m\right\}=1-\delta,
\end{displaymath} 
where $\psi^m$ is the Fr\'{e}chet mean, which we assumed to be unique. This is because, for $n \geq N(\epsilon, \delta)$, each expectation can be approximated with enough precision  that the optimization can be carried out without error with high probability.
\end{proof}

Proof of Proposition \ref{Prop:SER}
\begin{proof}
The objective is to prove that, for the CER with parameters $(\mathcal{G}^m,\gamma)$, the mode $\mathcal{G}^m$ coincides with the Fr\'{e}chet mean. We divide the proof into two parts: for the first part, we provide a condition for when the inner product between a $s-$dimensional vector $a$ with non-negative entries and a pmf $w$ is minimized, where the optimization is taken over all permutations of indices for the entries of $a$, \emph{i.e.,} $\left\{1,2,\dots,s \right\}$; for the second part, we prove that taking the expectation of the distances with respect to a graph $\mathcal{G}_k$ for the CER$(\mathcal{G}^m,\alpha)$ is an example of the setup described in the first part, even more, the permutations of the vector of distances involved in computing the expectation, correspond to different choices for $\mathcal{G}_k$. We conclude by proving that the minimum of the expectation is attained at $\mathcal{G}^m$.

\underline{Part 1}\\
Let $a$ and $w$ be vectors with $s$ entries, in addition, let $w$ be such that $w_i>0$ and $\sum w_i=1$ and $a$ be such that $a_i\geq 0$. Let $\sigma$ be a permutation of $\left\{1,2,\dots,s \right\}$ such that $w_{\sigma[i]}\geq w_{\sigma[j]}$ for every pair $\left\{i,j\right\}$ with $i<j$. Let $\tau$ be a permutation of $\left\{1,2,\dots,s \right\}$ such that $a_{\tau[i]}\leq a_{\tau[j]}$ for every pair $\left\{i,j\right\}$ with $i<j$. Therefore, $\tau$ fullfils:
\begin{displaymath}
\tau=\arg\min_{\varphi \in \text{Sym}(s)} \sum a_{\varphi[i]}\times w_{\sigma[i]}
\end{displaymath}
where $\text{Sym}(s)$ denotes the set of permutations over $s$ indices.\\

We proof the last statement by induction:
\begin{description}
\item[for $s=2$]:
Let us start with the case $a_{\tau[1]}< a_{\tau[2]}$ and $w_{\sigma[1]}> w_{\sigma[2]}$. Since the entries of $w$ are nonegative and add to $1$, it follows that $w_{\sigma[1]}>\frac{1}{2}> w_{\sigma[2]}$. Therefore
\begin{equation}\label{Eq:ConvexLin1}
a_{\tau[1]}\times w_{\sigma[1]}+a_{\tau[2]}\times w_{\sigma[2]} 
\end{equation}
is closer to $a_{\tau[1]}$ than it is to $a_{\tau[2]}$. If one permutes the indices of $a_{\tau}$ to obtain a new vector $a_{\tau'}$, then, it follows that
\begin{equation}\label{Eq:ConvexLin2}
a_{\tau'[1]}\times w_{\sigma[1]}+a_{\tau'[2]}\times w_{\sigma[2]} 
\end{equation}
is closer to $a_{\tau[2]}$ than it is to $a_{\tau[1]}$; one way to visualize this argument is to note that Expressions \ref{Eq:ConvexLin1} and \ref{Eq:ConvexLin2} correspond to convex linear combinations of two non-negative numbers, namely $(a_{\tau[1]},a_{\tau[2]})$ and the statements about closeness correspond to the size of the weights $(w_{\sigma[1]},w_{\sigma[2]})$ . Since  $a_{\tau[1]} <a_{\tau[2]}$, the condition is fulfilled. For the cases where either $a_{\tau[1]}=a_{\tau[2]}$ or $w_{\sigma[1]}=w_{\sigma[2]}$, it is trivial to show that the condition is fulfilled.

\item[for $s=k$]: Let us assume that the result holds for $s=k$.

\item[for $s=k+1$]: We consider two cases, which are defined in terms of the existence of fixed points of $\tau$.

Case 1. At least one entry in $a_\tau$ remains fixed.

WLOG we can assume that the entry of $a_\tau$ that remained invariant is the $(k+1)-$th. We can re-normalize the first $k$ entries of $w_{\sigma}$ by making
\begin{displaymath}
w'_{\sigma[i]}=\frac{w_{\sigma[i]}}{1-w_{\sigma[k+1]}}.
\end{displaymath}
Since $w'_{\sigma[i]}\geq w'_{\sigma[j]}$ for all $1\leq i <j \leq k$, we can apply the hypothesis of induction to the first $k$ entries of $a_\tau$, $a_{\tau'}$ and $w_{\sigma}$ to obtain  
\begin{equation}\label{Eq:IndexIneq}
\sum_{i=1}^k a_{\tau[i]}\times w'_{\sigma[i]} \leq \sum_{i=1}^k a_{\tau'[i]}\times w'_{\sigma[i]}.
\end{equation}
Equation \ref{Eq:IndexIneq} is valid for any permutation $\tau'$ that leaves the $k+1$ entry unchanged when compared to $\tau$. Therefore
\begin{displaymath}
\sum_{i=1}^k a_{\tau[i]}\times w_{\sigma[i]} \leq \sum_{i=1}^k a_{\tau'[i]}\times w_{\sigma[i]}
\end{displaymath}
since we only need to multiply $w'_\sigma$ by a positive constant, namely $1-w_{\sigma[k+1]}$. We assumed $a_{\tau[k+1]}=a_{\tau'[k+1]}$, it follows that
\begin{displaymath}
\sum_{i=1}^{k+1} a_{\tau[i]}\times w_{\sigma[i]} \leq \sum_{i=1}^{k+1} a_{\tau'[i]}\times w_{\sigma[i]}.
\end{displaymath}
We conclude that $\tau$ minimises $\sum a_{\tau[i]}\times w_{\sigma[i]}$ for all the permutations that leave at least one entry unchanged with respect to $\tau$.

Case 2. We now consider the case where no entry of $a_\tau$ was left invariant by a new indexing $\tau'$. Let $a_{\tau^\star}$ be the vector that results from permuting two entries of $a_{\tau'}$ so the $(k+1)-$th entry of $a_{\tau^\star}$ coincides with the $(k+1)-$th entry of $a_{\tau}$. By applying an argument analogous to the one made for $k=2$, we obtain:
\begin{displaymath}
\sum_{i=1}^{k+1} a_{\tau^\star[i]}\times w_{\sigma[i]} \leq \sum_{i=1}^{k+1} a_{\tau'[i]}\times w_{\sigma[i]}.
\end{displaymath}
Now, since at least the $(k+1)-$th entry of $a_{\tau^\star}$ coincides with the $(k+1)-$th entry of $a_{\tau}$, we have
\begin{displaymath}
\sum_{i=1}^{k+1} a_{\tau[i]}\times w_{\sigma[i]} \leq \sum_{i=1}^{k+1} a_{\tau\star[i]}\times w_{\sigma[i]},
\end{displaymath}
therefore the conclusion is valid for this case also.
\end{description}

\underline{Part 2}\\
We start by proving that, given $N\in \mathbb{N}$, the number of graphs in $\left\{ \mathcal{G}_{[N]}\right\}$ such that $d_H(\mathcal{G}, \mathcal{G}^m)=h$, 
where $h\in\left\{ 1,2,\dots, N_e\right\}$, is constant with respect to $\mathcal{G}^m$. Having $d_H(\mathcal{G}, \mathcal{G}^m)=h$ implies that $h$ entries of the adjacency matrix of $\mathcal{G}^m$ were modified. This is equivalent from choosing $h$ entries from the upper triangular of the adjacency matrix of $\mathcal{G}^m$. Since the graphs are labelled, the number of graphs such that $d_H(\mathcal{G}, \mathcal{G}^m)=h$ is $\binom{N_e}{h}$, which is constant with respect to $\mathcal{G}^m$. The same argument can be made for every value of $h\in\left\{1,2,\dots,N_e\right\}$. \\

The expectation $\mathbb{E}\left\{ d_H^2(\mathcal{G},\mathcal{G}_k)  \right\}$ for a CER with parameters $(\mathcal{G}^m,\alpha)$ can be computed as follows:
\begin{displaymath}
\mathbb{E}\left\{ d_H^2(\mathcal{G},\mathcal{G}_k)  \right\}=\sum_{i\in I} d_H^2(\mathcal{G}_i,\mathcal{G}_k)\times p(\mathcal{G}_i\mid \mathcal{G}^m,\alpha),
\end{displaymath}
where $I$ is an indexing for $\left\{ \mathcal{G}_{[N]}\right\}$ such that $p(\mathcal{G}_i\mid \mathcal{G}^m,\alpha)\geq p(\mathcal{G}_j\mid \mathcal{G}^m,\alpha)$ for $i<j$. Given the fact that the number of graphs in $\left\{ \mathcal{G}_{[N]}\right\}$ that fulfill $d_H(\mathcal{G}, \mathcal{G}^m)=h$ is constant with respect to $\mathcal{G}^m$ for every $h\in\left\{1,2,\dots,N_e\right\}$, the vector 
\begin{displaymath}
\left(d_H^2(\mathcal{G}_1,\mathcal{G}_k),d_H^2(\mathcal{G}_2,\mathcal{G}_k),\dots, d_H^2(\mathcal{G}_{|\left\{ \mathcal{G}_{[N]}\right\}|},\mathcal{G}_k)\right)
\end{displaymath}
is obtained from permuting the entries from  
\begin{displaymath}
\left(d_H^2(\mathcal{G}_1,\mathcal{G}^m),d_H^2(\mathcal{G}_2,\mathcal{G}^m),\dots, d_H^2(\mathcal{G}_{|\left\{ \mathcal{G}_{[N]}\right\}|},\mathcal{G}^m)\right),
\end{displaymath}
Now:
\begin{equation}\label{Eq:Expect}
\mathbb{E}\left\{ d_H^2(\mathcal{G},\mathcal{G}^m)  \right\}=\sum_{i\in I} d_H^2(\mathcal{G}_i,\mathcal{G}^m)\times p(\mathcal{G}_i\mid \mathcal{G}^m,\alpha).
\end{equation}
The vectors on the right side of Equation \ref{Eq:Expect} and the indexing $I$ fulfill the assumptions of Part 1 (see proof of Proposition 3.1). This implies that $\mathcal{G}^m$ is the Fr\'{e}chet mean for the CER with parameters $(\mathcal{G}^m,\alpha)$.

\end{proof}

\section{Diagnostics for Bayesian Models}

Posterior predictive checks (\cite{GelMeng}) are based on following the intuition: if the model assumptions are reasonable, the observed value of a statistic should, with low probability, be extreme with respect to the predictive distribution for that statistic. One way to translate this intuition to our context is the following: Let $\eta^{(0)}$ be a one-dimensional summary of the observed networks $\left\{  \mathcal{G}_1,\mathcal{G}_2, \dots ,\mathcal{G}_n \right\}$ (\emph{e.g.}, the average diameter, the average number of communities). Obtain $K$ Monte Carlo data sets $\left\{  \mathcal{G}^{(i)}_1,\mathcal{G}^{(i)}_2, \dots ,\mathcal{G}^{(i)}_n \right\}$, $1\leq i \leq K$,  from the posterior predictive distribution. For each of these data sets, we compute $\eta^{(i)}$, a realisation of the predictive distribution of the one-dimensional summary, $1\leq i \leq K$. If $\eta^{(0)}$ is extreme with respect to the Monte Carlo predictive distribution implied by $\left\{ \eta^{(i)}  \right\}_{1\leq i \leq K}$, then we can regard this as evidence for lack of fit.

The Bayesian $\chi^2$ was proposed by \cite{JohnV} and it is based on the following rationale: Each sample from the posterior $(\mathcal{G}^{m,(i)}, \gamma^{(i)})$ entails the distribution of a univariate summary $Y$, \emph{i.e.},
\begin{displaymath}
(\mathcal{G}^{m,(i)}, \gamma^{(i)})\to F_{Y}(\cdot \mid \mathcal{G}^{m,(i)}, \gamma^{(i)}).
\end{displaymath}
In the context of multivariate modelling for networks, such summary is a descriptive statistic that can be computed efficiently, \emph{e.g.}, the mean of the degree distribution. Let $(y_{1},y_{2},\dots,y_{n})$ be the observed values for this summary, with $\mathcal{G}_s\to y_{s}$. Given a partition 
\begin{displaymath}
0=a_0<a_1< \dots <a_{D-1}<a_{D}=1,
\end{displaymath}
of the interval $[0,1)$, we can compute the counts
\begin{displaymath}
C^{(i)}_k=\sum_{j=1}^n \mathbb{I}_{[a_{k-1},a_k)}(F_Y(y_j \mid \mathcal{G}^{m,(i)}, \gamma^{(i)})),
\end{displaymath}
for $k\in \left\{ 1,2,\dots, D  \right\}$. Let $p_k=a_k-a_{k-1}$, then
\begin{displaymath}
R^{B}( \mathcal{G}^{m,(i)}, \gamma^{(i)})=\sum_{k=1}^D \left( \frac{C^{(i)}_k-np_k}{\sqrt{np_k}} \right)^2
\end{displaymath}
 measures the discrepancy between the observed and expected counts for the bins $[a_{k-1},a_k)$, $k\in \left\{ 1,2,\dots, D  \right\}$. Goodness-of-fit is assessed via q\slash q plots of $R^{B}(\cdot)$ with respect to a $\chi^2$ with $D-1$ degrees of freedom.

\end{document}